\documentclass[aps,prl,twocolumn,superscriptaddress,notitlepage,footinbib,floatfix,10pt]{revtex4-2}
\usepackage{graphicx}
\usepackage{dcolumn}
\usepackage{indentfirst}
\usepackage[top=2.5cm,left=2.0cm,right=2.0cm,bottom=2.5cm]{geometry}

\usepackage{array}
\usepackage{physics}
\usepackage{amsfonts}
\usepackage{amsmath}
\usepackage{amssymb}
\usepackage{comment}
\usepackage{bm}
\usepackage{hyperref}
\usepackage{multirow}
\usepackage{xcolor}
\usepackage[english]{babel}

\begin{document}
\title{Elucidating the Inter-system Crossing of the Nitrogen-Vacancy Center up to Megabar Pressures}

\author{Benchen Huang}
\thanks{These authors contributed equally to this work.}
\affiliation{Department of Chemistry, University of Chicago, Chicago, IL 60637, USA}%

\author{Srinivas V. Mandyam}
\thanks{These authors contributed equally to this work.}
\affiliation{Department of Physics, Harvard University, Cambridge, MA 02135, USA}%

\author{Weijie Wu}
\affiliation{Department of Physics, Harvard University, Cambridge, MA 02135, USA}%

\author{Bryce Kobrin}
\affiliation{Department of Physics, University of California, Berkeley, CA 94720, USA}%
\affiliation{Materials Science Division, Lawrence Berkeley National Laboratory, Berkeley, CA 94720, USA}%

\author{Prabudhya Bhattacharyya}
\affiliation{Department of Physics, University of California, Berkeley, CA 94720, USA}%
\affiliation{Materials Science Division, Lawrence Berkeley National Laboratory, Berkeley, CA 94720, USA}%

\author{Yu Jin}
\affiliation{Pritzker School of Molecular Engineering, University of Chicago, Chicago, IL 60637, USA}

\author{Bijuan Chen}
\affiliation{Department of Physics, Harvard University, Cambridge, MA 02135, USA}%

\author{Max Block}
\affiliation{Department of Physics, Harvard University, Cambridge, MA 02135, USA}%

\author{Esther Wang}
\affiliation{Department of Chemistry and Chemical Biology, Harvard University, Cambridge, MA 02135, USA}%

\author{Zhipan Wang}
\affiliation{Department of Physics, Harvard University, Cambridge, MA 02135, USA}%

\author{Satcher Hsieh}
\affiliation{Department of Physics, University of California, Berkeley, CA 94720, USA}%
\affiliation{Materials Science Division, Lawrence Berkeley National Laboratory, Berkeley, CA 94720, USA}%

\author{Chong Zu}
\affiliation{Department of Physics, Washington University, St.~Louis, MO 63130, USA}

\author{Christopher R. Laumann}
\affiliation{Department of Physics, Boston University, Boston, MA 02215, USA}

\author{Norman Y. Yao}
\email{nyao@fas.harvard.edu}
\affiliation{Department of Physics, Harvard University, Cambridge, MA 02135, USA}
\affiliation{Department of Physics, University of California, Berkeley, CA 94720, USA}
\affiliation{Materials Science Division, Lawrence Berkeley National Laboratory, Berkeley, CA 94720, USA}%

\author{Giulia Galli}
\email{gagalli@uchicago.edu}
\affiliation{Department of Chemistry, University of Chicago, Chicago, IL 60637, USA}
\affiliation{Pritzker School of Molecular Engineering, University of Chicago, Chicago, IL 60637, USA}
\affiliation{Materials Science Division and Center for Molecular Engineering, Argonne National Laboratory, Lemont, IL 60439, USA}%

\date{\today}

\begin{abstract}
The integration of Nitrogen-Vacancy color centers into diamond anvil cells has opened the door to quantum sensing at megabar pressures. 
Despite a multitude of experimental demonstrations and applications ranging from quantum materials to geophysics, a detailed microscopic understanding of how stress affects the NV center remains lacking.
In this work, using a combination of first principles calculations as well as high-pressure NV experiments, we develop a complete description of the NV's optical properties under general stress conditions. 
In particular, our \emph{ab initio} calculations reveal the complex behavior of the NV's inter-system crossing rates under stresses that both preserve and break the defect's symmetry.
Crucially, our proposed framework immediately resolves a number of open questions in the field, including: (i) the microscopic origin of the observed contrast-enhancement in (111)-oriented anvils, and (ii) the surprising observation of NV contrast-inversion in certain high-pressure regimes. 
Our work lays the foundation for optimizing the performance of NV high-pressure sensors by controlling the local stress environment, and more generally, suggests that symmetry-breaking stresses can be utilized as a novel tuning knob for generic solid-state spin defects.
\end{abstract}

\maketitle

\begin{figure}[t]
    \centering
    \includegraphics[width=0.48\textwidth]{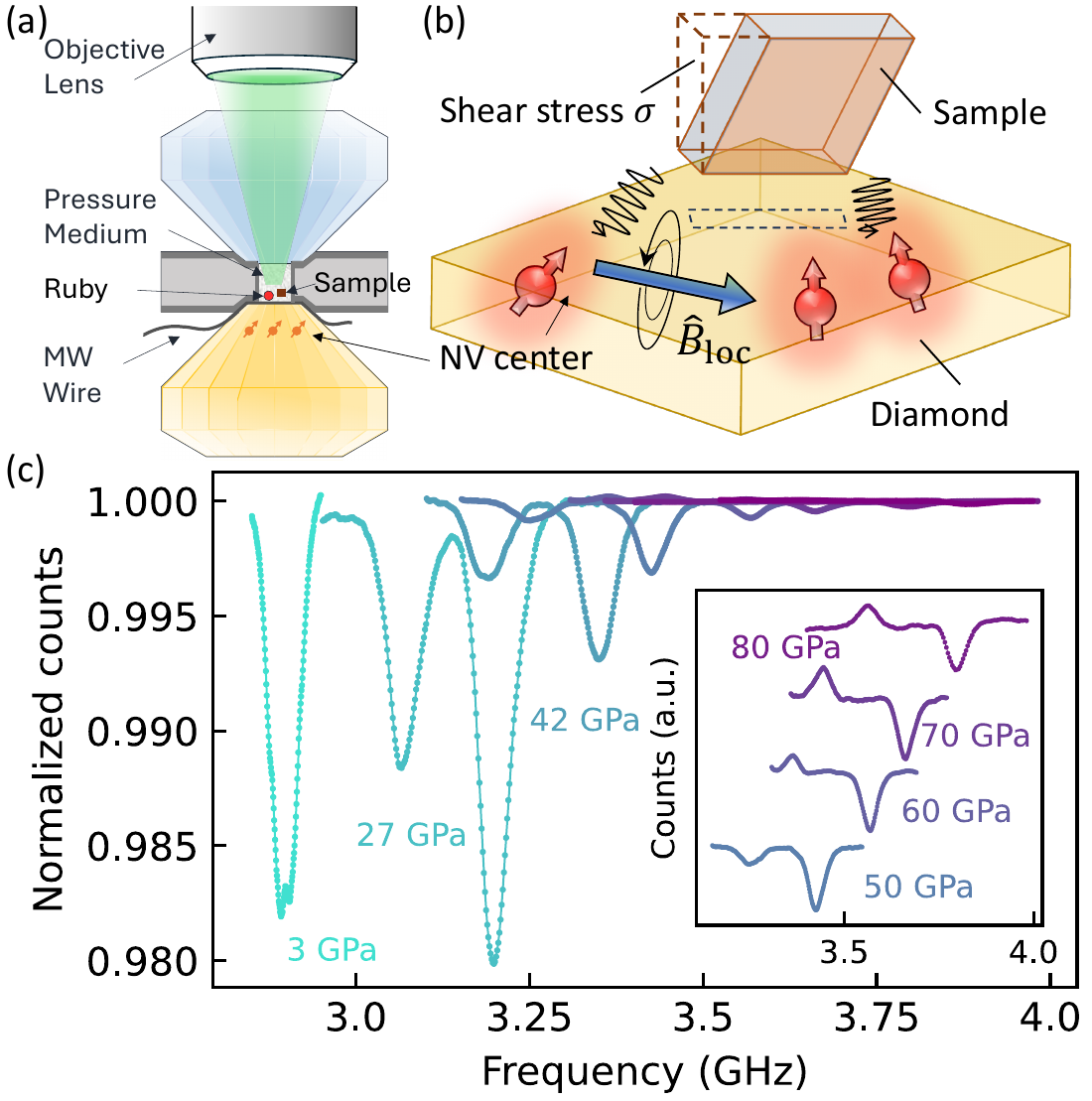}
    \caption{(a) Schematic of the diamond anvil cell (DAC) geometry. The DAC sample chamber is defined by the gasket-anvil assembly; it is loaded with the sample of interest, pressure-transmitting medium, and a ruby microsphere. A $\sim$50-nm layer of NV centers (about 1 ppm density) is embedded into the diamond anvil directly below the sample chamber. For ODMR measurements, a platinum wire is placed on the bottom culet to deliver microwaves.
    (b) Major quantum sensing applications using the NV center include magnetometry~\cite{rondin2014magnetometry} and sensing normal and shear (depicted) stresses $\mathbf{\sigma}$ in the sample~\cite{broadway2019microscopic, ho2021recent, kehayias2019imaging, suda2025gpa}.
    (c) Continuous-wave ODMR measurements of NV centers in the (100)-cut anvil exhibit a drastic reduction in contrast with increasing pressure. The dominant culet stresses have symmetry-preserving and breaking projections on all NV subgroups, thereby inducing both a blue shift, $\Pi_z$, and a splitting, $2\Pi_\perp$ in the ODMR peaks. Notably, a surprising inversion of contrast is observed on the left peak around 60 GPa, as shown in the inset.}
    \label{fig:nv_quantum_sensing}
\end{figure}

Pressure represents a powerful tuning knob for condensed matter systems, enabling access to novel physical states, ranging from record-high temperature superconductivity \cite{drozdov2019superconductivity} to exotic structural phases~\cite{levitas2018high}.
Access to megabar pressures~\cite{jayaraman1983diamond} in the laboratory is enabled by the diamond anvil cell (DAC) [Fig.~\ref{fig:nv_quantum_sensing}(a)], an apparatus consisting of two opposing diamond tips that compress a small sample within a gasketed chamber.
However, the DAC imposes severe constraints on metrology. 
Perhaps the most important is the inability to perform spatially-resolved local measurements of the physics inside the high-pressure chamber~\cite{ishizuka2000pressure, gregoryanz2002superconductivity, jackson2003magnetic}.

To this end, a tremendous amount of excitement has centered on the integration of nitrogen vacancy (NV) color centers into diamond anvil cells~\cite{hsieh2019imaging, yip2019measuring, lesik2019magnetic, dai2022optically, hilberer2023enabling, bhattacharyya2024imaging, wang2024imaging, mai2025megabar, hao2025diamond}.
By directly implanting such spin-defect sensors into the anvil tip (i.e.~culet) applying the pressure~\cite{hsieh2019imaging, yip2019measuring, lesik2019magnetic}, seminal recent experiments have demonstrated the ability to image local stresses and magnetism with sub-micron resolution~\cite{zhong2024high}.
This approach has had an almost immediate impact on our understanding of multiple families of materials under pressure, ranging from hydride~\cite{bhattacharyya2024imaging, song2025room, chen2025imaging} and nickelate~\cite{mandyam2025uncovering, liu2025evidence} superconductors to magnetic minerals~\cite{wang2024imaging}. It is now possible to resolve sub-micron scale heterogeneities in functional materials at high pressure, and to correlate them with their structural and stoichiometric sources \cite{mandyam2025uncovering}.

Despite these successes, our microscopic understanding of the NV center under pressure remains relatively nascent, with two broad sets of open questions.
First, it is generally believed that the stress environment must be carefully managed in order to enable high-pressure NV quantum sensing [Fig.~\ref{fig:nv_quantum_sensing}(b)]~\cite{batalov2009low, tetienne2012magnetic, ernst2023modeling, ernst2023temperature}.
As an example, the NV's optical contrast depends sensitively on its crystallographic orientation relative to the culet~\cite{bhattacharyya2024imaging, wang2024imaging, mai2025megabar}.
However, the underlying reasons for this sensitivity---and whether more optimal stress conditions exist---remain unclear.
Second, a multitude of experiments across a variety of conditions~\cite{hilberer2023enabling, bhattacharyya2024imaging}, have all observed the puzzling \textit{inversion} of NV contrast under pressure [Fig.~\ref{fig:nv_quantum_sensing}(c)]~\cite{bhattacharyya2024imaging}.
On one hand, this inversion complicates signal extraction from the NV center; on the other, it may offer metrological advantages of its own.
Taken together, these questions point to the importance of developing a microscopic framework for understanding and predicting the properties of NV centers in a generic stress environment.

\begin{figure}
    \centering
    \includegraphics[width=0.48\textwidth]{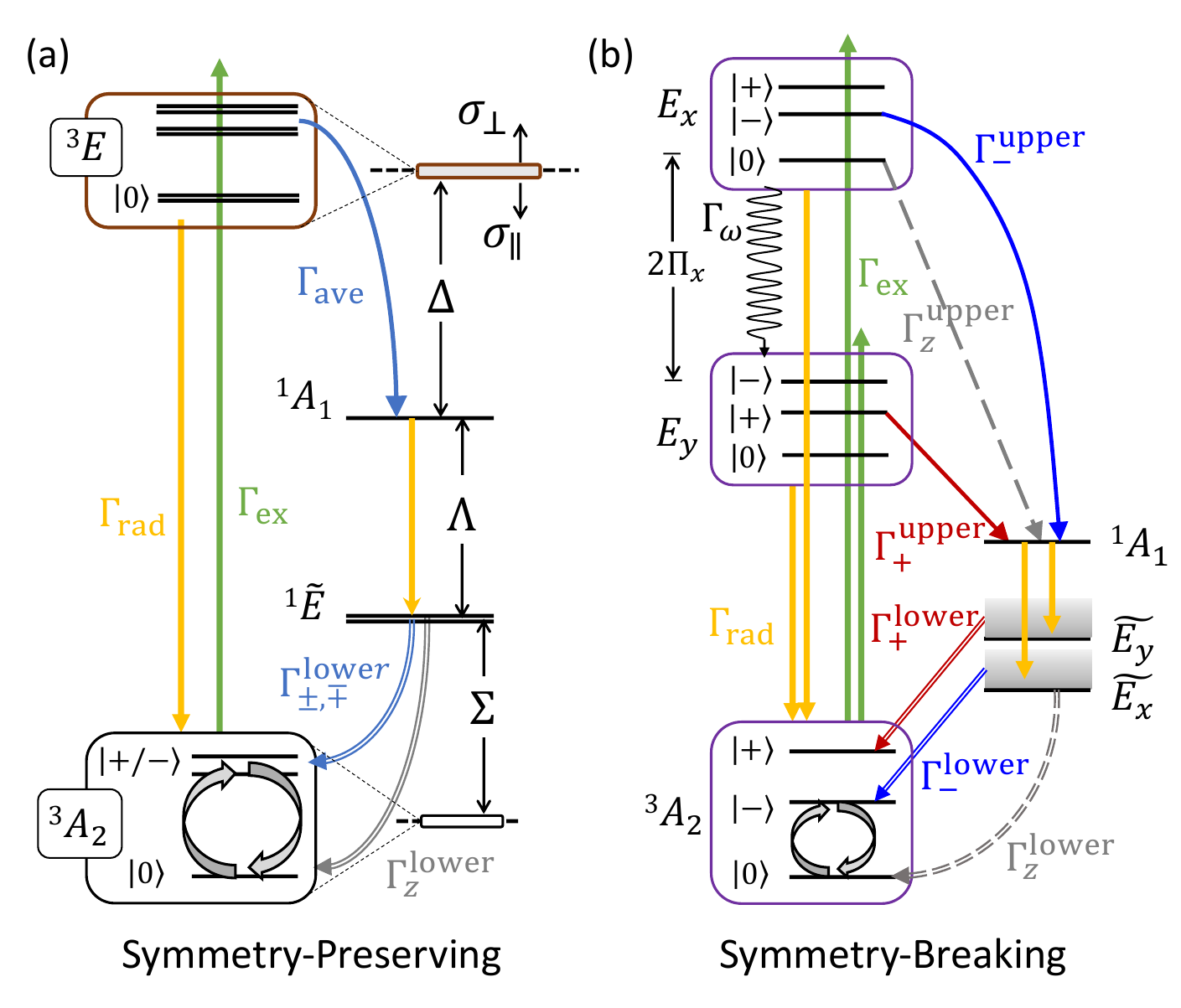}
    \caption{The negatively charged NV center's energy level diagram and its optical cycle under (a) symmetry-preserving and (b) symmetry-breaking stress (here $\Pi_x$ stress, defined in Supplemental Materials). The NV center's low-lying electronic states contain two spin triplets ${}^3\!A_2$ and ${}^3\!E$, and two spin singlets ${}^1\!A_1$ and ${}^1\!E$, with three energy gaps defined as $\Delta, \Lambda, \Sigma$ in (a). The spin-1 basis adopted here is $\left|m_s=0\right\rangle$, and $\left|m_s=\pm\right\rangle = \frac{1}{\sqrt{2}}\left(\left|m_s=1\right\rangle \pm \left|m_s=-1\right\rangle\right)$, which are the spin eigenstates under $\Pi_x$ stress. Notably, for symmetry-preserving stress, $\sigma_\perp=\frac{1}{2}\left(\sigma_{xx} + \sigma_{yy}\right)$ and $\sigma_\parallel=\sigma_{zz}$ play qualitatively different roles in shifting energy gaps.
    Symmetry-breaking stress, however, breaks the defect's point group symmetry and allows every spin state to participate in the optical cycle. The ISC rates are color coded for their spins, with blue, dark red, and grey for $\ket{m_s=-, +, 0}$, respectively in (b). The line styles denote the microscopic origin for these ISCs (see Supplemental Materials for details).}
    \label{fig:optical_cycle}
\end{figure}

\begin{figure}
    \centering
    \includegraphics[width=0.5\textwidth]{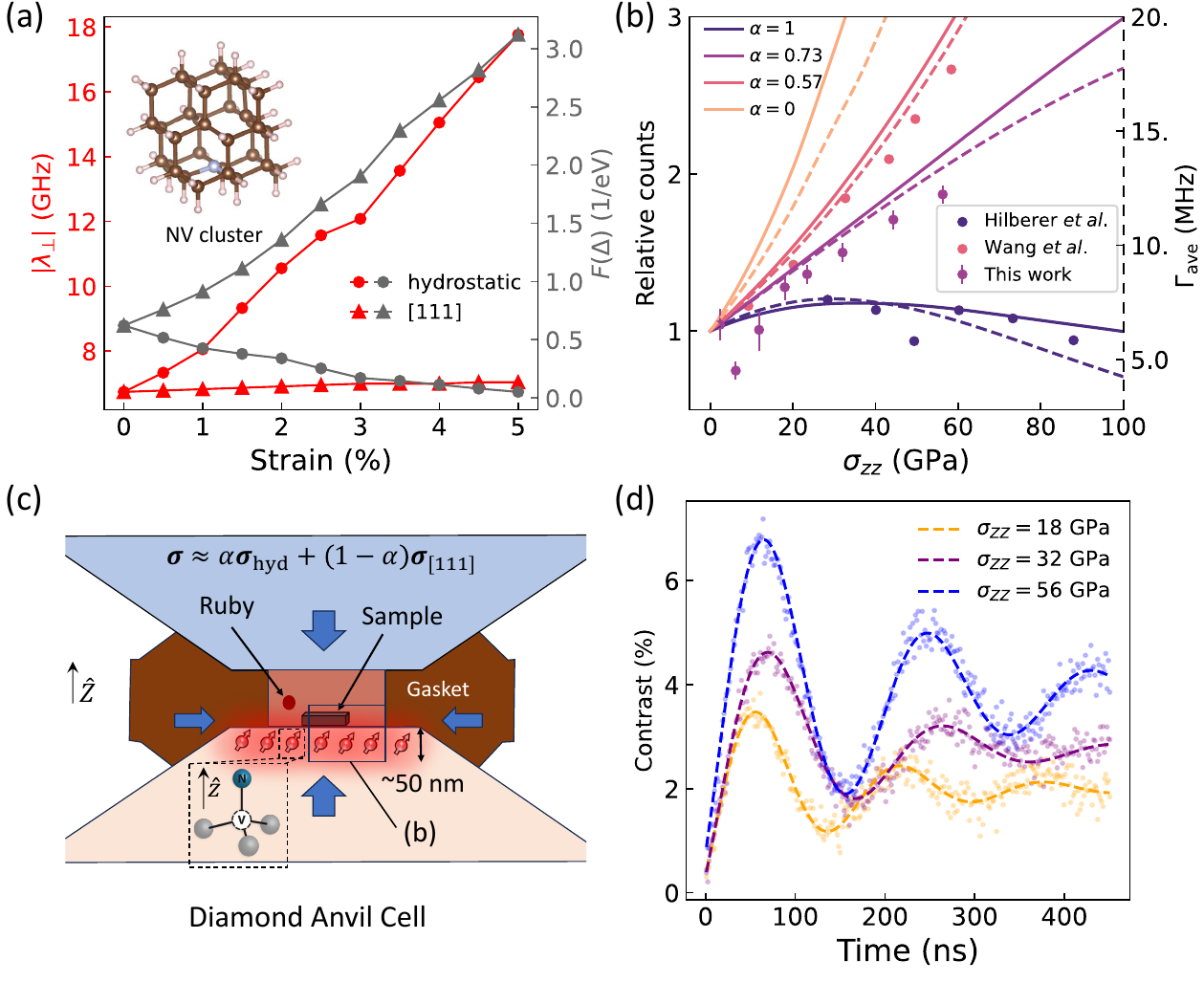}
    \caption{(a) Computations of the intermediate components of the upper ISC, i.e., transverse spin-orbit coupling $\lambda_\perp$ (red) and vibrational overlap $F(\Delta)$ between the ${}^3\!E$ and ${}^1\!A_1$ manifolds (gray) versus hydrostatic (circle) and uniaxial [111] strain (triangle). The inset shows a cluster model of the NV center, which we use as a basis for our ab initio calculations.
    (b) Upper ISC rate $\Gamma_\text{ave}$ assembled from $\lambda_\perp$ and $F(\Delta)$ (dashed), and comparison of the contrast (relative to that at the ambient condition) between simulation (solid) and experiments~\cite{hilberer2023enabling, wang2024imaging} (dots), where the color codes the hydrostaticity $\alpha$. Notably, the ISC rate exhibits a strong correlation with the relative contrast.
    (c) Schematic of a zoom-in DAC with (111)-cut diamond, and the embedded NV centers.
    (d) Rabi oscillations of the [111] NV at $\sigma_{ZZ}=18, 32, 56$ GPa respectively, from which contrast is extracted. The experimental data are fitted by damped sine waves and plotted by the orange, purple and blue dashed lines.}
    \label{fig:contrast_symmetry_preserving}
\end{figure}

\begin{figure}[t]
    \centering
    \includegraphics[width=0.45\textwidth]{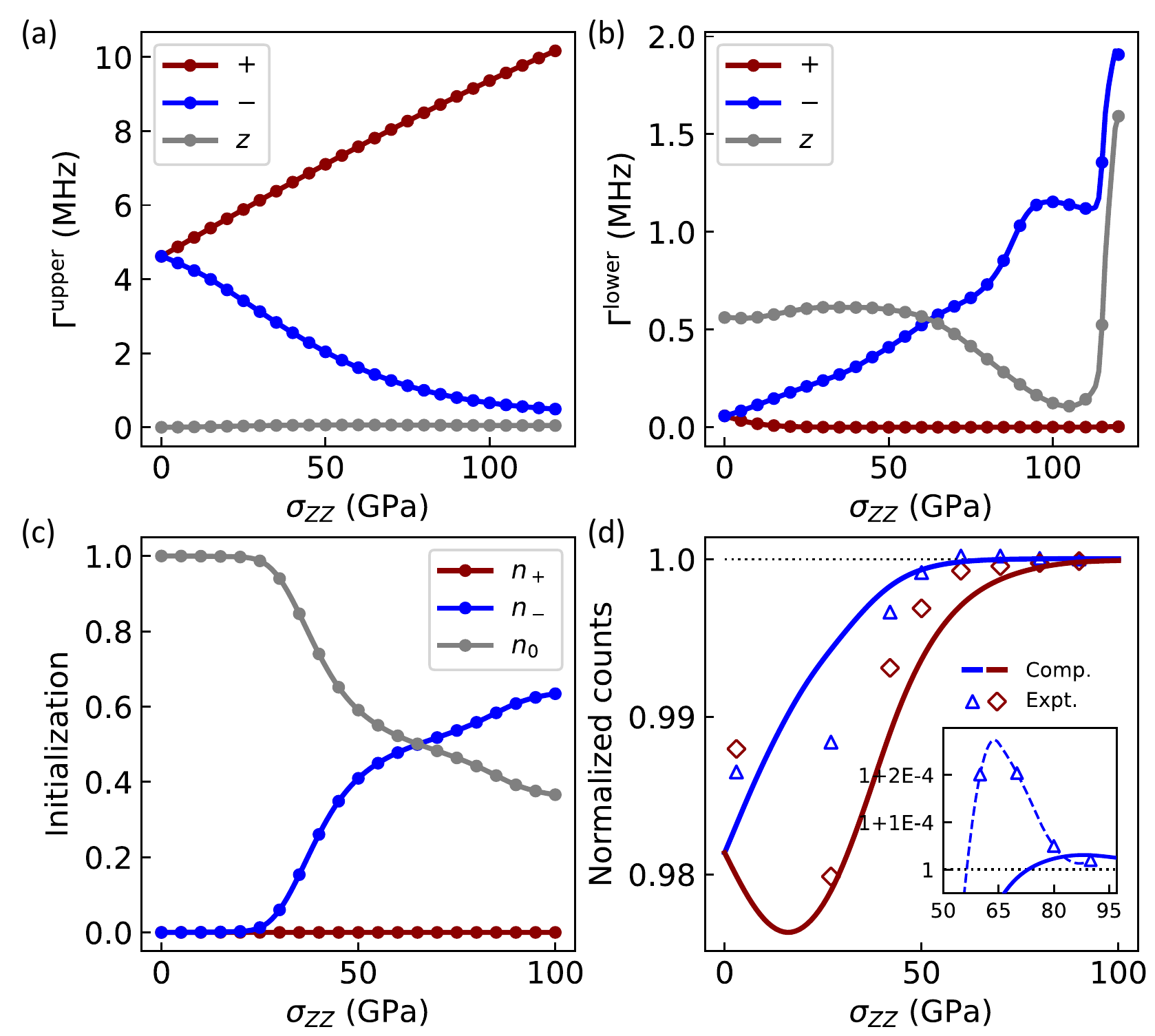}
    \caption{Computations of the ISC rates and ODMR contrast of NV centers in the (100)-cut diamond under stress. (a) Upper ISC rates versus stress with color codes for the three spins. (b) Lower ISC rates versus stress, with $\Gamma_z^\text{lower}$ exhibiting non-monotonic trend coming from negative interference between different ISC mechanisms (see main text and SM). (c) Ground state population distribution among the three spins. A gradual transfer from $n_0$ to $n_-$ begins around $25$ GPa and $n_-$ dominates the population from $65$ GPa. (d) Simulated ODMR contrast (solid) obtained by solving the rate model defined in the main text, with ISC rates acting as inputs. Notably, the contrast inversion in the left peak (representing transitions $\left|m_s=0\right\rangle \leftrightarrow \left|m_s=-\right\rangle$ driven by the MW) observed from experiments~\cite{bhattacharyya2024imaging} (discrete) is reproduced, as shown in the inset. The `predicted' onset of positive contrast is slightly later compared to experiments, and the magnitude is also smaller, with possible reasons for this discrepancy discussed in detail in the Supplemental Materials.}
    \label{fig:contrast_symmetry_breaking}
\end{figure}

\begin{figure}
    \centering
    \includegraphics[width=0.48\textwidth]{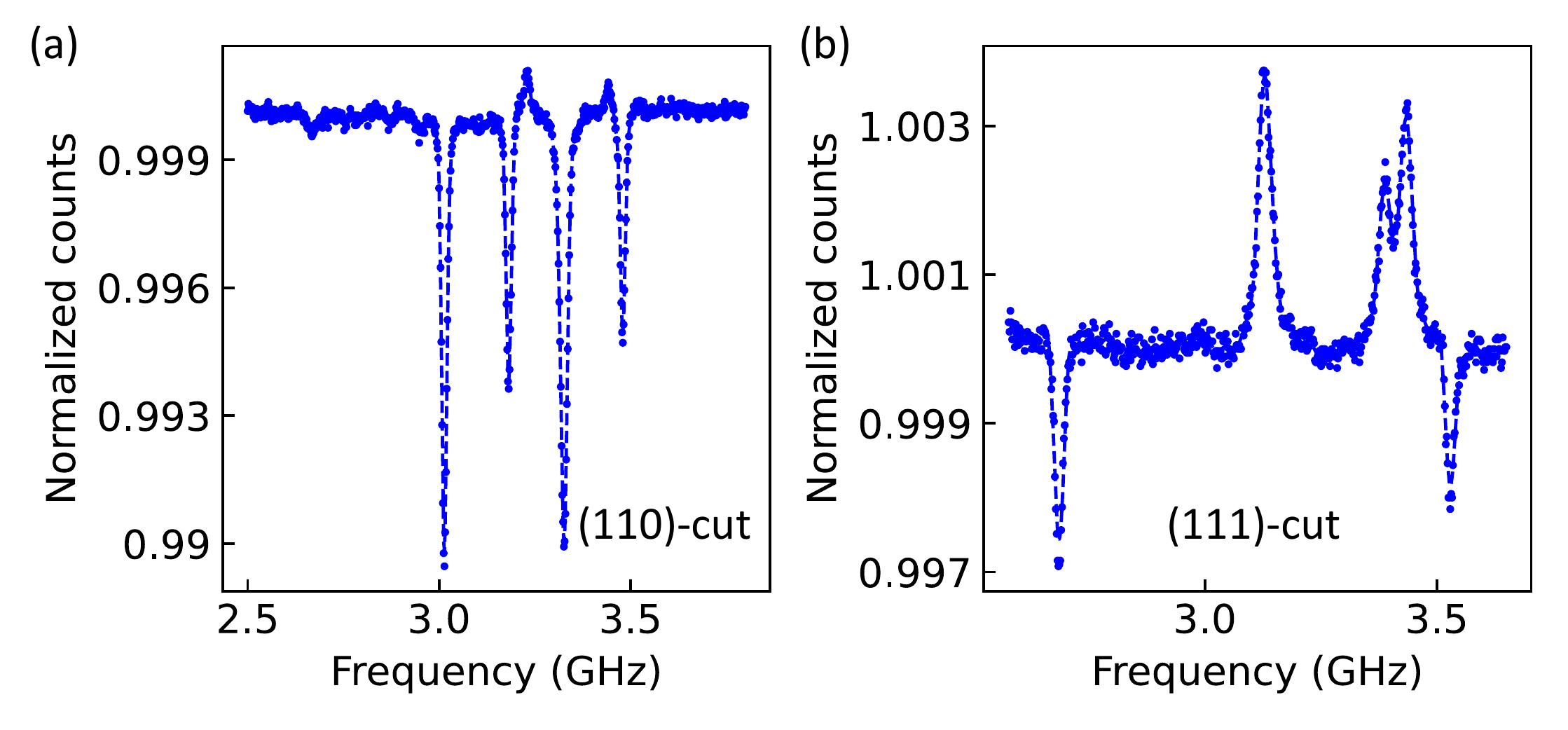}
    \caption{Positive contrast observed from ODMR measurements performed on NV centers in (a) (110)-cut anvil at 300 K, 25 GPa, $B_Z=85$ G~\cite{hsieh2019imaging}, and (b) (111)-cut anvil at 30 K, 28 GPa, $B_Z=150$ G. For the latter, the positive contrast originates from the non-[111] NV centers.}
    \label{fig:positive_contrasts}
\end{figure}

In this Letter, we combine an extensive set of first principles calculations with high-pressure NV experiments on three different culet orientations [i.e.~(100)-, (110)- and (111)-oriented anvils].
Our \emph{ab initio} simulations allow us to estimate two crucial sets of NV parameters as a function of the stress tensor: (i) the rates of inter-system crossing (ISC), which is a non-radiative transition between electronic states with different spin multiplicities and (ii) the spin polarization in the ground-state manifold.
This enables us to propose and analyze a microscopic model that characterizes the NV's optically-detected magnetic resonance (ODMR) contrast under general stress conditions.
%
Our main results are two fold. For stress environments which preserve the $C_{3v}$ symmetry of the NV center, we predict that the optical contrast is mainly determined by the ``upper'' inter-system crossing rate, $\Gamma_\text{ave}$ [Fig.~\ref{fig:optical_cycle}(a)].
To test these predictions, we directly compare to DAC measurements exhibiting a range of different hydrostaticities [Fig.~\ref{fig:contrast_symmetry_preserving}(b)]~\cite{hilberer2023enabling, dai2022optically, bhattacharyya2024imaging, wang2024imaging}.
For symmetry-breaking stresses, we uncover a subtle interplay between the stress-induced spin-orbit coupling (SOC) and the Jahn-Teller (JT) effects of the NV center.
This interplay causes a non-monotonic response of the NV center's ``lower'' ISC rate, $\Gamma^\text{lower}_z$ [Fig.~\ref{fig:optical_cycle}(b)] as a function of stress (Fig.~\ref{fig:contrast_symmetry_breaking}), and ultimately produces an unconventional polarization mechanism that yields the observed contrast inversion.
While our \emph{ab initio} simulations focus on the (100)-oriented culet, our proposed mechanism for contrast inversion should also apply to both (110)- and (111)-oriented culets.
To this end, we perform experiments on both of these anvil cell geometries and indeed observe the predicted contrast inversion (Fig.~\ref{fig:positive_contrasts}).

\emph{Microscopics of the NV's ODMR contrast}---Each NV center hosts a spin-1 electronic ground state that can be optically polarized and read out~\cite{maze2011properties, doherty2011negatively}.
Here, we will work in the spin triplet basis $\ket{m_s=0,+,-}$, where $\ket{m_s=\pm}=\frac{1}{\sqrt{2}}(\ket{+1}\pm\ket{-1})$ and $\ket{\pm1}$ are the familiar Zeeman eigenstates (where the quantization axis is defined along the NV axis). 

NV center metrology is primarily performed via ODMR spectroscopy, where a 532-nm laser first excites the NV center and polarizes its population into the $\ket{m_s=0}$ spin sublevel~\cite{maze2011properties, doherty2011negatively, robledo2011spin}. 
Microscopically, this polarization arises because the $\ket{m_s=0}$ sublevels in ${}^3\!E$ are forbidden to inter-system cross (into the ${}^1\!A_1$ manifold) at leading order~\cite{goldman2015phonon, goldman2015state}, while the lower inter-system crossing rates ($\Gamma^{\textrm{lower}}$, Fig.~\ref{fig:optical_cycle}) exhibit a weak spin dependence~\cite{thiering2018theory}.
Thus, during each optical cycle, population is preferentially transferred into the $\ket{m_s=0}$ ground state; this yields the conventional experimental observation of $\gtrsim 70\%$ spin polarization~\cite{waldherr2011dark, song2020pulse}. 

Crucially, the same optical pathway also naturally leads to spin-dependent fluorescence, enabling optical readout of the NV's magnetic resonance spectra.
In particular, since the $\ket{m_s=0}$ sublevels in ${}^3\!E$ exhibit an extremely small upper inter-system crossing rate, their dominant dynamics correspond to radiative relaxation directly to the ground state; since the laser excitation is spin-preserving, this immediately implies that the $\ket{m_s=0}$ ground state is `brighter' than the $\ket{m_s=+,-}$ states.
ODMR spectroscopy proceeds by measuring the magnitude of the NV’s fluorescence dip (i.e.~contrast) when a microwave field resonant with the spin transition is applied (compared to when it is off).
To this end, the NV center's contrast is controlled by two key ingredients, both determined by the ISC rates (Fig.~\ref{fig:optical_cycle}): (i) the degree of spin polarization and (ii) the relative brightness of the three spin sublevels.

We evaluate the upper ISC rates using Fermi's Golden rule~\cite{goldman2015state}, i.e., $\Gamma = \frac{2\pi}{\hbar} \left|\lambda\right|^2 F(\Delta)$, where the matrix element, $\lambda = \left\langle \psi_\text{final}\right| H_{\text{so}} \left|\psi_\text{init}\right\rangle$, arises from spin-orbit interactions and the vibrational overlap function $F(\Delta)$ characterizes the density of states at the gap $\Delta$ [Fig.~\ref{fig:optical_cycle}(a)].
By comparison, the lower ISC rates are significantly more complicated, since  they are forbidden at first order. 
In addition, ${}^1\!E$ exhibits non-negligible electron-phonon coupling due to Jahn-Teller effects~\cite{thiering2018theory, jin2022vibrationally}. 
Thus, we estimate the lower ISC rates by first solving for the vibronic wavefunction of the singlet states, i.e., $\left|\widetilde{{}^1\!E}\right\rangle, \left|\widetilde{{}^1\!A_1}\right\rangle$ from a Jahn-Teller model Hamiltonian~\cite{thiering2018theory, jin2022vibrationally}, and then evaluating their spin-orbit matrix elements and vibrational density of states with respect to ${}^3\!A_2$ (see Supplemental Materials for additional details).

\emph{Optimizing NV contrast for symmetry-preserving stresses}---We investigate a general symmetry-preserving stress of the form:
\begin{equation} \label{eq:symmetry_preserving_stress}
    \boldsymbol{\sigma} =\alpha \boldsymbol{\sigma}_\text{hyd} + (1-\alpha) \boldsymbol{\sigma}_{[111]},
\end{equation}
where $\alpha$ characterizes the degree of hydrostaticity, and $\boldsymbol{\sigma}_\text{hyd}, \boldsymbol{\sigma}_\text{[111]}$ represent the hydrostatic and uniaxial [111] stresses, respectively. 
Since the symmetry of the NV center is preserved, the optical cycle is qualitatively the same as the ambient case.

To estimate the inter-system crossing rates, we compute the transverse SOC $\lambda_{\perp}$ and $F(\Delta)$ [Fig.~\ref{fig:contrast_symmetry_preserving}(a)] as a function of strain (and then convert to stress), for both the uniaxial [111] and hydrostatic cases (i.e. $\alpha = 0, 1$ respectively).
A few remarks are in order.
First, $\lambda_{\perp}$ (red) increases with compression in both hydrostatic and uniaxial [111] environments, although the effect is significantly stronger for the former. 
Second, we find that $F(\Delta)$ (grey) exhibits opposite trends for the two types of stress environments, implying that the vibrational overlap increases significantly with uniaxial [111] strain, but is suppressed by hydrostatic strain~\footnote{We note that this result is in agreement with expectations for the effects of these strains on $\Delta$~\cite{davies1976optical}.}.

Using $\lambda_\perp$ and $F(\Delta)$, we now compute the upper ISC rates, $\Gamma_\text{ave}$, versus stress [dashed curves, Fig.~\ref{fig:contrast_symmetry_preserving}(b)].
For hydrostatic stress (dark purple, $\alpha=1$), the ISC rate exhibits a non-trivial trend, with a peak value at approximately $\sim 30$~GPa.
Interestingly, this is a manifestation of competition between the behaviors of $\lambda_\perp$ and $F(\Delta)$, where the former dominates at small stresses while the latter controls the large stress limit.
For smaller $\alpha$ (i.e. a larger uniaxial component), $\Gamma_\text{ave}$ exhibits a more monotonic behavior as a function of stress.

In order to predict how the NV's ODMR contrast changes versus stress, we directly solve the rate-equation model for the NV's optical cycle [Fig.~\ref{fig:optical_cycle}(a)] utilizing these computed ISC rates.
As depicted in Fig.~\ref{fig:contrast_symmetry_preserving}(b) (solid curves), we find that uniaxial $[111]$ stress ($\alpha=0$), although not achievable in the conventional DAC setup, yields the largest NV contrast~\footnote{Although the uniaxial $[111]$ stress is predicted to exhibit the best contrast, we find $\alpha \in [0.0, 0.1]$ all yield very similar contrast.} and that there exists a strong correlation between the upper ISC rates (dashed curves) and the predicted contrast (solid curves).

To investigate these predictions, we directly measure [111]-NV contrast [extracted from Rabi oscillations, Fig.~\ref{fig:contrast_symmetry_preserving}(d)] as a function of pressure in a (111)-cut diamond with NV centers implanted  $\sim50$~nm below the culet surface~[Fig.~\ref{fig:contrast_symmetry_preserving}(c)]. 
By carefully measuring the stress tensor at each pressure, we estimate the degree of hydrostaticity to be  $\alpha\approx0.73$.
In addition, we also compare our predictions to two other sets of experimental NV-DAC measurements with differing degrees of hydrostaticity: (i) a nearly hydrostatic ($\alpha \approx 1$) measurement using NV's contained within a nanopillar fabricated at the center of (100)-cut DAC~\cite{hilberer2023enabling}, and (ii) a measurement with $\alpha\approx0.57$~\cite{wang2024imaging} that also utilizes NVs in a (111)-cut DAC~\footnote{We note that in practice, $\alpha$ can deviate from the theoretical value of $0.61$ derived from a semi-infinite anvil~\cite{ruoff1991closing} due to the uniaxial bias imposed by the compression being done along the $Z$ axis, and that variations in $\alpha$ between experiments relates to the slightly different mechanical conditions of different experimental configurations (culet size, gasket thickness, etc).}.
As illustrated in Fig.~\ref{fig:contrast_symmetry_preserving}(c), all three sets of experimental measurements are in semi-quantitative agreement with our \emph{ab initio} predictions.
Interestingly, since our calculations consider only a single NV center, this suggests that the contrast enhancement from utilizing a (111)-cut anvil is \emph{intrinsic} to the [111]-oriented NV itself, ruling out previous interpretations based on the `darkening' of non-[111] NVs~\cite{wang2024imaging}.

\emph{Microscopic origin of stress-induced positive NV contrast}---Let us now turn to a second puzzle regarding the NV's contrast in high-pressure experiments, namely, the observation of contrast inversion [depicted in Fig.~\ref{fig:nv_quantum_sensing}(c) for a (100)-cut anvil]~\footnote{It has also been reported that local electric field around the NV center plus near resonance ODMR measurement (resonant optical excitation at zero phonon line wavelength) could also lead to positive contrast at cryogenic temperatures ($\sim 5$K)~\cite{block2021optically, akhmedzhanov2016optically}. The positive contrast that we investigate in this work is from off-resonant ODMR spectroscopy and therefore has a completely different mechanism.}.
As previously discussed, conventional `negative' contrast occurs because the NV becomes optically polarized to the `bright' $\ket{m_s=0}$ spin state, and the applied microwave transfers population into the comparatively `dark' $\ket{m_s=+,-}$ states. Conversely, `positive' contrast suggests that the NV is becoming polarized into a dark state. 
This hypothesis, as we will see, requires the lower ISC to develop a strong spin selectivity towards population transfer into the dark states. 
Crucially, this selectivity is made possible by symmetry-breaking stresses, which open new ISC transitions within the NV's optical cycle, e.g., $\Gamma_z^\text{upper}, \Gamma_z^\text{lower}$ [Fig.~\ref{fig:optical_cycle}(b)].

To this end, let us begin by understanding the effect of symmetry-breaking stresses on the upper ISC rates.
Under uniaxial [100] stress, the ${}^3\!E$ sextuplet is split into two well-separated triplets, as the $e$ orbital degeneracy is lifted [Fig.~\ref{fig:optical_cycle}(b)].
Figure~\ref{fig:contrast_symmetry_breaking}(a) depicts the  upper ISC rates [defined in Fig.~\ref{fig:optical_cycle}(b)] as a function of increasing [100] stress.
The transition rate from $\ket{m_s=+}$ ($\ket{m_s=-}$) monotonically increases (decreases) with stress and is still primarily driven by the vibrational overlap between ${}^3\!E_{y}$ (${}^3\!E_{x}$) and ${}^1\!A_1$.
This is because the ${}^3\!E_x$ branch is rapidly detuned from ${}^1\!A_1$ (and vice versa for ${}^3\!E_y$).
Meanwhile, the symmetry-breaking stress enables a non-zero $\Gamma^\text{upper}_z$, which connects $\ket{m_s=0}$ to ${}^1\!A_1$.
Most importantly, we find that $\Gamma^\text{upper}_z$ remains the smallest [Fig.~\ref{fig:contrast_symmetry_breaking}(a)] throughout the entire pressure range investigated, confirming that the $\ket{m_s=0}$ state remains the brightest.

Next, we turn to the fascinating case of the lower ISC. Much like ${}^3\!E$, the reduction of symmetry lifts the orbital degeneracy of ${}^1\!E$. Stress serves to progressively increase (decrease) the vibronic overlap between ${}^1\!E_x$ (${}^1\!E_y$) and ${}^3\!A_2$. In Fig.~\ref{fig:contrast_symmetry_breaking}(b), we plot all three rates from the ${}^1\!E$ singlet into the ${}^3\!A_2$ manifold.
Among them, the behavior of $\Gamma^\text{lower}_z$ is particularly intriguing owing to its non-monotonic behavior versus stress: It exhibits a significant drop-off beyond $\sigma_{ZZ}=50$ GPa, and then increases sharply again after $\sim100$ GPa.
Somewhat remarkably, this results from the emergence of a new stress-induced spin-orbit channel from ${}^1\!E_x$, which \textit{destructively interferes} with the existing Jahn-Teller-based channel (see Supplemental Materials).
Above $\sim100$~GPa, this new channel dominates, rapidly restoring the $\Gamma^\text{lower}_z$ transition rate. 
Comparatively, $\Gamma^\text{lower}_-$ progressively increases with stress and becomes dominant at $\sigma_{ZZ}\approx60$~GPa, making the lower ISCs favor the dark spin state $|m_s=-\rangle$.

To this end, we observe a reversal of the spin selectivity of the lower ISC with symmetry breaking stress.
By combining all of the aforementioned ISC rates, we compute the NV's spin polarization [Fig.~\ref{fig:contrast_symmetry_breaking}(c)] and fluorescence contrast [Fig.~\ref{fig:contrast_symmetry_breaking}(d)].
For small stresses, the NV remains initialized to the $\ket{m_s=0}$ state, while fraction initialized into the $\ket{m_s=-}$ state begins to increase at $\sigma_{ZZ}\sim25$~GPa.
At even larger stresses, $\sigma_{ZZ}\gtrsim65$~GPa, the NV becomes dominantly polarized to the $\ket{m_s=-}$ state.
This represents the complex interplay between several symmetry-breaking-stress-induced modifications to both the upper and lower ISCs.
From the perspective of contrast, once the NV becomes initialized into the dark $\ket{m_s=-}$ state, the resulting ODMR spectrum will naturally exhibit positive contrast [inset, Fig.~\ref{fig:contrast_symmetry_breaking}(d)].
Interestingly, our theory predicts that while the ODMR peaks may exhibit contrast inversion, their positions still encode the same spectral content of the ground-state spin sublevel splittings.

Our theoretical framework also implies the existence of positive contrast in more general stress conditions. 
In particular, the interplay between symmetry breaking stresses and/or transverse magnetic fields can serve to hybridize excited state spin sublevels, enhancing the effective $\Gamma^\text{upper}_z$ and thereby promoting ground state polarization inversion. 
To probe this generality, we perform high-pressure NV measurements in both a (110)-cut and a (111)-cut anvil.
As shown in Fig.~\ref{fig:positive_contrasts}, in the presence of an external magnetic field, we observe contrast inversion for both settings.

\emph{Conclusion and Outlook}---Our work opens the door to a number of intriguing future directions.
First, our computational framework can readily be generalized to accommodate a wide array of environmental conditions, e.g. temperatures, electrical/magnetic fields, and stress environments.
Moreover, our method can also be adapted to the vast emerging landscape of solid state spin defects~\cite{wolfowicz2021quantum, lee2025intrinsic, he2025probing, mohseni2025magneto}, and will help identify candidates that are viable under extreme conditions.
Second, while we have provided a general framework for the emergence of contrast inversion, the details of how and when such inversion occurs in different anvil cuts remains an open challenge. 
Third, we motivate the deliberate engineering of stress environments to promote metrological sensitivity. Our model suggests that achieving $\alpha=0$ maximizes contrast. This stress environment may be obtained with a uniaxial press~\cite{barson2017nanomechanical}. Meanwhile, a recent study makes use of NV center contrast inversion to perform measurements of the full stress tensor up to tens of GPa~\cite{mandyam2025uncovering}.
Finally, the phenomenon of positive contrast suggests the use of symmetry-breaking stresses as another tuning parameter for defect physics. Modifying the polarization dynamics of the NV center and other spin defects may prove useful in the context of both quantum information and quantum sensing.

\emph{Note Added}---During the completion of this work, a related manuscript appeared~\cite{liu2025strain}, which  explores the contrast inversion of the NV center under pressure in a (111)-cut anvil.
The authors also conclude that this inversion arises from a reversal in spin polarization.

\begin{acknowledgments}
\emph{Acknowledgements}---We gratefully acknowledge discussions with R.~Jeanloz, T.~Smart, and D.~Budker.
This work was supported in part by the Next Generation Quantum Science and Engineering (Q-NEXT) center that develops quantum science and engineering technologies, and by the Brown Institute for Basic Sciences. Q-NEXT is supported by the U.S. Department of Energy, Office of Science, National Quantum Information Science Research Centers.
S.V.M. acknowledges support from the National Science Foundation Graduate Research Fellowship under Grant No. DGE-1752814.
B.K. and M.B. acknowledge support from NSF QLCI program (grant no. OMA-2016245). 
C.Z. acknowledges the support from Enterprise Science Fund of Intellectual Ventures Management, LLC.
The computations performed in this research used resources from the University of Chicago Research Computing Center.
\end{acknowledgments}

\bibliography{bibliography}

\onecolumngrid
\section*{End Matter}

\twocolumngrid
The pursuit of accurate and efficient methods for solving the electronic structure problem has been the holy grail of computational chemistry and condensed matter physics. One central physical quantity investigated in this work is the spin orbit coupling (SOC), which is numerically obtained using the complete active space self-consistent field (CASSCF) method~\cite{roos1980complete} applied to a single-NV cluster~\cite{giannozzi2020quantum, bhandari2021multiconfigurational}, as shown in the inset of Fig.~\ref{fig:contrast_symmetry_preserving}(a). This model has been widely adopted in several recent studies~\cite{bhandari2021multiconfigurational, li2024excited, benedek2024comprehensive} due to its affordable computational cost and availability of codes~\cite{neese2020orca}.

During this research, a subset of the authors developed an alternative description~\cite{jin2025first} of the defect's SOC based on a quantum defect embedding theory (QDET)~\cite{chen2025advances} which partitions the problem into separate calculations of the defect center and of the host material. Since this approach uses periodically repeated cells, the surrounding solid-state environment is represented more accurately. QDET is based on a Green's function formalism, and has been applied to study several defects in semiconductors~\cite{otis2025strongly, somjit2025nv} for quantum technologies. In this End Matter, we shall make a comparison of these two computational approaches, showcasing their respective pros and cons, to guide future computational or applied research.

In Table.~S6 in the Supplemental Materials, we compare the upper ISC rates computed from these two methods with experimental measurements in ambient conditions, showing that the QDET approach~\cite{jin2025first} gives a better agreement with experiment, while the CASSCF@cluster approach gives results roughly one order of magnitude smaller than what is observed. The inaccuracy of the CASSCF approach can be traced back to an underestimation of $\lambda_\perp$ (see Table.~S4). By comparing the weights of different configurations in the many-body wavefunctions from Ref.~\cite{li2024excited} and Ref.~\cite{jin2025first}, we expect that this underestimation arises from the relatively large weight ($\sim20\%$) of the $e_x^2 e_y^2$ configuration in the ${}^1\!A_1$ wavefunction from CASSCF (see Sec.~S6.2 in the Supplemental Materials for details). However, in this work, the relavant quantities in our calculations are the strain/stress susceptibilities of SOCs, and CASSCF@cluster provides a consistent way to estimate them, as we reach a semi-quantitative agreement with experimental ODMR measurements.

Future work will employ the QDET approach to refine the estimation of susceptibilities.

\end{document}


\title{Supplemental Materials: Elucidating the Inter-system Crossing of the Nitrogen-Vacancy Center up to Megabar Pressures}

\author{Benchen Huang}
\thanks{These authors contributed equally to this work.}
\affiliation{Department of Chemistry, University of Chicago, Chicago, IL 60637, USA}%

\author{Srinivas V. Mandyam}
\thanks{These authors contributed equally to this work.}
\affiliation{Department of Physics, Harvard University, Cambridge, MA 02135, USA}%

\author{Weijie Wu}
\affiliation{Department of Physics, Harvard University, Cambridge, MA 02135, USA}%

\author{Bryce Kobrin}
\affiliation{Department of Physics, University of California, Berkeley, CA 94720, USA}%
\affiliation{Materials Science Division, Lawrence Berkeley National Laboratory, Berkeley, CA 94720, USA}%

\author{Prabudhya Bhattacharyya}
\affiliation{Department of Physics, University of California, Berkeley, CA 94720, USA}%
\affiliation{Materials Science Division, Lawrence Berkeley National Laboratory, Berkeley, CA 94720, USA}%

\author{Yu Jin}
\affiliation{Pritzker School of Molecular Engineering, University of Chicago, Chicago, IL 60637, USA}

\author{Bijuan Chen}
\affiliation{Department of Physics, Harvard University, Cambridge, MA 02135, USA}%

\author{Max Block}
\affiliation{Department of Physics, Harvard University, Cambridge, MA 02135, USA}%

\author{Esther Wang}
\affiliation{Department of Chemistry and Chemical Biology, Harvard University, Cambridge, MA 02135, USA}%

\author{Zhipan Wang}
\affiliation{Department of Physics, Harvard University, Cambridge, MA 02135, USA}%

\author{Satcher Hsieh}
\affiliation{Department of Physics, University of California, Berkeley, CA 94720, USA}%
\affiliation{Materials Science Division, Lawrence Berkeley National Laboratory, Berkeley, CA 94720, USA}%

\author{Chong Zu}
\affiliation{Department of Physics, Washington University, St.~Louis, MO 63130, USA}

\author{Christopher R. Laumann}
\affiliation{Department of Physics, Boston University, Boston, MA 02215, USA}

\author{Norman Y. Yao}
\email{nyao@fas.harvard.edu}
\affiliation{Department of Physics, Harvard University, Cambridge, MA 02135, USA}
\affiliation{Department of Physics, University of California, Berkeley, CA 94720, USA}
\affiliation{Materials Science Division, Lawrence Berkeley National Laboratory, Berkeley, CA 94720, USA}%

\author{Giulia Galli}
\email{gagalli@uchicago.edu}
\affiliation{Department of Chemistry, University of Chicago, Chicago, IL 60637, USA}
\affiliation{Pritzker School of Molecular Engineering, University of Chicago, Chicago, IL 60637, USA}
\affiliation{Materials Science Division and Center for Molecular Engineering, Argonne National Laboratory, Lemont, IL 60439, USA}%

\date{\today}

\maketitle
\tableofcontents
\newpage

\section{Experimental details} \label{sec:experiment}

\subsection{Sample preparation}
We used type Ib 16-sided (111)-cut diamond anvils (Almax-easyLab) with culets of 200 $\mu$m. We perform 12C$+$ ion implantation at an energy of 30 keV with a dosage of 5$\times10^{12}$ cm$^{-2}$ (CuttingEdge Ions, LLC) to generate a layer of vacancies up to 50 nm from the culet surface. Following implantation, we vacuum anneal the diamond anvils (at pressure below 10$^{-6}$ mbar) in a home built furnace at a temperature above 850$^{\circ}$C for 12 hours. During annealing, mobile vacancies diffuse and bind with substitutional nitrogen defects to form NV centers. Anvils are loaded into a miniBX80-type cell, with a microwave-compatible insulating gasket as in Hsieh et al.~\cite{hsieh2019imaging}. Sodium chloride is used as the pressure medium.

\subsection{Contrast measurement}
We perform optical measurements in a home-built confocal microscope. A 532-nm laser is used to excite the NV centers, and fluorescence counts are read out as a function of applied microwave frequency. A field of 100 Gauss along the [111]-NV is applied to split the resonances apart.
To measure the contrast of the [111]-NV resonance, we set microwave power equal to half the full width at half maximum (FWHM) of the resonance linewidth, and frequency to the resonance center, and perform a standard Rabi oscillation measurement. We extract contrast from an exponentially decaying sinusoid fit.

The advantage of measuring contrast via this Rabi oscillation method is that it circumvents the issue of microwave power inhomogeneity. Because the microwave transmission line has different transmission efficiencies at different frequencies, contrast (which is affected by microwave power) can in principle be spuriously affected by the frequency location of the resonance. In addition, by calibrating the microwave power to the resonance of the linewidth, we ensure that the same fraction of NV spins in the ensemble are being driven at each pressure point, which also affects contrast as non-driven spins contribute to fluorescence background.

\section{Group theory analysis} \label{sec:group_theory}
To understand how stress affects the optically detected magnetic resonance (ODMR) contrast, we need to first understand how stress affects the optical cycle of the NV center. Because the NV center has $C_{3v}$ symmetry in ambient conditions, we rely on group theory to investigate the potential couplings between different electronic states (in a \textit{perturbative} fashion), which is of vital importance to determine possible inter-system crossing\footnote{ISC, as we introduced in the main text, is a non-radiative transition between two electronic states with different spin multiplicities.} (ISC) routes in the optical cycle. To facilitate subsequent discussions, we briefly summarize in this section key information derived from group theory, and direct the readers to Ref.~\cite{hepp2014thesis} for a more detailed formulation. In this section, several operators related to the NV triplet excited manifold ${}^3\!E$ will be assumed; we will define them more rigorously in the next section.

Following a standard group theoretic approach, we consider all interaction terms invariant under symmetry transformations. To construct the stress-coupled Hamiltonian of the NV triplet excited manifold, we write the Hamiltonian as the product of orbital operators, spin operators, and the stress tensor. Take the spin-spin Hamiltonian as an example:
\begin{equation}
    H_{\text{ss}} = \sum_{ijk}\chi_{ijk} P_i^{(\Gamma_1)}\otimes \left(S^2\right)_j^{(\Gamma_2)}\sigma_k^{(\Gamma_3)},
\end{equation}
where $\chi_{ijk}$ is the susceptibility, $P_i^{(\Gamma_1)}$ is the orbital Pauli operator transforming as the irreducible representation (irrep) $\Gamma_1$, $\left(S^2\right)_j^{(\Gamma_2)}$ is the quadratic spin operator transforming as irrep $\Gamma_2$ and $\sigma_k^{(\Gamma_3)}$ is the stress component transforming as irrep $\Gamma_3$. The Hamiltonian should transform as $A_1$, so $\chi_{ijk}$ is not vanishing only if $A_1 \subset \Gamma_1 \otimes \Gamma_2 \otimes \Gamma_3$. We categorized the operators according to their irreps in Table~\ref{table:irrep_operators}.

\begin{table}[hbt!]
\centering
\caption{Orbital operators, spin operators and stress categorized into different irreps of $C_{3v}$, where the orbital Pauli operators $P_x = |E_x\rangle\langle E_y| + |E_y\rangle\langle E_x|$, $P_y = -i|E_x\rangle \langle E_y| + i|E_y\rangle\langle E_x|$ and $P_z = |E_x\rangle\langle E_x| - |E_y\rangle\langle E_y|$ are defined w.r.t. the two orbital branches of ${}^3\!E$.}
{\setlength{\extrarowheight}{5pt}%
 \begin{tabular}{c | c | p{5cm}<{\centering} | p{5cm}<{\centering}}
 \hline
 \hline
  & Orbital & Spin & Stress \\ [0.5ex] 
 \hline
 $A_1$ &  & $S_z^2$, $S_x^2+S_y^2$ & $\sigma_{zz}$, $\sigma_{xx} + \sigma_{yy}$ \\ 
 \hline
 $A_2$ & $P_y, L_z$ & $S_z$, $S_x S_y - S_y S_x$  & \\
 \hline
 $E$ & $\{P_z, -P_x\}$, $\{L_y, -L_x\}$ & $\{S_y^2-S_x^2, S_xS_y + S_yS_x\}$, $\{S_xS_z + S_zS_x,S_yS_z + S_zS_y\}$, $\{S_y, -S_x\}$ & $\{\sigma_{yy} - \sigma_{xx}, \sigma_{xy} + \sigma_{yx}\}$, $\{\sigma_{xz} + \sigma_{zx}, \sigma_{yz} + \sigma_{zy}\}$ \\
 \hline
 \hline
\end{tabular}}
\label{table:irrep_operators}
\end{table}

The stress components transforming as $A_1$ can couple with the following terms,
\begin{equation}
A_1 = \left\{
\begin{array}{l}
     I \otimes S_z^2  \\
     I \otimes (S_x^2 + S_y^2) \\
     P_z \otimes (S_y^2 - S_x^2) - P_x\otimes (S_xS_y + S_yS_x) \\ 
     P_z \otimes (S_xS_z + S_zS_x) - P_x\otimes(S_yS_z + S_zS_x) \\ 
\end{array}
\right.
\end{equation}
The stress components transforming as $E$ can couple with the following terms,
\begin{equation}
E = \left\{
\begin{array}{l} 
     \{I\otimes(S_y^2 - S_x^2),\;\; I\otimes(S_xS_y + S_yS_x)\} \\
     \{I\otimes(S_xS_z + S_zS_x),\;\; I\otimes(S_yS_z + S_zS_y)\} \\ 
     \{P_z\otimes S_z^2,\;\; -P_x\otimes S_z^2\}\\
     \{P_z\otimes (S_x^2 + S_y^2),\;\; -P_x\otimes (S_x^2 + S_y^2)\} \\ 
    \{P_y\otimes (S_xS_y + S_yS_x),\;\; -P_y\otimes (S_y^2 - S_x^2)\} \\ 
    \{P_y\otimes (S_yS_z + S_zS_y),\;\; -P_y\otimes (S_xS_z + S_zS_x)\} \\ 
    \{-P_x\otimes(S_xS_y + S_yS_x)-P_z\otimes(S_y^2 - S_x^2),\;\; P_z\otimes(S_xS_y + S_yS_x) - P_x\otimes(S_y^2 - S_x^2))\} \\ 
    \{-P_x\otimes(S_yS_z + S_zS_y) - P_z\otimes(S_xS_z + S_zS_x),\;\; P_z\otimes(S_yS_z + S_zS_y) - P_x\otimes(S_xS_z + S_zS_x) \}
\end{array}
\right.
\end{equation}
These terms are constructed using the following rule:
\begin{align*}
     A_1 \otimes A_1 \sim A_1,& \;\;A_2 \otimes A_2 \sim A_1,\\
     E_x \otimes E_x & + E_y \otimes E_y \sim A_1,\\
     A_2 \otimes E_y \sim E_x,& \;\;A_2 \otimes E_x \sim -E_y,\\
     \left(E_y \otimes E_y - E_x \otimes E_x\right) \sim E_x,& \;\;\left(E_x\otimes E_y + E_y\otimes E_x\right) \sim E_y.
\end{align*}
We will be using the above relations to derive other stress-coupled Hamiltonians.

\section{NV center interactions} \label{sec:interactions}
In this section, we discuss the various interactions of the NV center. These include spin-orbit coupling (SOC) and spin-spin coupling (SSC); we focus especially how stresses affect them. Understanding these interactions is crucial for understanding the variations of ODMR contrast under high pressure.

\begin{figure}
    \centering
    \includegraphics[width=0.75\textwidth]{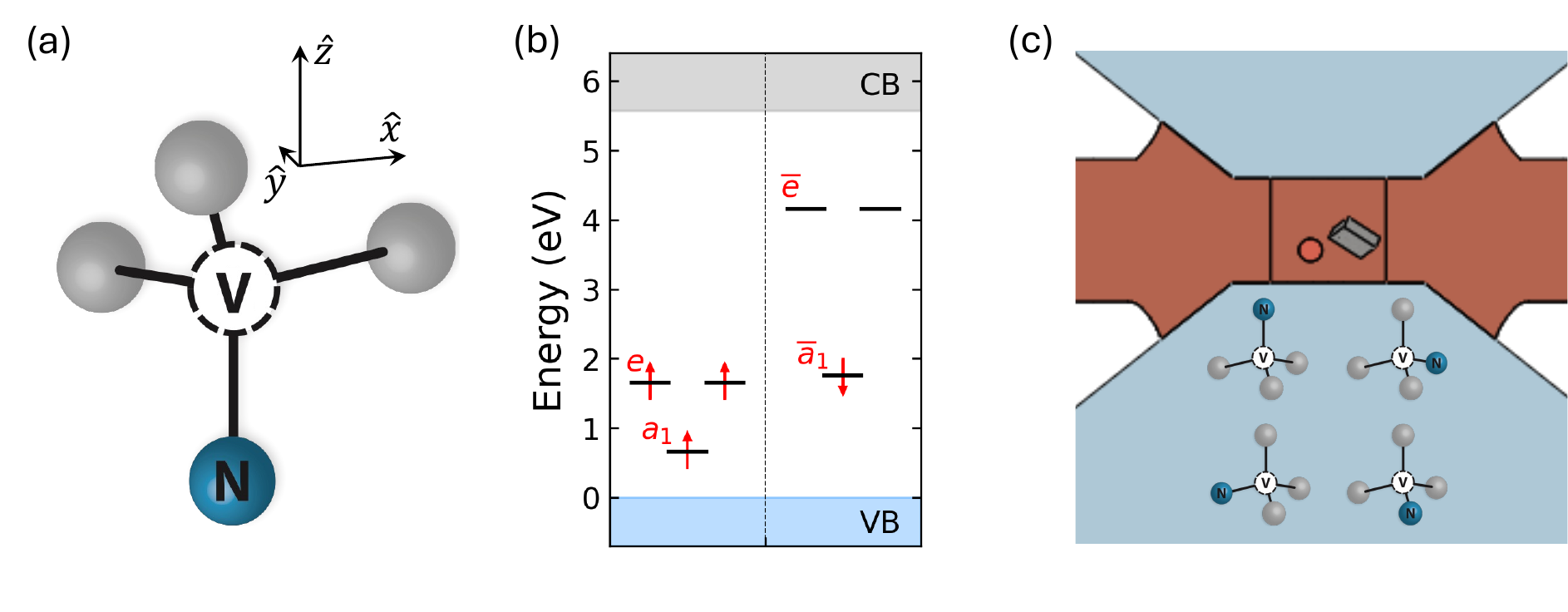}
    \caption{(a). Atomistic structure of the NV center with $C_{3v}$ symmetry and its local coordinate system. (b). Defect orbitals $\{a_1, e_x, e_y\}$ in the diamond band gap, computed from unrestricted density functional theory (DFT) using the SCAN functional~\cite{sun2015strongly}, with the left(right) panel showing the spin-up(down) channel. (c). NV centers implanted in the DAC culet, with the four different orientation groups of the (111)-cut . Only one group experiences purely uniaxial [111] stress, while the other three experience some amount of $\left[\overline{11}1\right]$ stress.}
    \label{fig:atomistic}
\end{figure}

The NV center is a crystallographic defect comprising a substitutional nitrogen atom adjacent to a lattice vacancy with $C_{3v}$ symmetry, as shown in Fig.~\ref{fig:atomistic}(a). Therefore, any stress $\pmb{\sigma}$ the NV experiences can be decomposed into a symmetry-preserving and a symmetry-breaking part. We shall introduce two spatial coordinate coordinate bases,
\begin{equation}
    \pmb{\sigma}_{xyz} = R \pmb{\sigma}_{XYZ} R^T,\;\; R = \begin{pmatrix}
        -\frac{1}{\sqrt{6}} & -\frac{1}{\sqrt{6}} & \sqrt{\frac{2}{3}}\\
        \frac{1}{\sqrt{2}} & -\frac{1}{\sqrt{2}} & 0\\
        \frac{1}{\sqrt{3}} & \frac{1}{\sqrt{3}} & \frac{1}{\sqrt{3}}
    \end{pmatrix},
\end{equation}
where $x, y, z$ describes the NV's local frame, while $X, Y, Z$ describes the crystal frame. This allows us to pick the coordinate most convenient for describing the specific stress we study. For symmetry-preserving stresses such as uniaxial [111] stress\footnote{We use Miller indices to notate the planes and directions in diamond throughout this work. Parentheses `()' denote surfaces, e.g., (111)-cut diamond and square brackets `[]' represent the surface norm, e.g., [100] uniaxial stress.}, we use the local frame and the stress tensor only has one non-zero element $\sigma_{zz} = \sigma$; for symmetry-breaking stresses such as uniaxial-[100] or $\left[\overline{11}1\right]$ stress, we use the crystal frame which may be transformed into the local frame as:
\begin{align}
    \pmb{\sigma}_{[100]} & = \frac{1}{3}\left(\sigma_{yy} + \sigma_{xx}\right) + \frac{1}{3}\sigma_{zz} - \frac{1}{3}\left(\sigma_{yy} - \sigma_{xx}\right) - \frac{\sqrt{2}}{3}\left(\sigma_{xz} + \sigma_{zx}\right), \label{eq:[001]_stress}\\
    \pmb{\sigma}_{\left[\overline{11}1\right]} & = \underbrace{\frac{4}{9}\left(\sigma_{yy} + \sigma_{xx}\right) + \frac{1}{9}\sigma_{zz}}_{\text{symmetry-preserving}} - \underbrace{\frac{4}{9}\left(\sigma_{yy} - \sigma_{xx}\right) + \frac{2\sqrt{2}}{9}\left(\sigma_{xz} + \sigma_{zx}\right)}_{\text{symmetry-breaking}}.
\end{align}
Since [100], [010], and [001] stress are all equivalent, we will be solely using [100] for simplicity in notations, although in our \emph{ab initio} calculations, the actual strain is applied in the $Z$ direction (therefore [001] strain/stress). Finally, we note that NV center can have different charge states. In this work, we only focus on the negatively charged state.

\subsection{Electronic structure}
We shall begin by discussing the NV's electronic structure. The NV center introduces an energy level with symmetry $a_1$, as well as a double-degenerate $e$-symmetry single-particle orbital energy level pair, into the diamond band gap; these levels are collectively occupied by four electrons in the defect's negatively charged state [Fig.~\ref{fig:atomistic}(b)]. The relevant electronic states studied here can be classified into two major configurations (without spin): $a_1^2 e^2$ and $a_1 e^3$; with the former generating the triplet ground-state manifold ${}^3\!A_2$ and two singlet excited-state manifolds ${}^1\!E, {}^1\!A_1$ in energy-ascending order and the latter generating triplet excited-state manifold ${}^3\!E$ and singlet ${}^1\!E'$. We refer the readers to Ref.~\cite{maze2011properties, doherty2011negatively} for a detailed discussion on their wavefunctions.

It is worth mentioning that $\{a_1, e_x, e_y\}$ only compose the so-called \textit{minimum model} of the NV center. In addition, this description only provides a qualitative picture of the NV's electronic wavefunctions. For a more accurate description, a multi-reference method is often required~\cite{bhandari2021multiconfigurational, li2024excited, jin2025first}, which we adopt in our simulations and will discuss in detail in Sec.~\ref{sec:simulations}.

\subsubsection{Ambient condition}
The electronic ground state of the NV center is an orbital singlet, spin triplet ($S=1$) manifold, usually denoted as ${}^3\!A_2$. In the absence of external perturbations, the ground-state spin Hamiltonian is given by $H = D_{\text{gs}} S_z^2$, where $D_{\text{gs}} = 2.87$~GHz is the temperature-dependent zero-field splitting (ZFS) between the $|m_s = 0\rangle$ sublevel and the degenerate $|m_s = \pm 1\rangle$ sublevels, and $\{S_x, S_y, S_z\}$ are spin-1 operators quantized along the N-V axis. The quantization axis may be oriented along any of the diamond bonds resulting in four subgroups of NV centers. A magnetic field $\vec{B}$ couples into the NV via the Hamiltonian $H_B = \gamma_B \vec{B}\cdot\vec{S}$, where $\gamma_B = (2\pi) \times 2.8$~MHz/G is the NV’s gyromagnetic ratio and $\vec{B}$ is expressed in the local frame of the NV center.

The triplet excited-state manifold ${}^3\!E$ is an orbital doublet with orbital states $|E_x\rangle, |E_y\rangle$, lying $1.945$ eV higher in energy than the ground state~\cite{davies1976optical}. At low temperatures, due to SOC and SSC (which will be discussed later), ${}^3\!E$ splits into four different levels~\cite{maze2011properties, doherty2011negatively}--$A_2, A_1, E_{x,y}, E_{1,2}$ according to their irreducible representation (irrep) in energy-descending order [Fig.~\ref{fig:3E_splittings}(a)]. At room temperature, orbital averaging leads the whole ${}^3\!E$ manifold to be an effective orbital singlet~\cite{batalov2009low} similar to the ground state, with an effective ZFS $D_{\text{es}}\approx1.42$~GHz.

Besides these triplet states, there are two low-lying singlet states participating in the optical cycle. They are denoted as ${}^1\!A_1$ and ${}^1\!E$ in energy-descending order. Historically, little was known about these states due to their darkness, except that they are $\sim 1.190$ eV apart in energy~\cite{rogers2015singlet}.

\subsubsection{Effects of stress}
\begin{figure}
    \centering
    \includegraphics[width=0.75\textwidth]{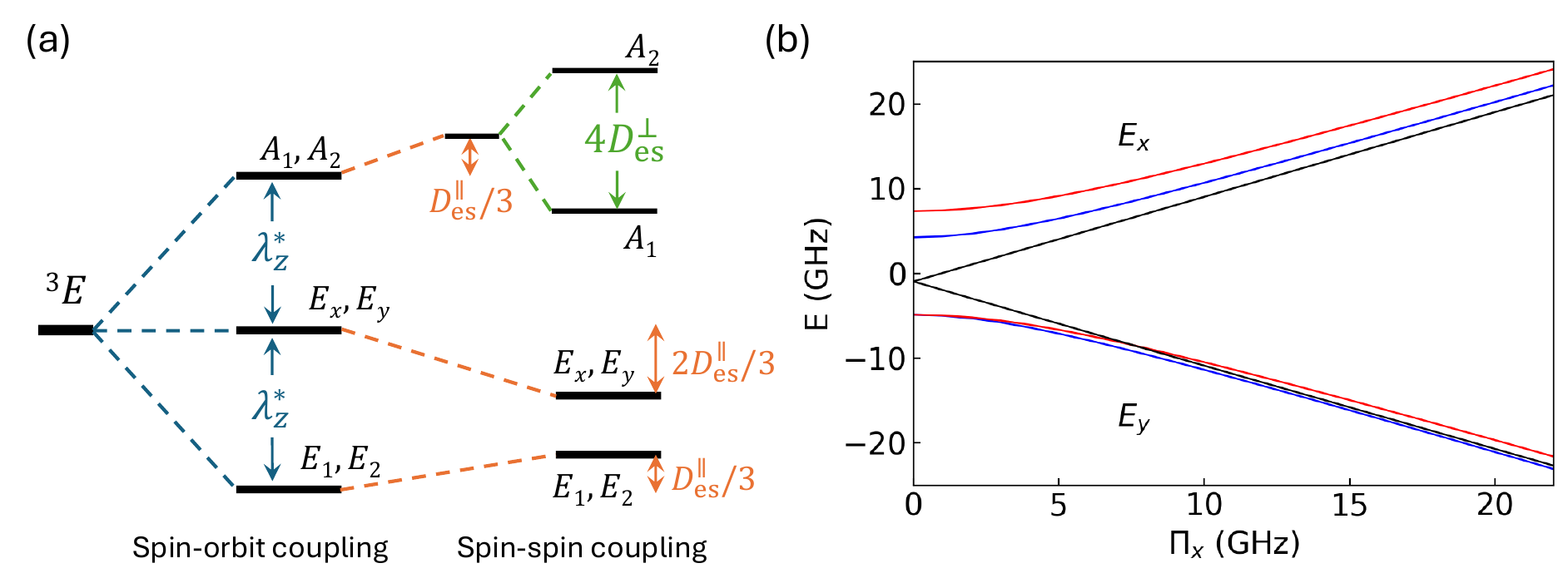}
    \caption{(a). The ${}^3\!E$ manifold is split by spin-orbit coupling $\lambda_z^* = \langle L_z S_z\rangle_{{}^3\!E}$ (blue) into three pairs and is further split by spin-spin coupling (orange and green) into four groups at low temperature. The $D$ parameters are discussed in Sec.~S3.3.2. $^*$The expectation value of diagonal SOC is quenched by the dynamic Jahn-Teller effect~\cite{thiering2017ab}. (b). The degeneracy of ${}^3\!E$ is lifted by $2\Pi_\perp = 2\Pi_x$ crystal strain and it splits into two branches.}
    \label{fig:3E_splittings}
\end{figure}

Stress can modify the energy gaps between these electronic states. Take ${}^3\!E$ as an example, its sublevels are coupled to stress via:
\begin{equation}
\begin{split}
    H_{\sigma} & = \Pi_z \left(\ket {E_x} \bra {E_x}+\ket {E_y} \bra {E_y} \right)+\Pi_x \left(\ket {E_y} \bra {E_y}-\ket {E_x} \bra {E_x} \right)+\Pi_y \left(\ket {E_x} \bra {E_y}+\ket {E_y} \bra {E_x} \right)\\
    & = \Pi_z I - \Pi_x P_z + \Pi_y P_x,
\end{split}
\end{equation}
with
\begin{subequations} \label{eq:susceptibilities}
\begin{align}
    \Pi_z &= \alpha_1^{\left({}^3\!E\right)} (\sigma_{yy}+\sigma_{xx}) + \beta_1^{\left({}^3\!E\right)} \sigma_{zz}, \\
    \Pi_x &= \alpha_2^{\left({}^3\!E\right)} (\sigma_{yy}-\sigma_{xx}) + \beta_2^{\left({}^3\!E\right)} (2\sigma_{xz}), \\
    \Pi_y &= \alpha_2^{\left({}^3\!E\right)} (2\sigma_{xy}) + \beta_2^{\left({}^3\!E\right)} (2\sigma_{yz}),
\end{align}
\end{subequations}
assuming a linear coupling between the electronic states and stress. These stress susceptibilities were measured near ambient conditions as $\left\{\alpha_1^{\left({}^3\!E\right)}, \beta_1^{\left({}^3\!E\right)}, \alpha_2^{\left({}^3\!E\right)}, \beta_2^{\left({}^3\!E\right)}\right\} = \left\{1295, -1523, -645, -89\right\}$~GHz/GPa~\cite{davies1976optical}. $\Pi_z$ represents the energy shift from symmetry-preserving stress, while $\Pi_x, \Pi_y$ represent energy splittings (within ${}^3\!E$) induced by symmetry-breaking stress, which lift the degeneracy of the two branches of ${}^3\!E$ by $2\Pi_\perp = 2\sqrt{\Pi_x^2+\Pi_y^2}$ [Fig.~\ref{fig:3E_splittings}(b)]. In the large stress limit, the states form two new orbital branches given by:
\begin{equation*}
\ket{E_{x^\prime}}\left \{
\begin{array}{l}
      \ket{0} \\
      \ket{\pm 1} 
\end{array} 
\right. , \quad \quad 
\ket{E_{y^\prime}}\left \{
\begin{array}{l}
      \ket{0} \\
      \ket{\pm 1}  
\end{array} 
\right.
\end{equation*}
where $\ket{E_{x^\prime}} = \cos \theta \ket {E_x}+ \sin \theta \ket {E_y}$ and $\ket {E_{y^\prime}} = -\sin \theta \ket {E_x}+ \cos \theta \ket {E_y}$. The phase $\theta$ is determined by the orientation of stress, i.e.~$\tan \theta = \Pi_y/\Pi_x$. Therefore, when $\Pi_y = 0$, which occurs for [100] stress, the orbital branches remain unmixed ($\theta = 0$). Stress will also alter the energy gaps between the singlet states and the ground state, which is difficult to measure. In Sec.~S5.2, we present these aformentioned susceptibilities computed from first principles.

\subsection{Spin-orbit coupling}
In this subsection, we discuss spin-orbit coupling. The spin-orbit Hamiltonian is defined as 
\begin{equation}
    H_{\text{so}} = \lambda_z L_z S_z + \lambda_\perp (L_x S_x + L_y S_y),
\end{equation}
where $S$ is a spin-1 operator and $L$ represents the angular momentum.

\subsubsection{Ambient condition}
\begin{figure}
    \centering
    \includegraphics[width=0.9\textwidth]{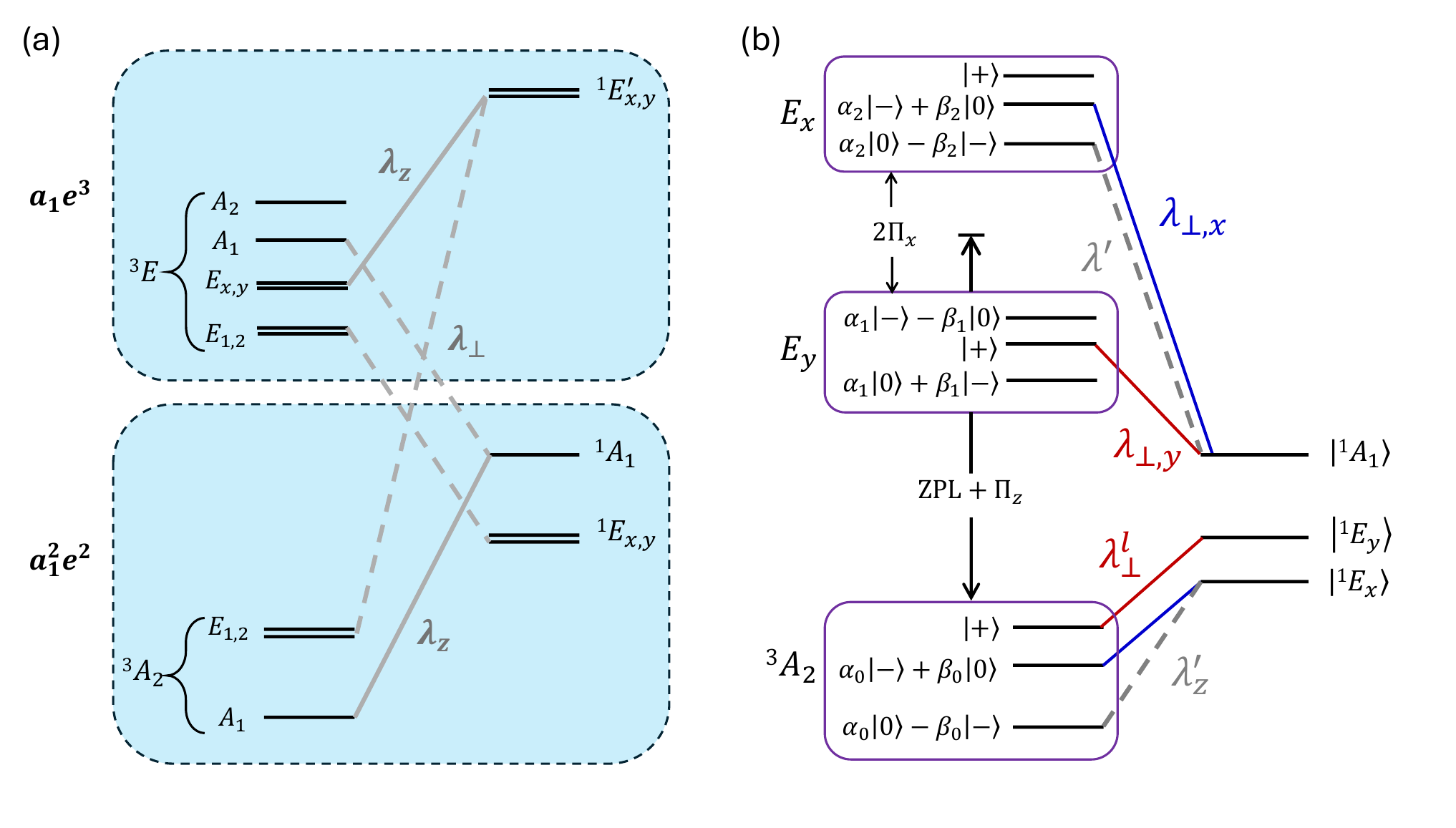}
    \caption{(a). Spin-orbit couplings between NV center electronic states in the minimum model (up to first order) at ambient conditions or under symmetry-preserving stress. The dashed lines represent coupling by $\lambda_\perp$ and the solid lines represent coupling by $\lambda_z$. We note that Ref.~\cite{maze2011properties} termed the two branches of the lowest singlet manifold as ${}^1\!E_{1,2}$. Here, we label them as ${}^1\!E_{x,y}$ for consistency. (b). SOCs between triplet and singlet levels relevant for ISC under $\Pi_x$ stress, where the color codes the triplet spin sublevel of these couplings. Stress-induced SOCs are represented as dashed lines. Notably, coupling between $\left|{}^1\!E_x\right\rangle$ and $\left|{}^3\!A_2^0\right\rangle$ is mediated by $\lambda_z^\prime$ which is beyond our group theoretical treatment. Stress-induced SSC is only between $|0\rangle$ and $|-\rangle$ sublevels, with $\beta$ defined by Eq.~\ref{eq:ssc_beta_0} and Eq.~\ref{eq:ssc_beta_12} and $|\alpha|^2 + |\beta|^2 = 1$. The SOCs shown do not take SSCs into consideration.}
    \label{fig:spin_orbit_coupling}
\end{figure}

The axial term $\lambda_z$, with experimentally measured magnitude $\sim 5.5$~GHz~\cite{batalov2009low, rogers2009time} at $T < 20$~K (which has been reduced by the dynamic Jahn-Teller (DJT) effect, to be discussed in the following subsection), splits the six sublevels in the ${}^3\!E$ manifold into three degenerate pairs in the absence of other perturbations [Fig.~\ref{fig:3E_splittings}(a)]. These states, in terms of spin and orbital degrees of freedom, are given by\footnote{What might be surprising at first glance is that $|E_x, 0\rangle$ transforms as $E_y$. This is because $|S=1, m_s=0\rangle$ transforms as $A_2$. Therefore, to avoid potential confusion, we use ket, i.e., $|E_x\rangle, |E_y\rangle$ to represent only the \textbf{orbital state} throughout, and we will refer to the six states within ${}^3\!E$ as sublevels.}:
\begin{equation} \label{eq:states}
\begin{array}{lll}
      A_1 &= \frac{1}{\sqrt{2}}\left(\ket{E_-}\ket{1}-\ket{E_+}\ket{-1}\right) & \quad \quad \quad(A_1 \textrm{ symmetry})\\
      A_2 &= \frac{1}{\sqrt{2}}\left(\ket{E_-}\ket{1}+\ket{E_+}\ket{-1}\right) & \quad \quad \quad(A_2) \\
      E_y &= -\ket{E_x}\ket{0} & \quad \quad \quad(E_y)\\
      E_x &= \ket{E_y}\ket{0} & \quad \quad \quad(E_x)\\
      E_1 &= \frac{1}{\sqrt{2}}\left(\ket{E_-}\ket{-1}-\ket{E_+}\ket{1}\right) & \quad \quad \quad(E_x)\\
      E_2 &= \frac{1}{\sqrt{2}}\left(\ket{E_-} \ket{-1} + \ket{E_+}\ket{1}\right) & \quad \quad \quad(E_y).
\end{array} 
\end{equation}
Here $\ket{E_{\pm}}$ ($\ket{E_{\pm}} = \frac{1}{\sqrt{2}} \left(\ket{E_x} \pm i\ket{E_y}\right)$) are the degenerate orbital states and $L_z \ket {E_\pm} = \pm \ket {E_\pm}$. $\lambda_z$ can also couple the triplet and singlet states--it couples the $E_{x,y}$ sublevels of ${}^3\!E$ and ${}^1\!E'$; it also couples the $|m_s=0\rangle$ sublevel of ${}^3\!A_2$ with ${}^1\!A_1$ [Fig.~\ref{fig:spin_orbit_coupling}(a)]. We have $\hbar\lambda_z = \frac{1}{2} \left\langle{}^1\!A_1|H_{\text{so}}|{}^3\!A_2^0\right\rangle$. However, no ISCs can happen between these states due to their energy gap being too large.

Before we move on to the transverse term $\lambda_{\perp}$, let us first officially define the orbital Pauli operators: $P_x = |E_x\rangle\langle E_y| + |E_y\rangle\langle E_x|$, $P_y = -i|E_x\rangle \langle E_y| + i|E_y\rangle\langle E_x|$, and $P_z = |E_x\rangle\langle E_x| - |E_y\rangle\langle E_y|$. Therefore, we have $P_x \sim L_yL_x + L_xL_y$, $P_y \sim L_z$ and $P_z \sim L_x^2 - L_y^2$. In the minimum model of NV center, $L$ operators are defined on the nine-dimensional space spanned by the single-particle orbital basis $\{a_1, e_x, e_y\}^{\otimes 4}$, while $P$ operators are defined on the two-dimensional subspace spanned by the orbital state basis $\left\{|E_x\rangle, |E_y\rangle\right\}$. By $\sim$ we mean that these operators are equivalent when restricted to the two-dimensional subspace. In this sense, the axial term of the spin-orbital interaction in the excited manifold can be written as: $H_{\text{so}}^\parallel = \lambda_z P_y S_z$~\cite{doherty2013nitrogen}.

The transverse term $\lambda_{\perp}$ couples the $A_1, E_{1, 2}$ sublevels of ${}^3\!E$ to the singlet ${}^1\!A_1, {}^1\!E$ states respectively\footnote{In the minimum model, we have $\hbar\lambda_\perp = \frac{1}{\sqrt{2}} \left\langle A_1\left|H_{\text{so}}\right| {}^1\!A_1\right\rangle = \frac{1}{\sqrt{2}} \left\langle E_{1,2}\left|H_{\text{so}} \right|{}^1\!E_{x,y}\right\rangle = \left\langle {}^3\!A_2^{\pm}\left|H_{\text{so}} \right|{}^1\!E^\prime_{x,y}\right\rangle $~\cite{doherty2011negatively}. Note that we only focus on the magnitude of these couplings and omit their phase in this work, unless otherwise stated.} [Fig.~\ref{fig:spin_orbit_coupling}(a)]. However, only the former pair leads to ISC due to the relatively small energy gap. Up to first order, there is no direct coupling between ${}^1\!E$ and the ${}^3\!A_2$ ground manifold. Therefore, higher order effects, e.g., multi-configurational interaction of electrons and the pseudo Jahn-Teller (PJT) interaction~\cite{thiering2018theory, li2024excited}, have to be considered when studying these ISCs to complete the optical cycle, which we will discuss in more detail in the following subsections.

\subsubsection{Effects of stress}
Under stress, we write the stress-coupled spin-orbit Hamiltonian using the principles discussed in Sec.~\ref{sec:group_theory} as:
\begin{equation}
\begin{split}
    H_{\text{so}} & = \left(\lambda_z + \Pi_z^{(1)}\right) L_z\otimes S_z + \left(\lambda_\perp + \Pi_z^{(2)}\right) \left(L_y \otimes S_y + L_x\otimes S_x\right)\\
    & + \Pi_x^{(1)} L_x\otimes S_z + \Pi_y^{(1)} L_y \otimes S_z\\
    & + \Pi_x^{(2)} L_z\otimes S_x + \Pi_y^{(2)} L_z \otimes S_y\\
    & + \Pi_x^{(3)} \left(L_y \otimes S_y - L_x \otimes S_x\right) + \Pi_y^{(3)} \left(L_y \otimes S_x + L_x \otimes S_y\right).
\end{split}    
\end{equation}
From this Hamiltonian, we can calculate the possible stress-induced SOCs between states in different manifolds directly. We summarize the results as follows. For couplings between the ${}^3\!E$ manifold and ${}^1\!A_1$, there are three possibilities: 
\begin{enumerate}
    \item $\Pi_z^{(2)}$ represents the stress dependency of $\lambda_{\perp}$.
    \item $\Pi_{x(y)}^{(1)}$ can induce couplings from sublevels $E_{x(y)}$\footnote{Note that sublevels $E_{1, 2}$ also follow the same irrep as $E_{x, y}$. However, couplings are only possible for $E_{x, y}$ because $S_z$ operator cannot couple two states with different spin projections and $E_{1,2}$ have spin $\pm$.}.
    \item $\Pi_{x(y)}^{(3)}$ can induce couplings from sublevels $E_{1(2)}$.
\end{enumerate}
For couplings between the ${}^3\!A_2$ manifold and ${}^1\!E$, there is one pair\footnote{Although the $L, S$ operators are defined w.r.t. ${}^3\!E$, we generalize it to the couplings between ${}^1\!E$ and ${}^3\!A_2$. This, however, cannot guarantee a complete search of all non-vanishing SOC matrix elements, which would require first principles calculations.}:
\begin{enumerate}
    \item $\Pi_{x}^{(2)}$ can induce couplings between $\left|{}^1\!E_{x(y)}\right\rangle$ and $\left|{}^3\!A_2^{-(+)}\right\rangle$\footnote{Here we have assumed $|\pm\rangle = \frac{1}{\sqrt{2}}\left(|m_s=1\rangle \pm |m_s=-1\rangle\right)$, which will be officially defined in the next subsection.}; $\Pi_{y}^{(2)}$ can induce couplings between $\left|{}^1\!E_{x(y)}\right\rangle$ and $\left|{}^3\!A_2^{+(-)}\right\rangle$.
\end{enumerate}
Note that these results are not solely from group theory, because we use the information of the explicit orbital and spin operators. Nevertheless, we see that only the $E$ stress can couple a state with $E$ symmetry to a state with $A_1$ symmetry. Likewise, only the $A_1$ stress can couple a state with $A_1$ symmetry to a state with $A_1$ symmetry. One can also understand that there is no coupling for $A_2$ since we do not have a stress component transforming as $A_2$.

In the large [100] stress limit, $\Pi_y = 0$ and the two orbital branches tend to separate and not mix. To facilitate discussions of ISCs in the following section, we explicitly write out the  stress perturbation part of the spin-orbit Hamiltonian from [100] stress:
\begin{equation} \label{eq:soc_001}
\begin{split}
   H_{\text{so}}' & = \underbrace{\left[\chi_1 \left(\sigma_{xx}+\sigma_{yy}\right) + \chi_1' \sigma_{zz}\right]}_{\Pi_z^{(1)}} L_z\otimes S_z\\ 
    & + \underbrace{\left[\chi_2 \left(\sigma_{xx}+\sigma_{yy}\right) + \chi_2' \sigma_{zz}\right]}_{\Pi_z^{(2)}} \left(L_y \otimes S_y + L_x\otimes S_x\right)\\
    & + \underbrace{\left[\chi_3 (\sigma_{yy} - \sigma_{xx}) + \chi_3'(\sigma_{xz} + \sigma_{zx})\right]}_{\Pi_x^{(1)}} L_x \otimes S_z\\ 
    & + \underbrace{\left[\chi_4 (\sigma_{yy} - \sigma_{xx}) + \chi_4' (\sigma_{xz} + \sigma_{zx})\right]}_{\Pi_x^{(2)}} L_z\otimes S_x\\ 
    & + \underbrace{\left[\chi_5 \left(\sigma_{yy} - \sigma_{xx}\right) + \chi_5' \left(\sigma_{xz} + \sigma_{zx}\right)\right]}_{\Pi_x^{(3)}} \left(-L_x\otimes S_x + L_y\otimes S_y\right),
\end{split}
\end{equation}
where the $\chi$ represents the SOC stress susceptibility. Under large [100] stress, the excited state orbital degeneracy is lifted and ${}^3\!E$ becomes two orbital branches separated by $2\Pi_x$. Therefore, the parallel spin-orbit term is greatly suppressed, leading to $\left\{|E_x\rangle, |E_y\rangle\right\}$ being a good basis. We identify out which electronic sublevels undergo ISC into the singlet state by rewriting the original ${}^3\!E$ manifolds in this basis. For example, the $A_1$ sublevel of ${}^3\!E$ can be rewritten as
\begin{equation}
    A_1 = \frac{1}{\sqrt{2}}\left(|E_y\rangle|-\rangle - |E_x\rangle|+\rangle\right).
\end{equation}
We see that the $|E_{x}, +\rangle$ and $|E_y, -\rangle$ sublevels would still undergo the ISC process mediated by $\left(\lambda_{\perp} + \Pi_z^{(2)}\right)$. The $E_{1,2}$ sublevel can also be rewritten as the following:
\begin{align}
    E_1 & = -\frac{1}{\sqrt{2}}\left(|E_x\rangle|+\rangle + |E_y\rangle|-\rangle\right),\\
    E_2 & = \frac{1}{\sqrt{2}}\left(|E_y\rangle|+\rangle + |E_x\rangle|-\rangle\right).
\end{align}
Since $\Pi_x^{(3)}$ couplings include SOC from the $E_1$ sublevel, we see that they act on the same branches as those with $\lambda_{\perp}$. $\Pi_x^{(1)}$ induces ISC from $|E_y, 0\rangle$ to $|{}^1\!A_1\rangle$, which is a \textbf{new route} induced by stress. 

The ISCs from ${}^1\!E$ to $|{}^3\!A_2^{\pm}\rangle$ also become first order upon [100] stress, as induced by $\Pi_x^{(2)}$. Finally, we found through \emph{ab initio} calculations another emerging SOC matrix element, $\lambda_z'$ that couples $|{}^1\!E_x\rangle$ and $\left|{}^3\!A_2^0\right\rangle$. This matrix element escapes our group theory, and we term its susceptibilities as $\chi_6, \chi_6^\prime$. These susceptibilities have never been measured or calculated before, and we provide the first numerical estimation for them in Sec.~S5.2.
\clearpage

\subsection{Spin-spin coupling}

\subsubsection{Ground state}
We have discussed before that the ground state is split by $D_{\text{gs}}$ between $|m_s=0\rangle$ and $|m_s=\pm\rangle$. With applied stress, the spin-spin interaction in the ground state can be expressed as:
\begin{equation}
\begin{split}
    H_{\text{ss}} & = \left(D_{\text{gs}}+\Pi^{(1)}_z\right) S_z^2 + \Pi^{(1)}_x \left(S_y^2-S_x^2\right) + \Pi^{(1)}_y \left(S_x S_y+S_y S_x\right) \\ 
    & + \Pi^{(2)}_x (S_x S_z+S_z S_x)+\Pi^{(2)}_y (S_y S_z+S_z S_y) + \dots,
\end{split}
\end{equation}
with $\Pi^{(i)}_{x, y, z}$ having analogous definitions to Eqs.~\ref{eq:susceptibilities}, with $\{\alpha^{(1)}_1, \beta^{(1)}_1, \alpha^{(1)}_2 , \beta^{(1)}_2\}= 2\pi \times \{8.6(2), -2.5(4), 1.95(9), 4.50(8)\}$ MHz/GPa being the stress susceptibilities~\cite{udvarhelyi2018spin}. The spin eigenvectors are $|\pm\rangle = \frac{1}{\sqrt{2}}\left(|m_s=1\rangle \pm e^{i\phi_{\Pi}}|m_s=-1\rangle\right)$, with $\phi_{\Pi} = \arctan (\Pi_y/\Pi_x)$, up to first order. Note that the sign of the last two susceptibilities is not yet known from experiments, but can be obtained from first principles~\cite{udvarhelyi2018spin}.

The SSC determines the order in energy of the three spin sublevels of ${}^3\!A_2$. Under [100] stress, we only need to consider $\Pi_x$. $\Pi_x^{(1)}$ is responsible for mixing $|m_s=\pm1\rangle$ into $|\pm\rangle$ to form the eigenstates of the Hamiltonian, and this term looks like $\begin{pmatrix}
    0 & 0 & -\Pi_x^{(1)}\\
    0 & 0 & 0\\
    -\Pi_x^{(1)} & 0 & 0
\end{pmatrix}$. Since $\Pi_x^{(1)} < 0$, we have $|+\rangle$ being higher in energy compared to $|-\rangle$ and the gap is $2\left|\Pi_x^{(1)}\right|$. $\Pi_x^{(2)}$ is responsible for inducing spin-mixing. Since $\left|\Pi_x^{(2)}\right| \ll \left|\Pi_x^{(1)}\right|$, we can use perturbation theory to estimate the magnitude of spin-mixing as:
\begin{equation} \label{eq:ssc_beta_0}
    \beta_0 \approx \frac{\Pi_x^{(2)}}{\triangle}\Big\langle -\big|S_x S_z+S_z S_x\big|0\Big\rangle = -\frac{\Pi_x^{(2)}}{D_{\text{gs}}+\Pi^{(1)}_z + \Pi_x^{(1)}}.
\end{equation}
Note that spin mixing is only between $|-\rangle$ and $|0\rangle$.

\subsubsection{Excited state}
For excited states, the SSC at ambient condition can also be expressed in the following form \cite{doherty2013nitrogen}, 
\begin{equation}
\begin{split}
H_{\text{ss}} & = D_{\text{es}}^\parallel (S_z^2 - 2/3) + D_{\text{es}}^\perp\left[P_z\otimes\left(S_y^2 - S_x^2\right) - P_x\otimes\left(S_xS_y + S_yS_x\right)\right] \\ 
& + D_{\text{es}}^{\perp'}\left[P_z\otimes\left(S_xS_z + S_zS_x\right) - P_x\otimes\left(S_yS_z + S_zS_y\right)\right],
\end{split}
\end{equation}
where $D_{\text{es}}^\parallel = 3\Delta$, $D_{\text{es}}^\perp = \Delta'$ and $D_{\text{es}}^{\perp'} = -\Delta''/\sqrt{2}$ compared with Eq.(8) in Ref.~\cite{maze2011properties} [Fig.~\ref{fig:3E_splittings}(a)]. With stress applied, the Hamiltonian can be expressed as
\begin{equation}
\begin{split}
    H_{\text{ss}} & = \left(D_{\text{es}}^\parallel + \Pi_z^{(1)}\right) \left(S_z^2 - 2/3\right) + \left(D_{\text{es}}^\perp + \Pi_z^{(2)}\right) \left[P_z\otimes\left(S_y^2 - S_x^2\right) - P_x\otimes\left(S_xS_y + S_yS_x\right)\right] \\ 
    & + \left(D_{\text{es}}^{\perp'} + \Pi_z^{(3)}\right) \left[P_z\otimes\left(S_xS_z + S_zS_x\right) - P_x\otimes\left(S_yS_z + S_zS_y\right)\right] \\
    & + \Pi^{(1)}_x (S_y^2-S_x^2) + \Pi^{(1)}_y (S_x S_y+S_y S_x) \\
    & + \Pi^{(2)}_x (S_x S_z+S_z S_x) + \Pi^{(2)}_y (S_y S_z+S_z S_y) \\
    & + \Pi^{(3)}_x P_z\otimes S_z^2 - \Pi^{(3)}_y P_x\otimes S_z^2 + \dots,
\end{split}
\end{equation}
which is truncated at third order. The higher order terms can also be derived based on Table~\ref{table:irrep_operators}.

In the large [100] stress limit, we can again resort to perturbation theory to analyze how different spin sublevels within each orbital branch are ordered, where the spin-spin Hamiltonian is treated as a perturbation. The $\left(D_{\text{es}}^{\parallel} + \Pi_z^{(1)}\right)$ term separates the $|\pm 1\rangle$ state from the $|0\rangle$ state and it is independent of the orbital branch. The $\left(D_{\text{es}}^{\perp} + \Pi_z^{(2)}\right)$ term is responsible for mixing $|\pm1\rangle \rightarrow |\pm \rangle$ to form the eigenstates and the reason is the same as the ground state case. Note that $P_x\otimes\left(S_xS_y + S_yS_x\right)$ in the square bracket of equation S18 involves the $P_x$ operator, which mixes the two orbital branches and is therefore greatly suppressed by stress; we therefore ignore it. As for the third term, since $P_z = |E_x\rangle\langle E_x| - |E_y\rangle\langle E_y|$, we can draw the conclusion that the order of $|+\rangle, |-\rangle$ spin sublevels will be different for the two orbital branches. $D_\text{es}^{\perp}\sim1.55/2$ GHz at ambient conditions, and we assume for now that $\left(D_{\text{es}}^{\perp} + \Pi_z^{(2)}\right)$ does not change sign under stress. Then, for the $|E_x\rangle$ branch (lower in energy than $|E_y\rangle$), $|+\rangle$ will be lower in energy than $|-\rangle$ and vice versa for the $|E_y\rangle$ branch, with the same reasoning as used for the ground state. The fourth and fifth terms do not explicitly involve the orbital operator and function in a similar way as in the ground state. Finally, the last term will modify the gap between $|0\rangle$ and $|\pm\rangle$ sublevels. The spin mixing between $|-\rangle$ and $|0\rangle$ can therefore be estimated as:
\begin{equation} \label{eq:ssc_beta_12}
    \beta_{1,2} \approx - \frac{D_{\text{es}}^{\perp'} + \Pi_z^{(3)} \mp \Pi_x^{(2)}}{D_{\text{es}}^{\parallel} + \Pi_z^{(1)} \pm \Pi_x^{(3)} \pm \left(D_{\text{es}}^{\perp} + \Pi_z^{(2)} \pm \Pi_x^{(1)}\right)}.
\end{equation}
The effects of SSC on the wavefunction of the NV's triplet states under [100] stress are shown in Fig.~\ref{fig:spin_orbit_coupling}(b). We will present a first numerical estimation of the $\beta$s in Sec.~S5.2.

\subsection{Jahn-Teller effects}
As we have seen, the NV center has high point group symmetry. Therefore, it is liable to undergo spontaneous symmetry breaking due to the Jahn-Teller (JT) interaction~\cite{abtew2011dynamic}. The ${}^3\!E$ and ${}^1\!E$ manifolds are degenerate, the dominant JT interactions are also referred to as dynamic and static respective~\cite{benedek2024comprehensive}. In addition, because of the relatively small gap between singlet states, there exists the pseudo-JT (PJT) interaction. In this subsection, we will review the important conclusions related to the JT effects, with a focus on how they affect the ISCs and the optical cycle.

\begin{figure}
    \centering
    \includegraphics[width=0.9\textwidth]{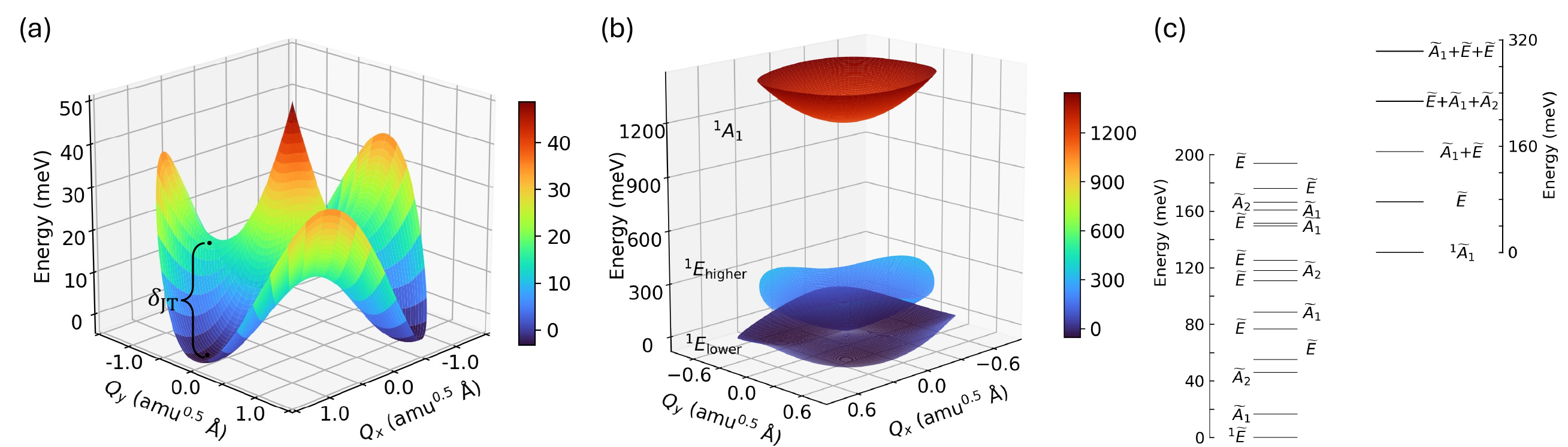}
    \caption{(a). Adiabatic potential energy surface (PES) of the ${}^3\!E$ manifold under the dynamic Jahn-Teller interaction in the two-effective-phonon model. The PES has a Mexican-hat shape with three local minima separated by a barrier $\delta_{\text{JT}}$. (b). Adiabatic PES of ${}^1\!A_1, {}^1\!E$ under the pseudo Jahn-Teller interaction in the two-effective phonon model. (c). Vibronic states of the singlets from solving the pseudo Jahn-Teller Hamiltonian. All the numerical calculations in this figure are obtained at ambient condition, and therefore would change with stress.}
    \label{fig:Jahn_Teller}
\end{figure}

\subsubsection{Dynamic Jahn-Teller effects of the triplet excited manifold}
Since the triplet excited-state manifold ${}^3\!E$ is an orbital doublet, it would couple to an $e$ vibrational mode, resulting in an $E\otimes e$ JT system. Within the quadratic vibronic model~\cite{abtew2011dynamic, thiering2017ab}, the JT interaction mixes the two orbital branches of ${}^3\!E$, and leads to three local minima away from the $C_{3v}$ geometry, forming a Mexican-hat-like adiabatic potential energy surface (PES) [Fig.~\ref{fig:Jahn_Teller}(a)]. These local minima are separated by the JT barrier $\delta_{\text{JT}}$. One notable consequence of JT interaction is the damping of $\lambda_z$, producing a so called Ham reduction factor $p\sim 0.304$~\cite{thiering2017ab}. For ISCs, the DJT effect enables sublevels without direct SOC to ${}^1\!A_1$ to inter-system cross, which agrees with experiments at low temperatures~\cite{jin2025first}.

For nuclear motion, there exists the zero-point vibrational mode which cannot be ignored. For vibrational modes with $h\nu_e > \delta_{\text{JT}}$, the zero-point motion has enough energy to drive the atoms to flip between the three local minima (regardless of the temperature) of the PES, which is usually referred to as the DJT effect. Therefore, the net geometry of the system should still preserve the highest symmetry. This is the case of the ${}^3\!E$ manifold of the NV center. For elevated temperatures, higher vibrational modes get unfrozen, and the electronic and nuclear motion gradually decouple. Since the contrast measurements that we are trying to elucidate in this work are all taken at room temperature, we will not consider the DJT effect in the simulations and we comment on potential errors this approximation might have in Sec.~\ref{sec:errors}.

\subsubsection{Pseudo Jahn-Teller effects of the singlet states}
From Sec.~S3.2.1, we have seen that there is no SOC between ${}^1\!E$ and the ${}^3\!A_2$ manifold up to first order at ambient conditions, leaving the optical cycle incomplete. Therefore, we have to look for higher order effects that contribute to these ISCs, and that is the PJT effects of singlet states.

The PJT effects have been extensively studied in Ref.~\cite{thiering2018theory}, and we mostly followed their derivations. Since ${}^1\!E$ and ${}^1\!A_1$ have different irreps, only the symmetry-distorting $e$ vibration modes may couple the two states. We took a two-effective phonon modes approximation and the PJT interaction can be described by the following Hamiltonian\footnote{Strictly speaking, this Hamiltonian also includes DJT effect within ${}^1\!E$ up to the linear coupling. For sake of simplicity, we will be referring the interaction described by this Hamiltonian as PJT interaction throughout this SI.}:
\begin{equation} \label{eq:PJT_Hamiltonian}
\begin{split}
    H_{\text{PJT}} & = H_{\text{e}} + H_{\text{ph}} + H_{\text{e-ph}}\\
    & = \sum_{i \in {}^1\!E, {}^1\!A_1} E_i a_i^\dagger a_i + \sum_{k=x, y} \hbar \omega_e \left(b_k^\dagger b_k + \frac{1}{2}\right) + \sum_{ij}\sum_{k=x, y} g_{ij,k} a_i^\dagger a_i\left(b_k^\dagger + b_k\right),
\end{split}
\end{equation}
where we have assumed $\omega_e$ to be the effective phonon frequency and defined $g$ to be the linear electron-phonon coupling strength between electronic state $i, j$ and phonon mode $k$. We direct the readers interested in learning how to compute the $g$ matrices to Ref.~\cite{jin2022vibrationally} for more details.

The effect of PJT interaction is to mix the electronic and vibrational degrees of freedom and, therefore, mix singlet states into a series of vibronic states [Fig.~\ref{fig:Jahn_Teller}(c)]. We can classify them according to their irrep, and each irrep exhibits a general wavefunction format as:
\begin{gather}
    \left|\widetilde{E_{x}}\right\rangle = \sum_i^{\infty} \left[c_i |{}^1\!E_{x}\rangle |\chi_i(A_1)\rangle + d_i|{}^1\!A_1\rangle|\chi_i(E_{x})\rangle + \frac{f_i}{\sqrt{2}}\left(|{}^1\!E_x\rangle|\chi_i(E_x)\rangle - |{}^1\!E_y\rangle|\chi_i(E_y)\rangle\right) - g_i|{}^1\!E_y\rangle|\chi_i(A_2)\rangle\right], \label{eq:PJT_wfc_Ex}\\
    \left|\widetilde{E_{y}}\right\rangle = \sum_i^{\infty} \left[c_i |{}^1\!E_{y}\rangle |\chi_i(A_1)\rangle + d_i|{}^1\!A_1\rangle|\chi_i(E_{y})\rangle - \frac{f_i}{\sqrt{2}}\left(|{}^1\!E_x\rangle|\chi_i(E_y)\rangle + |{}^1\!E_y\rangle|\chi_i(E_x)\rangle\right) + g_i|{}^1\!E_x\rangle|\chi_i(A_2)\rangle\right], \label{eq:PJT_wfc_Ey}\\
    \left|\widetilde{A_1}\right\rangle = \sum_i^{\infty} \left[c'_i |{}^1\!A_1\rangle |\chi_i(A_1)\rangle + \frac{d'_i}{\sqrt{2}}\left(|{}^1\!E_x\rangle|\chi_i(E_x)\rangle + |{}^1\!E_y\rangle|\chi_i(E_y)\rangle\right)\right], \label{eq:PJT_wfc_A1}
\end{gather}
where the tilde hat represents the vibronic state. In the equations above, we have used $|\chi_i(\Gamma)\rangle$ to represent the $i$th phonon wavefunctions with irrep $\Gamma$. These symmetric phonon wavefunctions are constructed from the two effective phonon modes (that transform as $E_x, E_y$ respectively), and we summarize them in the Tab.~\ref{tab:phonon_wfc_irrep} with small phonon occupation numbers.

\begin{turnpage}
{\renewcommand{\arraystretch}{2.0}
\begin{table}[hbt!]
    \centering
    \caption{Symmetry-adapted phonon wavefunctions under the two-effective-phonon approximation. The empty block means the corresponding wavefunction does not exist.}
    \begin{tabular}{c|c|c|c}
    \hline
    \hline
    $n_i$ & $A_1$ & $A_2$ & $E = \{E_x, E_y\}$ \\
    \hline
    $0$ & $|0,0\rangle$ & & \\
    \hline
    $1$ & & & $\left\{|x\rangle=b_{x}^{+}|0,0\rangle, |y\rangle=b_{y}^{+}|0,0\rangle\right\}$\\
    \hline
    $2$ & $|x^2 + y^2\rangle$ & & $\left\{|x^2 - y^2\rangle, |-2xy\rangle\right\}$\\
    \hline
    $3$ & $|x(x^2 - 3y^2)\rangle$ & $|y(3x^2 - y^2)\rangle$ & $\left\{|x(x^2 + y^2)\rangle, |y(x^2 + y^2)\rangle\right\}$\\
    \hline
    \multirow{2}{*}{$4$} & $|(x^2 + y^2)^2\rangle$ & & $\left\{|x^4 - y^4\rangle, |-2xy(x^2 + y^2)\rangle\right\}$\\
    & & & $\left\{|x^4 - 6x^2 y^2 + y^4\rangle, |4xy(x^2 - y^2)\rangle\right\}$\\
    \hline
    \multirow{2}{*}{$5$} & $|x(x^2 + y^2)(x^2-3y^2)\rangle$ & $|y(x^2 + y^2)(3x^2-y^2)\rangle$ & $\left\{\left|x(x^2 + y^2)^2\right\rangle, \left|y(x^2 + y^2)^2\right\rangle\right\}$\\
    & & & $\left\{\left|x^5 - 10x^3 y^2 + 5xy^4\right\rangle, \left|-y^5 + 10x^2 y^3 - 5x^4y\right\rangle\right\}$\\
    \hline
    \multirow{2}{*}{$6$} & $\left|(x^2 + y^2)^3\right\rangle$ & $\left|xy(x^2 - 3y^2)(3x^2 - y^2)\right\rangle$ & $\left\{\left|(x^2+y^2)^2(x^2-y^2)\right\rangle, \left|-2xy(x^2 + y^2)^2\right\rangle\right\}$\\
    & $\left|(x^2 - y^2)(x^4 - 14x^2 y^2 + y^4)\right\rangle$ & & $\left\{\left|(x^2+y^2) (x^4 - 6x^2y^2 + y^4)\right\rangle, \left|4xy(x^2 + y^2)(x^2 - y^2)\right\rangle\right\}$\\
    \hline
    \multirow{3}{*}{$7$} & $\left|x(x^2 + y^2)^2 (x^2 -3y^2)\right\rangle$ & $\left|y(x^2 + y^2)^2 (3x^2 -y^2)\right\rangle$ & $\left\{\left|x(x^2 + y^2)^3\right\rangle, \left|y(x^2 + y^2)^3\right\rangle\right\}$\\
    & & & $\left\{\left|x(x^2 - y^2) (x^4 - 14x^2 y^2 + y^4)\right\rangle, \left|y(x^2 - y^2) (x^4 - 14x^2 y^2 + y^4)\right\rangle\right\}$\\
    & & & $\left\{\left|(x^2 + y^2)(x^5 - 10x^3 y^2 + 5xy^4)\right\rangle, \left|-(x^2 + y^2)(y^5 - 10x^2 y^3 + 5x^4y)\right\rangle\right\}$\\
    \hline
    \multirow{3}{*}{$8$} & $\left|(x^2 + y^2)^4\right\rangle$ & $\left|xy(x^2 + y^2)(x^2 - 3y^2)(3x^2 - y^2)\right\rangle$ & $\left\{\left|(x^2 + y^2)^3(x^2 - y^2)\right\rangle, \left|-2xy(x^2 + y^2)^3\right\rangle\right\}$\\
    & $\left|(x^4 - y^4) (x^4 -14 x^2 y^2 + y^4)\right\rangle$ & & $\left\{\left|(x^2 + y^2)^2 (x^4 - 6x^2 y^2 + y^4)\right\rangle, \left|4xy(x^2 + y^2)^2 (x^2 - y^2)\right\rangle\right\}$\\
    & & & $\left\{\left|(x^2 - y^2)^2 (x^4 - 14x^2 y^2 + y^4)\right\rangle, \left|-2xy(x^2 - y^2)(x^4 -14 x^2 y^2 + y^4)\right\rangle\right\}$\\
    \hline
    \hline
    \end{tabular}
    \label{tab:phonon_wfc_irrep}
\end{table}}
\end{turnpage}

Under symmetry-preserving stress, the above derivations would still hold except that the parameters in the Hamiltonian would be altered by stress. Under symmetry-breaking stress, however, new non-vanishing elements might appear in the linear coupling term in the Hamiltonian. With pressure increasing, the $E$ degeneracy would be lifted, and the system deviates away from a perfect $E\otimes e$ system, leading to the PJT effect being weakened. Nevertheless, we stick to the above analysis for all stress considered as an approximation, keep the phonon-related parameters as constants, and only modify the electronic energies in $H_e$ under stress.

\section{Optical cycle and contrast} \label{sec:optical_cycle}
\begin{figure}
    \centering
    \includegraphics[width=0.7\textwidth]{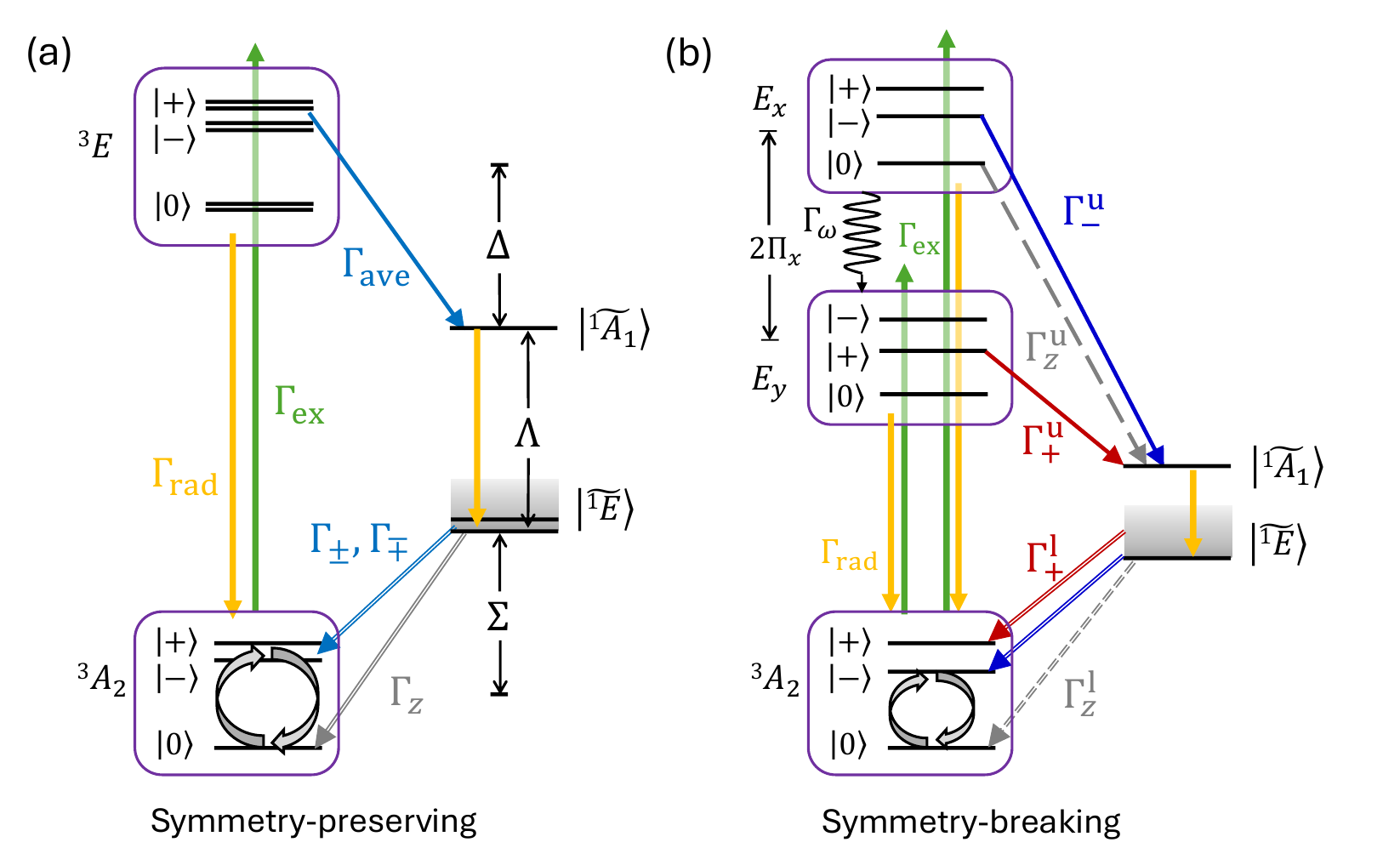}
    \caption{Optical cycle of the NV center under (a) symmetry-preserving stress and (b) symmetry-breaking stress at room temperature. Under symmetry-preserving stress, the degeneracies are preserved, and the upper ISC with rate $\Gamma_{\text{ave}}$ is only from $|\pm\rangle$ sublevels. The lower ISCs are not spin-selective, with $\Gamma_z, \Gamma_{\pm}, \Gamma_{\mp}$ defined in Eqs.~\ref{eq:low_isc_gammaz_symmetrypreserving} and Eqs.~\ref{eq:low_isc_gammaperp_symmetrypreserving}. The energy gaps between ${}^3\!E \leftrightarrow {}^1\!A_1$, ${}^1\!A_1 \leftrightarrow {}^1\!E$, and ${}^1\!E \leftrightarrow {}^3\!A_2$ are denoted as $\Delta, \Lambda, \Sigma$, respectively. Under symmetry-breaking stress, the degeneracies are lifted, with only leading-order ISC processes plotted in (b). The upper ISC rates are defined in Eqs.~\ref{eq:up_isc_gamma}, and the lower ISC rates are defined in Eqs.~\ref{eq:low_isc_gammaz_symmetrybreaking_e} to Eqs.~\ref{eq:low_isc_gammaplus_symmetrybreaking}. The colors in those transitions codes the spin sublevel of the initial/final state for upper/lower ISCs.}
    \label{fig:optical_cycle}
\end{figure}

In this section, we assemble what we have gone through in the previous section to investigate the ISC processes of the NV between the triplet and singlet states, which serve as the building blocks of the optical cycle and spin contrast. There exist two major ISCs in the optical cycle [Fig.~\ref{fig:optical_cycle}(a)]: first,  transitions from ${}^3\!E$ to ${}^1\!A_1$, which we refer to as the ``upper'' ISCs, and  second, transitions from ${}^1\!E$ back to ${}^3\!A_2$, which we refer to as the ``lower'' ISCs. We start with the former.

\subsection{Upper ISC rates}

\subsubsection{Ambient condition}
\begin{figure}
    \centering
    \includegraphics[width=0.9\textwidth]{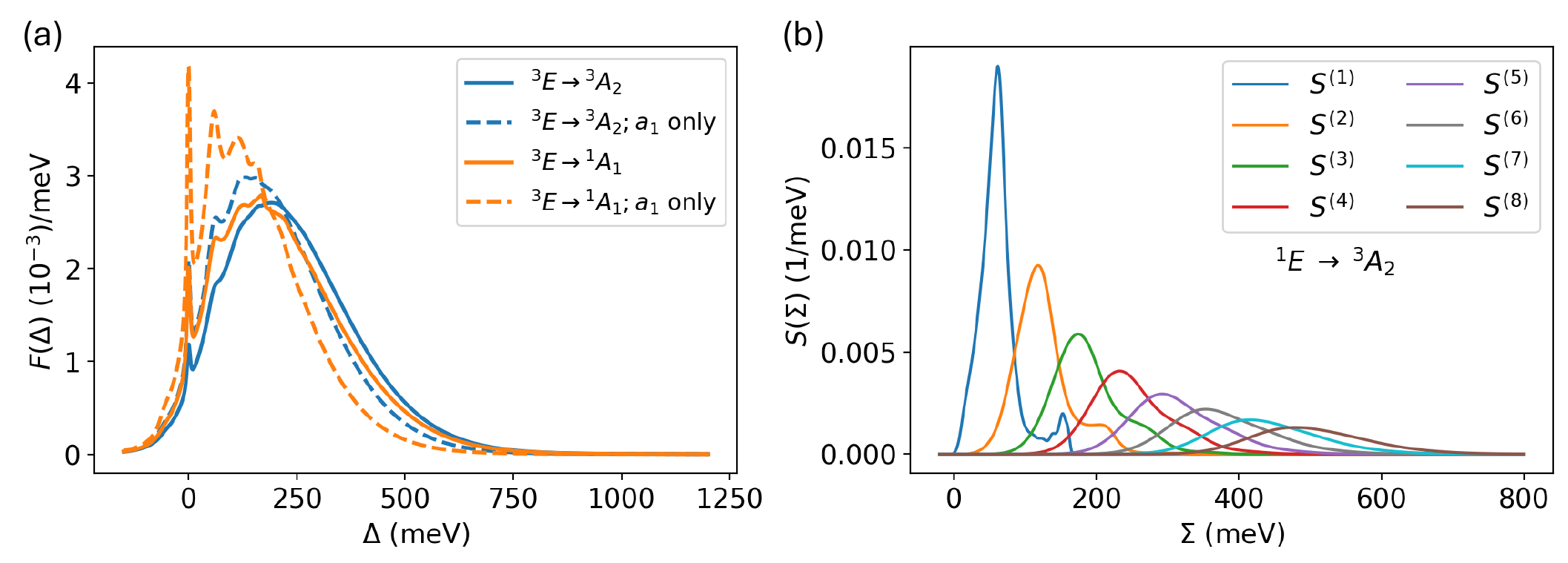}
    \caption{(a). Phonon sideband of the ${}^3\!E \rightarrow {}^3\!A_2$ radiative transition and ${}^3\!E \rightarrow {}^1\!A_1$ transition computed using Huang-Rhys theory~\cite{jin2022vibrationally} at room temperature. The two line-shapes being similar indicates a similar geometry of ${}^3\!A_2$ and ${}^1\!A_1$. For the upper ISCs, only the $a_1$ mode phonons contribute, plotted as the dashed lines. (b). Convoluted phonon overlap spectral functions $S^{(i)}, i\in\{1, 2, \dots, 8\}$ (with the $E$ subscript omitted for simplicity) for the ${}^1\!E \rightarrow {}^3\!A_2$ transition at room temperature, with $S^{(1)}$ being the standard spectral density~\cite{thiering2018theory}.}
    \label{fig:vibration_overlap}
\end{figure}

In the zero-stress limit, the upper ISC is generally believed to be mediated by the transverse spin-orbit interaction $\lambda_{\perp}$ between the $A_1$ sublevel of ${}^3\!E$ and the singlet ${}^1\!A_1$ state~\cite{goldman2015phonon, goldman2015state}. The rate can be computed using Fermi's golden rule~\cite{goldman2015state}:
\begin{equation}
    \Gamma_{A_1} = \frac{2\pi}{\hbar} \big|\langle A_1|H_{\text{so}}|{}^1\!A_1\rangle\big|^2 F(\Delta) = 4\pi \hbar \left|\lambda_{\perp}\right|^2 F(\Delta),
\end{equation}
where we have $\hbar\lambda_{\perp} = \frac{1}{\sqrt{2}} \langle A_1|H_{\text{so}}|{}^1\!A_1\rangle$, and $F(\Delta)$ represents the vibrational overlap function at detuning $\Delta$ between the triplet and singlet states. The $F$ function can be approximated by a fictitious photoluminescence spectrum between ${}^3\!E$ and ${}^1\!A_1$. Historically, this $F$ function is further approximated by the photoluminescence spectrum (from ${}^3\!E$ to ${}^3\!A_2$) from experiments~\cite{goldman2015state}. We verify it to be a reasonable approximation from ab initio calculations, as the blue and orange curves in Fig.~\ref{fig:vibration_overlap}(a) closely resemble each other. Note that only $a_1$ mode phonons contribute to this vibrational overlap, since both the initial and final electronic states transform as $A_1$.

At elevated temperatures, $e$-symmetric phonons could also mediate spin-conserving transitions within the ${}^3\!E$ branch~\cite{ernst2023temperature, ernst2023modeling}, inducing a second-order ISC route from the $E_{1,2}$ sublevels as~\cite{goldman2015state}:
\begin{equation}
    \Gamma_{E_{1,2}} = 8\hbar^2 \big|\lambda_{\perp}\big|^2 \eta \int_0^{\Omega} \omega \Big\{\left[n(\omega) + 1\right]F(\Delta-\omega) + n(\omega)F(\Delta + \omega)\Big\} d\omega.
\end{equation}
The $n(\omega)$ function denotes the thermal occupation of a phonon mode at frequency $\omega$; $\eta=2\pi\times(44.0\pm2.4)$ MHz meV$^{-3}$ parametrizes the coupling strength between the sublevels of ${}^3\!E$ and $e$-symmetric acoustic phonons; and $\Omega = 80$ meV sets a cutoff of acoustic phonons~\cite{goldman2015state}. There are some disputes about the choices of $\Omega$~\cite{ernst2023modeling}, and how these two parameters change under stress remains an open question. Nevertheless, we will assume these two parameters as constants under stress, and since they only affect a second-order rate, this approximation will not qualitatively change our conclusions about spin contrast under stress.

As we have discussed in Sec.~S3.1.1, the ${}^3\!E$ manifold is effectively two levels, with $D_{\text{es}}\sim 1.42$ GHz separating the $|m_s=0\rangle$ and $|m_s=\pm\rangle$ sublevels due to orbital averaging. Therefore, the net ISC rate is computed as $\Gamma_{\text{ave}} = \frac{1}{4}\left(\Gamma_{A_1} + 2\Gamma_{E_{1,2}}\right)$ [Fig.~\ref{fig:optical_cycle}(a)], where the temperature effects are also taken into account in $F(\Delta)$~\cite{goldman2015state, jin2025first}.

\subsubsection{Effects of stress}
As can be easily seen, the above argument naturally carries over to cases when the system experiences symmetry-preserving stress. The only change to be made is to alter the SOC parameters $\lambda_\perp$, and detuning $\Delta$.

The case of symmetry-breaking stress is more complicated. On the one hand, there are two detunings, namely $\Delta_x, \Delta_y$ associated with each branch respectively. On the other hand, the states within the two branches couple to ${}^1\!A_1$ differently via SOC, as we have seen in Sec.~S3.2.2. In addition, with pressure increasing, both orbital averaging and the DJT effect would be greatly suppressed by the increasing gap $2\Pi_\perp$ between the two branches~\cite{ernst2023modeling}. Therefore, we will only keep the first-order ISC processes in this case [Fig.~\ref{fig:spin_orbit_coupling}(b)]. We stick to Fermi's Golden rule to compute these rates as:
\begin{subequations} \label{eq:up_isc_gamma}
\begin{align}
    \Gamma_+ & = 2\pi\hbar \left|\lambda_{\perp, y}\right|^2 F(\Delta_y),\\
    \Gamma_- & = 2\pi\hbar \left|\lambda_{\perp, x}\right|^2 F(\Delta_x),\\
    \Gamma_z & = 2\pi\hbar \left|\lambda^\prime\right|^2 F(\Delta_x),
\end{align}
\end{subequations}
which are also sketched in Fig.~\ref{fig:optical_cycle}(b).

\subsection{Lower ISC rates}

\subsubsection{Ambient condition}
The lower ISCs are more complicated than their upper ISC counterparts. At ambient conditions, the lower ISC involves three distinct rates, i.e., $\Gamma_\pm$ representing rate from $\left|{}^1\!E_{x(y)}\right\rangle \rightarrow \left|{}^3\!A_2^{-(+)}\right\rangle$, $\Gamma_\mp$ for $\left|{}^1\!E_{x(y)}\right\rangle \rightarrow \left|{}^3\!A_2^{+(-)}\right\rangle$, and $\Gamma_z$ for ${}^1\!E \rightarrow \left|{}^3\!A_2^0\right\rangle$. Experiments have measured that these rates are not zero, but roughly one order of magnitude smaller compared to the upper ISCs~\cite{robledo2011spin}. This is because up to first-order, the transition ${}^1\!E \rightarrow |{}^3\!A_2^0\rangle$ is forbidden by symmetry, and the transitions ${}^1\!E \rightarrow |{}^3\!A_2^{\pm}\rangle$, although allowed by symmetry, are also forbidden because the $L_{x, y}$ operator excites an electron, changing the electronic configuration from $a_1^2 e^2$ into $a_1 e^3$. These transition probabilities are non-zero due to higher-order effects. For $\Gamma_{\pm}$ and $\Gamma_{\mp}$, the ${}^1\!E$ many-body wavefunctions contain a portion of the $a_1 e^3$ configurations~\cite{li2024excited} (which is the dominant configuration of the higher ${}^1\!E'$ manifold), and these configurations could couple to $|{}^3\!A_2^{\pm}\rangle$ via $\lambda_{\perp}$. As for $\Gamma_z$, the PJT interaction mixes the singlet states as written in Eqs.~\ref{eq:PJT_wfc_Ex} to~\ref{eq:PJT_wfc_A1}, and $\left|\widetilde{{}^1\!E}\right\rangle$ wavefunctions contain $|{}^1\!A_1\rangle$, which couples to $\left|{}^3\!A_2^0\right\rangle$ via $\lambda_z$. Therefore, ISC rates into the $\left|{}^3\!A_2^0\right\rangle$ sublevel, depending on the irreducible representation of the initial vibronic state, can be computed using Fermi's Golden rule as
\begin{subequations} \label{eq:low_isc_gammaz_symmetrypreserving}
\begin{align}
    \Gamma_z^{\widetilde{E_{x/y}}} & = \frac{2\pi}{\hbar} \sum_j \left|\langle\chi_j|\otimes \left\langle {}^3\!A_2^0\left| H_{\text{so}} \right|\widetilde{E_{x/y}} \right\rangle\right|^2 \delta\left(\Sigma + \varepsilon - E_{\chi_j}\right) \nonumber\\
    & = 8\pi \hbar |\lambda_z|^2 \sum_i^{\infty} d^2_i \underbrace{\sum_j \left|\langle \chi_j\left|\chi_i(E_{x/y})\right\rangle\right|^2 \delta(\Sigma + \varepsilon - n_j \hbar\omega_e)}_{\approx S_E^{(n_i)}(\Sigma + \varepsilon)},\\
    \Gamma_z^{\widetilde{A_1}} & = 8\pi \hbar |\lambda_z|^2 \sum_i^{\infty} c^2_i \underbrace{\sum_j \left|\langle\chi_j\left|\chi_i(A_1)\right\rangle\right|^2 \delta(\Sigma + \varepsilon - n_j\hbar\omega_e)}_{\approx S_{E}^{(n_i)}(\Sigma + \varepsilon)},
\end{align}
\end{subequations}
where $|\chi\rangle$ denotes the phonon wavefunction, and $\varepsilon$ represents the energy of the specific initial vibronic state considered relative to the ground vibronic state. The rate components of $\Gamma_{\pm}, \Gamma_{\mp}$ can be computed as
\begin{subequations} \label{eq:low_isc_gammaperp_symmetrypreserving}
\begin{align}
    \Gamma_{\pm}^{\widetilde{E_{x/y}}} & = 2\pi \hbar |\lambda_{\perp}^l|^2 \sum_i^{\infty} c_i^2 \sum_j \left|\langle\chi_j|\chi_i(A_1)\rangle\right|^2 \delta(\Sigma + \varepsilon - n_j \hbar \omega_e) \nonumber\\
    & + 2\pi \hbar |\lambda_{\perp}^l|^2 \sum_i^{\infty} \frac{f_i^2}{2} \sum_j \left|\langle\chi_j|\chi_i(E_x)\rangle\right|^2 \delta(\Sigma + \varepsilon - n_j \hbar \omega_e),\\
    \Gamma_{\pm}^{\widetilde{A_1}} & = 2\pi \hbar |\lambda_{\perp}^l|^2 \sum_i^{\infty} \frac{d_i^2}{2} \sum_j \left|\langle\chi_j|\chi_i(E_{x})\rangle\right|^2 \delta(\Sigma + \varepsilon - n_j\hbar\omega_e),\\
    \Gamma_{\mp}^{\widetilde{E_{x/y}}} & = 2\pi \hbar |\lambda_{\perp}^l|^2 \sum_i^{\infty} \frac{f_i^2}{2} \sum_j \left|\langle\chi_j|\chi_i(E_y)\rangle\right|^2 \delta(\Sigma + \varepsilon - n_j \hbar\omega_e),\\
    \Gamma_{\mp}^{\widetilde{A_1}} & = \Gamma_{\pm}^{\widetilde{A_1}},
\end{align}
\end{subequations}
where we have defined\footnote{We can also write $\lambda_\perp^l = \sqrt{w} \left\langle {}^1\!E_{x(y)}^\prime \left|H_{\text{so}}\right| {}^3\!A_2^{-(+)}\right\rangle$, where the coefficient $w$ represents the weight of the $a_1e^3$ configuration in the ${}^1\!E$ wavefunction. Note that sometimes people write $\lambda_\perp^l = \sqrt{w}\lambda_\perp$~\cite{thiering2018theory, li2024excited}, which is only true in the minimum model. Since the calculations we will perform go beyond the minimum model, in our notation we will explicitly write $\lambda_\perp^l$.} $\lambda_\perp^l = \frac{1}{\hbar}\left\langle {}^1\!E_{x(y)}\left|H_{\text{so}}\right|{}^3\!A_2^{-(+)}\right\rangle$. Clearly, to compute these rates, a PJT-modulated phonon overlap function is required. These phonon-occupation-number-dependent spectral functions can be obtained by convoluting the standard spectral density as in~\cite{kehayias2013infrared, thiering2018theory}
\begin{equation}
    S_E^{(n)}(\omega) = \left(S_E^{(n-1)} * S_E\right)(\omega),\;\; S_E^{(0)}(\omega) = \delta(\omega),
\end{equation}
where $S_E(\omega)$ is the phonon overlap spectral density, as we show in Fig.~\ref{fig:vibration_overlap}(b).

Finally, considering the effect of temperature, the initial state should be modified by a Boltzmann distribution among the vibronic states and so a prefactor $\frac{\exp(-\epsilon_i/k_BT)}{\sum_i \exp(-\varepsilon_i/k_BT)}$ should be associated with each vibronic state.

\subsubsection{Effects of stress}
All of the above analyses still hold true when the system experiences symmetry-preserving stresses, except that the parameters, e.g. $\Sigma$, will be altered by stress.

The case of symmetry-breaking stress is again much more complicated. This complexity comes from the existance of multiple SOCs for each transition, and that these mechanisms \textit{interfere} with each other. For sake of simplicity, we will focus on only $\Pi_x$ stress here as the symmetry-breaking component, but the analysis generalizes to arbitrary stress profiles. To see how interference happens, we examine $\Gamma_z^\text{lower}$ as an example. At ambient conditions, we have already learned that the leading-order mechanism is a SOC from the $|{}^1\!A_1\rangle$ electronic component in $\left|\widetilde{{}^1\!E} \right\rangle$ via the PJT effect. With $\Pi_x$ stress, a new SOC matrix element $\lambda_z'$ emerges that directly couples $|{}^1\!E_x\rangle$ and $\left|{}^3\!A_2^0\right\rangle$, as we discussed in Sec.~S3.2.2, and therefore contributes to $\Gamma_z^\text{lower}$. These two mechanisms come from different electronic components, and they interfere if the initial vibronic state transforms as $E$:
\begin{subequations} \label{eq:low_isc_gammaz_symmetrybreaking_e}
\begin{align}
    \Gamma^{\widetilde{E_x}}_z & = 2\pi\hbar \sum_{i}^{\infty} \left|\left(2d_i\lambda_z + \frac{f_i}{\sqrt{2}}\lambda_z'\right)\right|^2 \sum_j^{\infty} \left|\langle \chi_j|\chi_i(E_x)\rangle \right|^2 \delta(\Sigma + \varepsilon - n_j\hbar\omega_e) \nonumber\\
    & + 2\pi\hbar \left|\lambda_z'\right|^2 \sum_{i}^{\infty} c_i^2 \sum_j^{\infty} \left|\langle\chi_j|\chi_i(A_1)\rangle\right|^2 \delta(\Sigma + \varepsilon - n_j\hbar\omega_e),\\
    \Gamma_z^{\widetilde{E_y}} & = 2\pi\hbar \sum_{i}^{\infty} \left|\left(2d_i\lambda_z - \frac{f_i}{\sqrt{2}}\lambda_z'\right)\right|^2 \sum_j^{\infty} \left|\langle \chi_j|\chi_i(E_y)\rangle \right|^2 \delta(\Sigma + \varepsilon - n_j\hbar\omega_e).
\end{align}
\end{subequations}
Depending on the relative sign of the two terms in the parenthesis, these two mechanisms either positively or negatively interfere, leading to either an increase or a decrease in the ISC rate. For an initial state of$\left|\widetilde{A_1}\right\rangle$, the rate contribution is
\begin{equation} \label{eq:low_isc_gammaz_symmetrybreaking_a1}
\begin{split}
    \Gamma_z^{A_1} & = 8 \pi\hbar \left|\lambda_z\right|^2 \sum_{i}^{\infty}c_i^2 \sum_j^{\infty} \left|\langle \chi_j|\chi_i(A_1)\rangle \right|^2 \delta(\Sigma + \varepsilon - n_j\hbar\omega_e)\\
    & + 2\pi\hbar |\lambda_z'|^2 \sum_{i}^{\infty}\frac{d_i^2}{2} \sum_j^{\infty} \left|\langle \chi_j|\chi_i(E_x)\rangle \right|^2 \delta(\Sigma + \varepsilon - n_j\hbar\omega_e).
\end{split}
\end{equation}

Fortunately, this interference does not exist for either $\Gamma_+$ or $\Gamma_-$, as there is only one SOC matrix element mediating the corresponding ISC processes. Those rate components can be computed as
\begin{subequations} \label{eq:low_isc_gammaminus_symmetrybreaking}
\begin{align}
    \Gamma_-^{\widetilde{E_x}} & = 2\pi\hbar |\lambda_{\perp, x}^l|^2 \sum_i^{\infty} c_i^2 \sum_j |\langle\chi_j |\chi_i(A_1)\rangle|^2 \delta(\Sigma + \varepsilon - n_j\hbar\omega_e) \nonumber\\
    & + 2\pi\hbar |\lambda_{\perp, x}^l|^2 \sum_i^{\infty} \frac{f_i^2}{2}\sum_j |\langle\chi_j|\chi_i(E_x)\rangle|^2 \delta(\Sigma + \varepsilon - n_j\hbar\omega_e),\\
    \Gamma_-^{\widetilde{E_y}} & = 2\pi\hbar |\lambda_{\perp, x}^l|^2 \sum_i^{\infty} \frac{f_i^2}{2}\sum_j |\langle\chi_j |\chi_i(E_y)\rangle|^2 \delta(\Sigma + \varepsilon - n_j\hbar\omega_e),\\
    \Gamma_-^{\widetilde{A_1}} & = 2\pi\hbar |\lambda_{\perp,x}^l|^2 \sum_i^{\infty} \frac{d_i^2}{2}\sum_j |\langle\chi_j |\chi_i(E_x)\rangle|^2 \delta(\Sigma + \varepsilon - n_j\hbar\omega_e),
\end{align}
\end{subequations}
and
\begin{subequations} \label{eq:low_isc_gammaplus_symmetrybreaking}
\begin{align}
    \Gamma_+^{\widetilde{E_x}} & = 2\pi\hbar |\lambda_{\perp,y}^l|^2 \sum_i^{\infty} \frac{f_i^2}{2}\sum_j |\langle\chi_j |\chi_i(E_y)\rangle|^2 \delta(\Sigma + \varepsilon - n_j\hbar\omega_e),\\
    \Gamma_+^{\widetilde{E_y}} & = 2\pi\hbar |\lambda_{\perp,y}^l|^2 \sum_i^{\infty} c_i^2 \sum_j |\langle\chi_j |\chi_i(A_1)\rangle|^2 \delta(\Sigma + \varepsilon - n_j\hbar\omega_e) \nonumber\\
    & + 2\pi\hbar |\lambda_{\perp,y}^l|^2 \sum_i^{\infty} \frac{f_i^2}{2}\sum_j |\langle\chi_j |\chi_i(E_x)\rangle|^2 \delta(\Sigma + \varepsilon - n_j\hbar\omega_e),\\
    \Gamma_+^{\widetilde{A_1}} & = 2\pi\hbar |\lambda_{\perp,y}^l|^2 \sum_i^{\infty} \frac{d_i^2}{2}\sum_j |\langle\chi_j |\chi_i(E_y)\rangle|^2 \delta(\Sigma + \varepsilon - n_j\hbar\omega_e),   
\end{align}
\end{subequations}
where $\lambda_{\perp}^l$ is affected by $\Pi_x^{(2)}$ stress. Finally, taking the temperature into account, the initial state is modified by a Boltzmann distribution over the low-lying vibronic states. The numerical computations for these rates will be presented in the next section.

\subsection{Other rates}
In this section, we cover the additional rates relevant to the optical cycle. These rates include the laser excitation rate, the rates of the microwave-driven spin transitions within ${}^3\!A_2$, the spontaneous emission rate from ${}^3\!E$, and phonon-induced transition rates between the two orbital branches of ${}^3\!E$ and ${}^1\!E$.

\subsubsection{Laser excitation and microwave drive}
The laser excitation rate and microwave driving rate largely depend on the laser and microwave power used in the experiments and cannot be determined from first principles. We will discuss how we choose these rate parameters in Sec.~S5.4.

\subsubsection{Spontaneous emission}
The spontaneous emission rate from ${}^3\!E$ is independent of spin and can be computed as
\begin{equation}
    \Gamma_{\text{rad}} = \frac{n_D E_{\text{ZPL}}^3 |{\mu}_{eg}|^2}{3\pi \varepsilon_0 \hbar^4 c^3},
\end{equation}
where $n_D \sim 2.4$ is the refractive index of diamond; $\varepsilon_0$ is the vacuum permittivity; $E_{\text{ZPL}}$ is the zero phonon line; $\vec{\mu}_{eg}$ is the transition dipole moment vector; and $c$ is the speed of light. Under symmetry-breaking stress, the two branches of ${}^3\!E$ would radiate at different rate, as both $E_{\text{ZPL}}$ and $|\mu_{eg}|$ would be different.

\subsubsection{Phonon-induced transitions}
Next, we consider phonon-driven population dynamics within ${}^3\!E$. The relevant effect is population hopping between the two orbital branches. Ref.~\cite{ernst2023modeling} carefully studied these transitions and we mostly follow their analysis. At ambient conditions or under small stress, this hopping arises from a coupling to the phonon bath, where one- and two-phonon processes drive transitions (i.e. both upwards and downwards in energy) between the orbital branches of the NV center. An upward and a downward rate can be defined as $\Gamma_{\uparrow/\downarrow}$ and they relate to each other via $\frac{\Gamma_{\uparrow}}{\Gamma_{\downarrow}} = \text{exp}\left(-\frac{2\Pi_\perp}{k_B T}\right)$. Detailed expressions of the one and two-phonon process rates can be found in Ref.~\cite{ernst2023modeling}. At room temperature, these rates are estimated to be $\Gamma_{\uparrow/\downarrow} \sim$ THz at around $\Pi_\perp \sim 80$ GHz. Ref.~\cite{ulbricht2016jahn} also noted that these dynamics are ultrafast (at the femtosecond timescale). This is the origin of orbital averaging and the reason why an effective $D_{\text{es}}$ can be observed~\cite{batalov2009low}. These conclusions also apply to the case of symmetry-preserving stress.

In the large symmetry-breaking stress limit, however, the above picture is no longer valid, since $\Pi_x$ can be as large as $\sim 400$ meV. First, $\Gamma_{\uparrow}\rightarrow 0$ as $\text{exp}\left(-\frac{2\Pi_\perp}{k_B T}\right)\rightarrow 0$. Therefore, orbital averaging is greatly suppressed and the two orbital branches must be separately treated. Second, $\Gamma_{\downarrow}$ is not easy to characterize, since $\Pi_x$ can be too large to allow for one- or two- phonon processes, necessitating the consideration of higher-order phonon mechanisms. Nevertheless, we assume that $\Gamma_{\downarrow}$ remains orders of magnitude larger than other rates, e.g., $\Gamma_{\text{rad}}, \Gamma_{\text{ISC}}$. We model this rate implicitly by adopting a Boltzmann distribution of the population on the two orbital branches of ${}^3\!E$. We apply the same logic in the case of the ${}^1\!E$ manifold.

\subsection{ODMR contrast}
We finally arrive at the matter of ODMR contrast. Contrast measures the photoluminescence difference of different spin states by taking advantage of the spin-selective property of the defect's optical cycle. It serves as the core “observable” for most quantum metrology applications that employ the NV center as the probe. Conceptually, contrast measurement involves two processes. First, the incident laser drives the NV's optical cycle and initializes the NV's population distribution. Second, the applied microwave drive serves to redistribute population and modify fluorescence \footnote{There are essentially two experimental modes of contrast measurement (see Sec.~\ref{sec:experiment}). In continuous wave ODMR, the laser and microwave are enabled simultaneously, and an equilibrium population is reached for each microwave frequency. Conversely, pulsed methods such as Ramsey spectroscopy perform these two processes serially.}. In this subsection, we discuss how to model contrast.

As previously stated, contrast is based on photoluminescence, which can be estimated as $\bar{I} = \sum_{i\in {}^3\!E} \bar{n}_i \Gamma_{\text{rad}}$, where the overline denotes the steady-state solution and $n_i$ represents the population of sublevel $i$ in ${}^3\!E$. Contrast can be computed as $C = 1 - \frac{\bar{I}^{\text{MW}}}{\bar{I}}$, where the superscript “MW” corresponds to the scenario when the resonant microwave is on. These steady-state populations can be obtained by solving a rate model,\footnote{Note that we assume a steady-state solution for contrast, but in practical experimental conditions, the initialization/readout cycle takes a finite amount of time, typically $0.1$ to $10$ ms. While this time is generally chosen to achieve a steady state, one must still be careful when comparing experimental data to numerical simulations.}~\cite{li2024excited}
\begin{equation} \label{eq:rate_model}
    \frac{dn_i(t)}{dt} = \sum_j \left[\Gamma_{ji} n_j(t) - \Gamma_{ij} n_i(t)\right],
\end{equation}
where $\Gamma_{ij}$ represents a transition rate from state $i$ to state $j$.

At ambient conditions, as we have seen, $\Gamma_{\text{ave}}$ is only active for $|m_s=\pm\rangle$, making $|0\rangle$ the brightest spin. Taking the lower ISC rates into consideration with $\Gamma_{\pm}, \Gamma_{\mp}, \Gamma_z$ being on the same order of magnitude, the initialization process results in polarization into $|0\rangle$ in the ground state. This leads to the ODMR contrast peak being negative ($\bar{I}^{\text{MW}} < \bar{I}$), which is also a phenomenon ubiquitously observed when the stress is symmetry-preserving or only slightly symmetry-breaking.

When the polarization is no longer into the brightest spin, however, we expect to see positive contrast peaks~\cite{bhattacharyya2024imaging} ($\bar{I}^{\text{MW}} > \bar{I}$). This could be understood in the following way: when the resonant microwave is on, the NV population gets driven from a dark spin into a brighter spin, which increases the photoluminescence.

\section{Simulation details} \label{sec:simulations}
In the previous sections, we formulated the theory for estimating the ISC rates and the ODMR contrast of the NV, identifying a number of quantities which must now be computed from first principles. Specifically, we need to compute the susceptibilities of various energy gaps, SOCs that couple triplet and singlet states, SSCs that mix the spin, and lastly the PJT effects that couple the electronic and phononic degrees of freedom. In this section, we present the computational details of all the simulations we have performed in this work. Then, we compare the computed rates and lifetimes of different states with experimental data. Finally, we conclude by commenting on the errors associated with our computations as well as potential improvements we would like to implement in the future.

We used density functional theory (DFT)~\cite{kohn1965self} to compute zero-phonon lines (ZPLs) and optimize the underlying geometry of the NV center under strain. We adopted a hydrogen-terminated cluster model~\cite{bhandari2021multiconfigurational} based on the optimized DFT geometries, and computed its fine-structure properties including SOC and SSC with a well-established quantum chemistry method--- the complete active space self-consistent field (CASSCF) method~\cite{roos1980complete} with relativistic corrections~\cite{douglas1974quantum}. We carefully benchmark these properties against a series of computational parameters and link our results to other similar computational works. As for electron-phonon couplings, we employed time-dependent density functional theory (TDDFT)~\cite{jin2023excited} to solve the PJT effects among the singlets and employ the Huang-Rhys theory~\cite{alkauskas2014first, jin2021photoluminescence} to compute the phonon sidebands, following Ref.~\cite{jin2022vibrationally}.

\subsection{DFT for ZPL}
As the NV center is a solid-state defect in the diamond crystal lattice, we optimized its ground-state geometry under strain with periodic-boundary conditions. We applied the SCAN functional~\cite{sun2015strongly} to a 511-atom supercell ($4\times4\times4$ unit cells), with a plane-wave basis, 75 Ry energy cutoff, the ONCV pseudopotential~\cite{schlipf2015optimization}, and $\Gamma$-point sampling over the Brillouin zone. The strain is applied by adjusting the lattice vector of the supercell. The Quantum Espresso code~\cite{giannozzi2017advanced, giannozzi2020quantum} is used for all DFT calculations in this work.

\begin{figure}
    \centering
    \includegraphics[width=\textwidth]{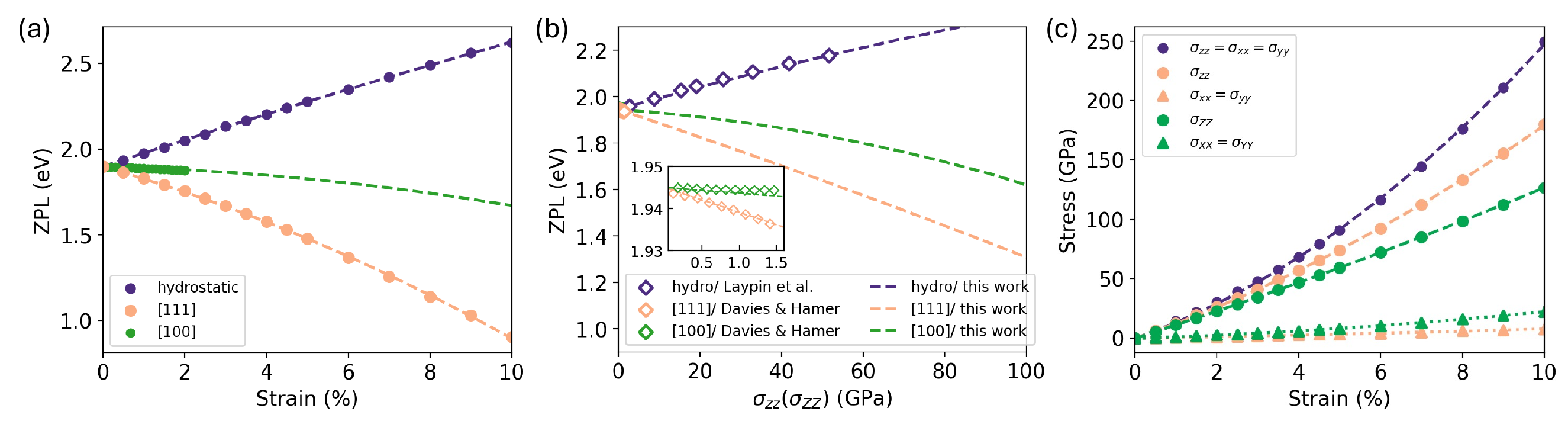}
    \caption{Zero-phonon lines (ZPL) as a function of strain/stress. (a). ZPL versus hydrostatic, [111] and [100] strain, computed from the SCAN@512-atom supercell level using constrained DFT (solid dots). Note that [100] strain breaks the NV symmetry and there are two ZPLs coming from the two orbital branches of ${}^3\!E$, and constrained DFT captures the lower branch. The dashed curves are fitted by a quadratic function. (b). Comparison between the computed ZPLs (dashed lines) and experimental results (diamonds). The experimental results are taken from Lyapin et al.~\cite{lyapin2018study} for hydrostatic stress, and Davies and Hamer~\cite{davies1976optical} for [111] and [100] stress. The computed ZPLs are converted from strain into stress according to (c), and all shear stress components are zero. Note that the ZPL at the zero-stress limit from SCAN is 1.898 eV, which is slightly smaller than the experimental observed 1.945 eV. To better compare the susceptibilities with experiments, we have applied a rigid shift to those computed ZPLs to align their values to experiments at zero stress.}
    \label{fig:ZPL_DFT}
\end{figure}

The ${}^3\!E$ energy is obtained by occupation-number constrained DFT~\cite{kaduk2012constrained}, where an electron is excited from the $a_1$ orbital into the $e$ orbital. The ZPL can be computed as $E_{\text{ZPL}} = E_{{}^3\!E}^{\text{DFT}} - E_{{}^3\!A_2}^{\text{DFT}}$. We plotted the computed ZPL versus different strains in Fig.~\ref{fig:ZPL_DFT}(a), and converted it to stress and compared it with experimental measurements~\cite{davies1976optical, lyapin2018study} in Fig.~\ref{fig:ZPL_DFT}(b). The extracted linear susceptibilities of ${}^3\!E$ from DFT reach a good agreement with experiments, as documented in Table~\ref{tab:energy_susceptibilities}.

\renewcommand{\arraystretch}{1.8}
\begin{table*}[ht]
\centering
\caption{The linear susceptibilities of ${}^3\!E$ with stress, defined in Eqs.~\ref{eq:susceptibilities}, from DFT and experiment.}
{\begin{tabular}{m{2cm}<{\centering} m{1.5cm}<{\centering} m{1.5cm}<{\centering} m{1.5cm}<{\centering} m{1.5cm}<{\centering}}
 \hline\hline
 GHz/GPa & $\alpha_1^{\left({}^3\!E\right)}$ & $\beta_1^{\left({}^3\!E\right)}$ & $\alpha_2^{\left({}^3\!E\right)}$ & $\beta_2^{\left({}^3\!E\right)}$\\
 \hline
 Ref.~\cite{davies1976optical} & 1295 & $-1523$ & $-645$ & $-89$ \\
 This work & 1356 & $-1417$ &  & \\
 \hline\hline
\end{tabular}}
\label{tab:energy_susceptibilities}
\end{table*}

\subsection{CASSCF for VEE, SOC, and SSC}
As we see from the previous subsection, DFT gives a very good agreement with experiments regarding the ZPLs of the triplet states. However, Kohn Sham DFT is a mean-field theory~\cite{kohn1965self}, and it cannot be applied to study the singlet states of the NV center, which are multi-reference in nature~\cite{maze2011properties, doherty2011negatively}. Studying these states in the framework of plane-wave basis and supercell is still possible, e.g., by employing quantum embedding theories~\cite{sheng2022green, lopez2024quantum, chen2025advances}, but this approach can be quite costly and it lacks implementation for these fine properties. Broadly speaking, multi-reference methods, e.g., CASSCF and CASPT2~\cite{andersson1992second}, have been developed for decades and widely used in the quantum chemistry community for calculating the fine properties of molecules. In recent years, there have been growing efforts in applying these quantum chemistry methods to study the electronic structure of spin defects~\cite{bhandari2021multiconfigurational, li2024excited, benedek2024comprehensive}, by employing a terminated cluster model of the original defects. Therefore, we followed this approach to study how the NV center responds to strain.

\begin{figure}
    \centering
    \includegraphics[width=0.7\textwidth]{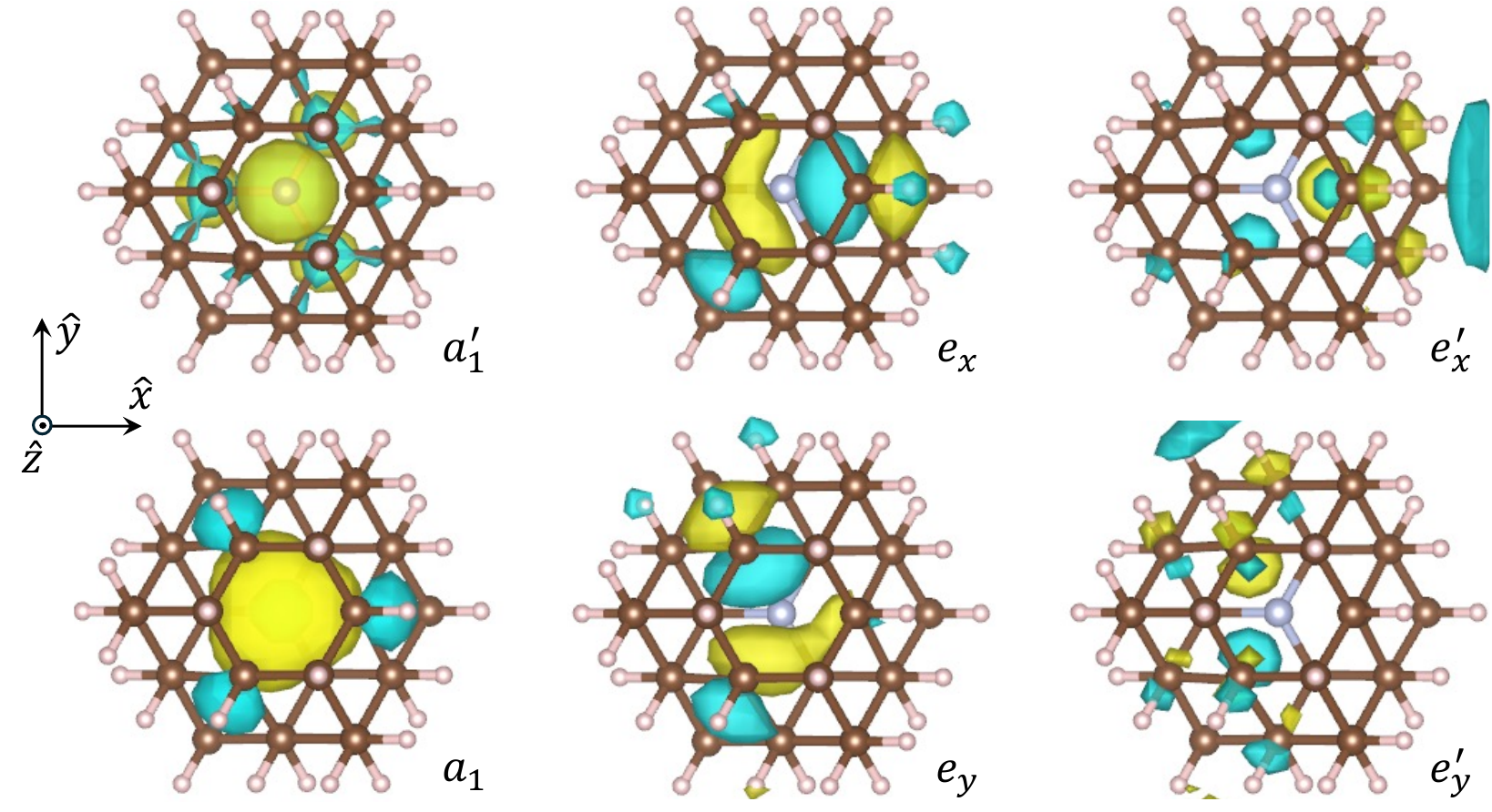}
    \caption{Orbitals $a_1^\prime, a_1, e_x, e_y, e_x^\prime, e_y^\prime$ of NV cluster $\text{C}_{33}\text{H}_{36}\text{N}^-$ from the CASSCF calculation. The iso-surface value of the orbitals is chosen to be 0.04. The orbital wavefunctions are plotted using VESTA~\cite{momma2011vesta}.}
    \label{fig:active_space}
\end{figure}

We considered two hydrogen-terminated clusters consisting of 70 ($\text{C}_{33}\text{H}_{36}\text{N}^-$) and 162 ($\text{C}_{85}\text{H}_{76}\text{N}^-$) atoms~\cite{bhandari2021multiconfigurational}, respectively. These clusters are cut from SCAN-optimized, 512-atom cubic supercells and no further geometry optimization is performed. The $z$-axis of the cluster coordinate is aligned along the [111] direction of diamond\footnote{For symmetry-breaking strain, the NV axis is strictly speaking no longer well defined, however, we ignore this (minor) deformation and utilize our previously defined spatial transformations.}. We employed the CASSCF method to compute the vertical excitation energies (VEEs), SOCs, and SSCs of the NV center under strain in a state-averaged fashion to extract relavant susceptibilities, using the ORCA software package~\cite{neese2020orca}. We also include relativistic effects by using the Douglas-Kroll-Hess (DKH) Hamiltonian~\cite{douglas1974quantum, hess1986relativistic}. The CASSCF method has dependencies on its computational setup, e.g., size of the active space and basis set, so we first benchmark and discuss results for the NV at ambient conditions against these parameters. The active spaces we constructed in this work consists of (at most) six defect orbitals\footnote{For the (4e, 6o) active space, this replaces the doubly-occupied $a_1'$ orbital in Fig.~\ref{fig:active_space} with an virtual $a_1$ orbital, which is not shown.} (localized around the defect and lying in the band gap, Fig.~\ref{fig:active_space}).

\subsubsection{Ambient condition}
The results of VEEs and SOCs are summarized in Tab.~\ref{tab:VEE_ambient}. By comparing the reported values from our work and other similar works, we obtain some qualitative results. First, increasing the active space or using a larger basis set for a specific cluster size would lower the VEEs, but not necessarily the detuning $\Delta$. Applying perturbation theory on top of CASSCF has similar effects. At the 70-atom cluster level, using a (6e, 6o) active space with a double-zeta basis gives a reasonable description of the VEEs. By increasing the cluster size, these computed VEEs significantly increase. Notably, Ref.~\cite{benedek2024comprehensive} studied VEEs using the n-electron valence state perturbation theory (NEVPT)~\cite{angeli2001introduction} with a cluster size up to 294 atoms and a basis set ranging from double zeta to sextuple zeta. Their findings confirm our previous observations based on relatively small cluster sizes. At larger clusters, applying perturbation theory to obtain a reasonable estimation of VEEs becomes necessary. However, using a large cluster or a big basis set substantially increases the computational cost.

Compared with the VEEs, the SOCs appear to be less sensitive to the computational setup. For $\lambda_z$, CASSCF is able to yield a reasonable estimation across all active spaces and clusters; in contrast, $\lambda_\perp$ is significantly underestimated compared to experiments, despite a mild increase with cluster sizes. We comment on possible reasons for this discrepancy in Sec.~\ref{sec:errors}. Now, we would like to address the difference between the SOCs deduced from group theory and those computed from first-principles here. According to Table~2 of Ref.~\cite{doherty2011negatively}, we have in the minimum model of the NV center
\begin{equation}
    \hbar\lambda_\perp = \frac{1}{\sqrt{2}} \left\langle A_1\left|H_{\text{so}}\right| {}^1\!A_1\right\rangle = \frac{1}{\sqrt{2}} \left\langle E_{1,2}\left|H_{\text{so}} \right|{}^1\!E_{x,y}\right\rangle = \left\langle {}^3\!A_2^{\pm}\left|H_{\text{so}} \right|{}^1\!E^\prime_{x,y}\right\rangle.
\end{equation}
The first equality is how we define $\lambda_\perp$. The second and third equalities do not necessarily hold when going beyond the minimum model. In fact, we have $\frac{\left\langle E_{1,2}\left|H_{\text{so}} \right|{}^1\!E_{1,2}\right\rangle}{\left\langle A_1\left|H_{\text{so}}\right| {}^1\!A_1\right\rangle} \approx2.69$ from our CASSCF calculation at the 70-atom cluster level, and Ref.~\cite{li2024excited} reported this ratio to be $\sim2.94$. The last term in equation S37 is the source of coupling for the lower ISCs due to the Coulomb interaction, as we discussed in Sec.~S4.2.1. For the lower ISC, an effective SOC connecting ${}^1\!E$ and $\left|{}^3\!A_2^{\pm}\right\rangle$ is  in the minimum model often written as $\lambda_\perp^l = \sqrt{w}\lambda_\perp (< \lambda_\perp) $~\cite{thiering2018theory, li2024excited}. However, we observed from CASSCF calculations that the case of $\lambda_\perp^l > \lambda_\perp$ could occur. It is currently hard to comment which case is a more realistic description of the SOC of the system.

Since we care more about the susceptibilities of energies and SOCs than their absolute values, we select the (6e, 6o) active space, and the cc-pVDZ-DK~\cite{dunning1989gaussian} basis to proceed with strain calculations. Such a computational setup balances accuracy and computational cost, and has been verified to give a reasonable agreement with experiments in prior work~\cite{li2024excited}.

\renewcommand{\arraystretch}{1.8}
\begin{table*}
\caption{Vertical excitation energies (VEEs) and spin-orbit coupling (SOC) matrix elements computed in this work compared to previous computational studies using the CASSCF/PT2 methods with varying setups. “SA(3,3)” represents a state-averaged CASSCF calculation over the lowest three triplet states and three singlet states. “(6e,6o)” represents the active space containing six electrons and six orbitals. “DZ/TZ” denotes the correlation-consistent double/triplet zeta basis set with the “DK” suffix standing for relativistic contraction. All energies are reported in eV while SOCs are in GHz.}
\label{tab:VEE_ambient}
\centering
\setlength{\tabcolsep}{4pt}
\begin{tabularx}{\textwidth}{c|c|cccccc}
 \hline\hline
 \multicolumn{2}{c|}{} & ${}^1\!E$ or $\Sigma$ & ${}^1\!A_1$ & ${}^3\!E$ & $\Delta$ & $|\lambda_z|$ & $|\lambda_{\perp}|$\\
 \hline
 \multicolumn{2}{c|}{Expt.~\cite{davies1976optical, goldman2015state}} & 0.341–0.434\textsuperscript{a} & 1.531-1.624\textsuperscript{a} & 2.180 & 0.321-0.414\textsuperscript{b} & $17.5\pm0.1$\textsuperscript{c} & $21.1\pm3.6$\textsuperscript{d}\\
 \hline
 \multirow{11}{3em}{70-atom} & SA(3,3)-(4e,6o) @ DZ-DK (this work) & 0.76 & 2.48 & 2.70 & 0.22 & 14.69 & 5.05\\
 & SA(3,3)-(4e,6o) @ TZ-DK (this work) & 0.62 & 1.88 & 2.23 & 0.35 & 14.78 & 5.09\\
 & SA(3,3)-(6e,6o) @ DZ-DK (this work) & 0.68 & 1.99 & 2.38 & 0.39 & 15.54 & 6.36\\
 & SA(3,3)-(6e,6o) @ TZ-DK (this work) & 0.50 & 1.35 & 1.82 & 0.47 & 16.07 & 5.26\\
 & SA(3,3)-(6e,6o) @ DZ-DK~\cite{bhandari2021multiconfigurational}\textsuperscript{e} & 0.34 & 1.41 & 1.93 & 0.52 & 6.50 & \\
 & SA(3,3)-(6e,6o) @ DZ-DK~\cite{li2024excited} & 0.66 & 1.96 & 2.30 & 0.34 & 14.21 & 3.96\\
 & SA(3,3)-CASPT2 @ DZ-DK~\cite{li2024excited}\textsuperscript{f} & 0.55 & 1.57 & 2.22 & 0.65 & &\\
 & SA(5,5)-(4e,6o) @ DZ-DK~\cite{li2024excited} & 0.59 & 1.68 & 2.05 & 0.36 & 7.56 & 2.04\\
 & SA(5,5)-CASPT2 @ DZ-DK~\cite{li2024excited}\textsuperscript{f} & 0.60 & 1.71 & 2.43 & 0.72 & & \\
 & SA(5,5)-(6e,6o) @ DZ-DK~\cite{li2024excited} & 0.64 & 1.67 & 2.04 & 0.37 & &\\
 & SA(5,5)-CASPT2 @ DZ-DK~\cite{li2024excited}\textsuperscript{f} & 0.60 & 1.86 & 2.46 & 0.60 & & \\
 \hline
 \multirow{4}{3em}{162-atom} & SA(3,3)-(6e,4o) @ DZ-DK (this work) & 0.93 & 3.16 & 3.26 & 0.10 & 23.04 & 7.12\\
 & SA(3,3)-(4e,6o) @ DZ-DK (this work) & 0.84 & 2.88 & 3.00 & 0.12 & 13.82 & 5.72\\
 & SA(3,3)-(6e,6o) @ DZ-DK (this work) & 0.78 & 2.42 & 2.72 & 0.30 & 14.86 & 6.74\\
 & SA(3,3)-(6e,6o) @ DZ-DK~\cite{bhandari2021multiconfigurational}\textsuperscript{e} & 0.25 & 1.60 & 2.14 & 0.54 & 8.1 & \\
 \hline
 \multirow{2}{3em}{294-atom} & SA(3,3)-NEVPT2 @ DZ~\cite{benedek2024comprehensive}\textsuperscript{g} & 0.62 & 1.77 & 2.35 & 0.58 & 18.7 & \\
 & SA(5,8)-NEVPT2 @ DZ~\cite{benedek2024comprehensive}\textsuperscript{g} & 0.56 & 1.60 & 2.18 & 0.58 & & \\
 \hline\hline
\end{tabularx}

\vspace{1mm}
\begin{minipage}{\textwidth}
\raggedright
\hspace{0mm}\textsuperscript{a} The VEEs of ${}^1\!E$ and ${}^1\!A_1$ are estimated by subtracting $\Delta$ from the ZPL of ${}^3\!E$.\\
\hspace{0mm}\textsuperscript{b} The range of $\Delta$ was indirectly estimated by matching the computed ISC rate with experiments~\cite{goldman2015state}.\\
\hspace{0mm}\textsuperscript{c} $\lambda_z = 17.53 \pm 0.10$ GHz is obtained using the experimentally derived $p\lambda_z = 5.33 \pm 0.03$ GHz~\cite{batalov2009low} together with Ham reduction factor $p = 0.304$~\cite{thiering2017ab}.\\
\hspace{0mm}\textsuperscript{d} $\lambda_\perp = 21.06 \pm 3.62$ GHz is obtained using the approximated relation $\lambda_\perp = (1.2 \pm 0.2) \lambda_z$~\cite{maze2011properties, goldman2015state}.\\
\hspace{0mm}\textsuperscript{e} The VEEs from Ref.~\cite{bhandari2021multiconfigurational} are significantly smaller than the rest due to their removal of surface orbitals.\\
\hspace{0mm}\textsuperscript{f} The CASPT2 calculation is on-top of the CASSCF calculation above it.\\
\hspace{0mm}\textsuperscript{g} The NEVPT2 calculation is on-top of a (6e,4o) CASSCF calculation.\\
\end{minipage}
\end{table*}

\subsubsection{Effects of symmetry-preserving strain}
\begin{figure}
    \centering
    \includegraphics[width=\textwidth]{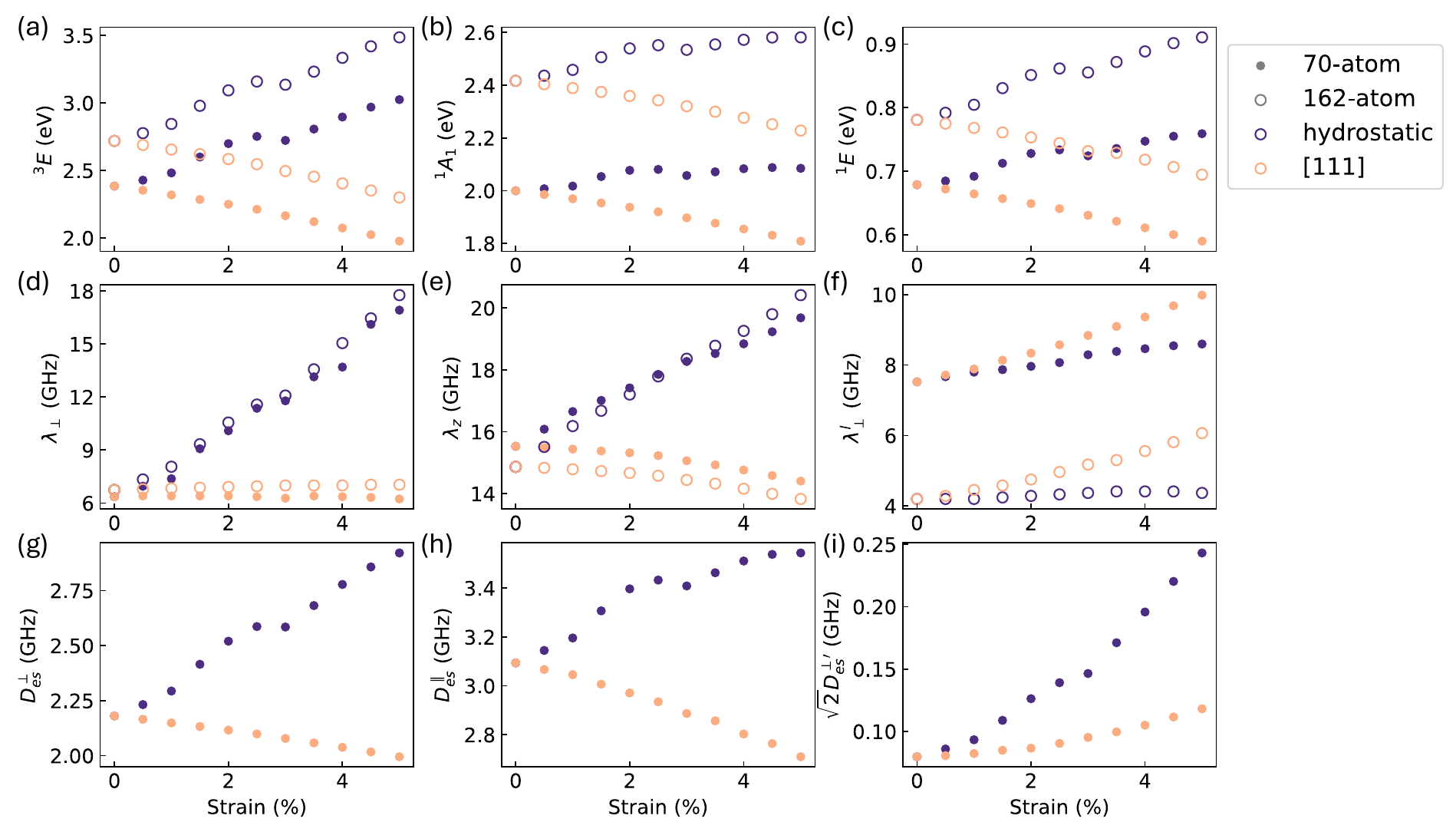}
    \caption{Vertical excitation energies (VEEs), spin-orbit couplings (SOCs) and spin-spin couplings (SSCs) versus symmetry-preserving strains computed from the SA(3,3)-CAS(6e, 6o) protocol on NV clusters using the cc-pVDZ-DK basis. Solid (empty) markers denote the $\text{C}_{33}\text{H}_{36}\text{N}^-$ ($\text{C}_{85}\text{H}_{76}\text{N}^-$) cluster; and dark purple (coral) markers represent hydrostatic (uniaxial [111]) strain. (a-c) plot VEEs for ${}^3\!E$, ${}^1\!A_1$, and ${}^1\!E$ respectively. (d-f) contain SOCs, specifically $\lambda_\perp$, $\lambda_z$, and $\lambda^l_\perp$ defined in Sec.~S3.2.1. SSCs are shown in (g-i), with $D_{\text{es}}$ parameters only deduced from the $\text{C}_{33}\text{H}_{36}\text{N}^-$ cluster.}
    \label{fig:cas_symm_preserve}
\end{figure}

Now we examine how these VEEs and SOCs respond to various strains. As always, we first consider symmetry-preserving strain. The results are plotted in Fig.~\ref{fig:cas_symm_preserve}. 
We observe that hydrostatic and uniaxial [111] strain affect VEEs and SOCs qualitatively differently. In Fig.~\ref{fig:cas_symm_preserve}(a-c), we plot the VEEs of ${}^3\!E$, ${}^1\!A_1$, and ${}^1\!E$ manifolds, respectively. Hydrostatic strain (dark circle) tends to enlarge these VEEs, while [111] strain tends to reduce them, with the ${}^3\!E$ manifold having the largest susceptibility, regardless of the cluster size. Notably, there is a small discontinuity for the hydrostatic curves at $\sim3\%$ strain for both clusters while no discontinuities are observed for the [111] strain case.

In Fig.~\ref{fig:cas_symm_preserve}(d-f), we plot the variations of SOCs versus strain. We see that hydrostatic strain tends to enhance $\lambda_\perp$ and $\lambda_z$ while [111] strain tends to not affect/slowly reduce them, regardless of the cluster size. We observe that $\lambda_\perp^l$, which connects ${}^1\!E$ and $\left|{}^3\!A_2^{\pm}\right\rangle$, has a more subtle dependence on the strain---[111] strain enhances it more than hydrostatic strain does. This dependence can be interpreted (by borrowing the minimum model here with $\lambda_\perp^l = \sqrt{w} \lambda_\perp$) as relating to the competition between $\lambda_\perp$ and $\sqrt{w}$. 
Hydrostatic strain increases both $\lambda_\perp$ and the ${}^1\!E \leftrightarrow {}^1\!E^\prime$ energy gap. This energy gap tends to reduce $w$, leading to a mild increase in $\lambda_\perp^l$. Conversely, [111] strain decreases the gap and roughly leaves $\lambda_\perp$ unchanged, leading to a larger enhancement with strain.

We also present the excited state SSCs in Fig.~\ref{fig:cas_symm_preserve}(g-i). Compared with experimental measurements, the SSC at zero-strain is overestimated from CASSCF, with $D_\text{es, 0}^\parallel = 3.09$ GHz and $D_\text{es, 0}^\perp = 2.18$ GHz, while experimentally reported values are $1.42$ and $1.55/2$ GHz~\cite{batalov2009low}, respectively. Lastly, the spin mixing parameter is calculated according to $\sqrt{2} D_\text{es}^{\perp'} \approx \beta\left(\lambda_z - D_{\text{es}}^\parallel\right)$, as the spin mixing is between $E_{1,2}$ and $E_{x,y}$ sublevels~\cite{maze2011properties}. Note that we are only able to extract the SSCs from the 70-atom cluster, as the $C_{3v}$ symmetry in the wavefunction is less well-preserved for the larger 162-atom cluster and $D$ gets severely suppressed.

We fit these computed data with quadratic functions and extract linear strain susceptibilities. We also convert strain to stress according to Fig.~\ref{fig:ZPL_DFT}(c), and extract their stress susceptibility counterparts. Both stress and strain susceptibilities are reported in Table~\ref{tab:susceptibilities}. Note that these susceptibilities are only valid in the small strain/stress limit and quadratic effects cannot be ignored with finite strain/stress.

\subsubsection{Effects of symmetry-breaking strain}
\begin{figure}
    \centering
    \includegraphics[width=\textwidth]{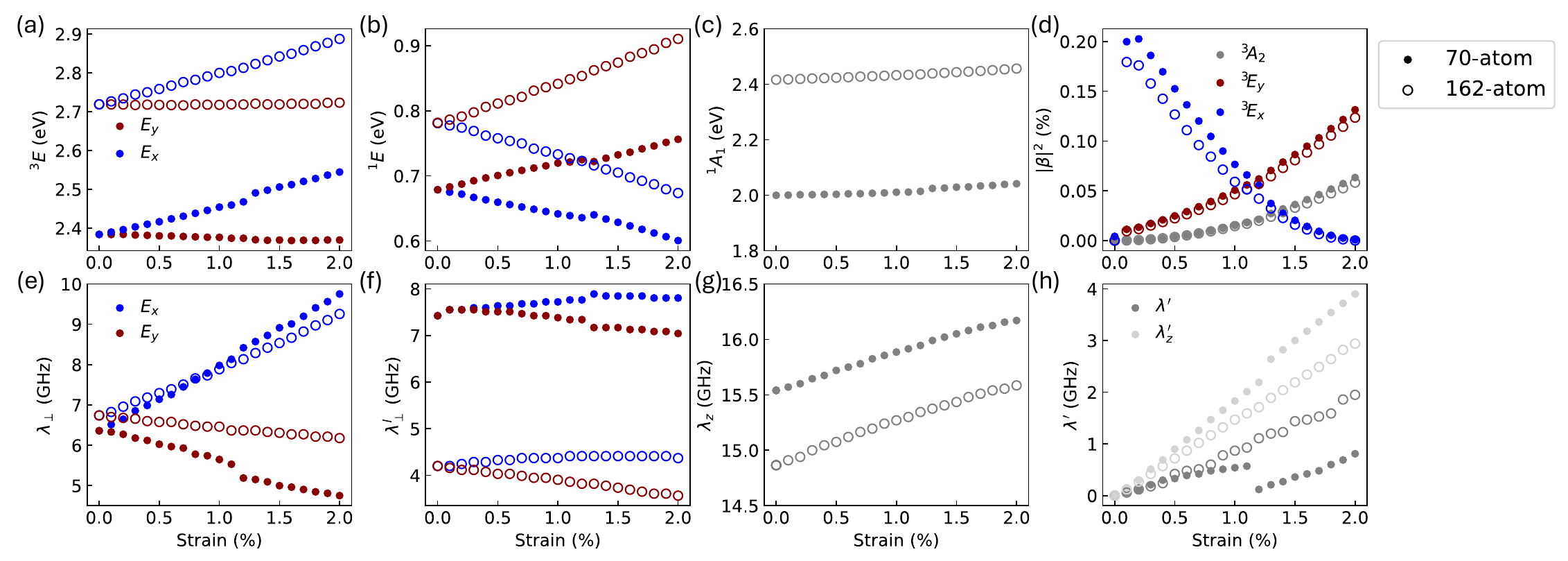}
    \caption{Vertical excitation energies (VEEs), spin-orbit couplings (SOCs) and spin-spin couplings (SSCs) versus uniaxial [100] strain computed from the SA(3,3)-CAS(6e, 6o) protocol on NV clusters using the cc-pVDZ-DK basis. Solid (empty) markers denote the $\text{C}_{33}\text{H}_{36}\text{N}^-$ ($\text{C}_{85}\text{H}_{76}\text{N}^-$) cluster. (a-c) plot VEEs of ${}^3\!E$, ${}^1\!E$, and ${}^1\!A_1$ respectively and (d) plots the spin mixing parameters $|\beta|^2$, as defined in Eq.~\ref{eq:ssc_beta_0} and Eq.~\ref{eq:ssc_beta_12}. Note that [100] strain lifts the $E$ degeneracies, and ${}^3\!E, {}^1\!E$ separate into two (color coded as blue and red for the $x, y$ branch). (e-h) plot SOCs, specifically $\lambda_\perp, \lambda^l_\perp, \lambda_z$, and finally two new (strain/stress-induced) matrix elements $\lambda^\prime, \lambda_z^\prime$, defined in Sec.~S3.2.2.}
    \label{fig:cas_symm_break}
\end{figure}

Next, we investigate the case of uniaxial [100] strain. Similarly to our presentation of the symmetry-preserving case, we plotted VEEs, SSCs, and SOCs in Fig.~\ref{fig:cas_symm_break}. Under [100] strain, the degeneracy of the ${}^3\!E$ and ${}^1\!E$ manifolds is lifted, and therefore their VEEs branch into two--$E_x$ ($E_y$) color coded as blue (red) [Fig.~\ref{fig:cas_symm_break}(a)]. The ${}^1\!A_1$ state only weakly depends on strain, since only the symmetry-preserving component affects it, see Eq.~\ref{eq:[001]_stress}. These susceptibilities all have little dependence on the cluster size. We observe again that there exists a small discontinuity from the small cluster at $\sim1.3\%$ strain. Fig.~\ref{fig:cas_symm_break}(d) shows the spin-mixing coefficients $\beta$ in the ground and excited triplet states.

We then report SOCs under [100] strain in Fig.~\ref{fig:cas_symm_break}(e-h). $\lambda_\perp, \lambda_\perp^l$ also split into two branches and we apply the same color code to distinguish them [Fig.~\ref{fig:cas_symm_break}(e, f)]. The branching of $\lambda_\perp$ is due to strain susceptibilities $\chi_5, \chi_5^\prime$, as we first introduced in Sec.~S.3.2.2, while the branching of $\lambda_\perp^l$ is more tricky. $\lambda_\perp^l$ has more dependencies, such as $\lambda_\perp$, gaps between ${}^1\!E$ and ${}^1\!E^\prime$, and the strain susceptibilities $\chi_4, \chi_4^\prime$. Nevertheless, $\lambda_{\perp, x/y}^l$ exhibits a similar trend as $\lambda_{\perp, x/y}$. Contrary to the transverse terms, the diagonal term $\lambda_z$ only weakly depends on strain [Fig.~\ref{fig:cas_symm_break}(g)], since only the symmetry-preserving component affects it in a similar fashion to ${}^1\!A_1$. 
Finally, [100] strain induces new SOC matrix elements: $\lambda^\prime$ and $\lambda_z^\prime$, which play an important role in both the upper and lower ISCs. We report their values in Fig.~\ref{fig:cas_symm_break}(h). As we expect, they are zero at the zero-strain limit and exhibit a linear dependence on strain. We note a conspicuous discontinuity in $\lambda^\prime$ from the 70-atom cluster at $\sim1.3\%$ strain, and other similar, albeit less pronounced, discontinuities can also be spotted for $\lambda_\perp$ from the 70-atom cluster. For the SOCs, we observe some dependencies on the cluster size--the slopes for $\lambda_\perp$ and $\lambda_z^\prime$ from the 70-atom cluster are slightly larger in magnitude than corresponding values obtained from the 162-atom cluster. The linear strain and stress susceptibilities are extracted from quadratic fitting and are recorded in Table~\ref{tab:susceptibilities}.

\renewcommand{\arraystretch}{1.8}
\begin{table*}[ht]
\caption{Linear strain and stress susceptibilities of the vertical excitation energies (VEEs) and spin-orbit couplings (SOCs), extracted from the SA(3,3)-CAS(6e, 6o) calculations performed on the 162-atom clusters. The VEE susceptibilities are defined in Eq.~\ref{eq:susceptibilities} and SOC's defined in Eq.~\ref{eq:soc_001}. The stress susceptibilities are indirectly estimated based on strain susceptibilities and the strain-stress relation plotted in Fig.~\ref{fig:ZPL_DFT}(c).}
\label{tab:susceptibilities}
\centering
\setlength{\tabcolsep}{6pt}
\begin{tabularx}{\textwidth}{ccc|ccc|ccc|ccc}
 \hline\hline
  & meV/\% & meV/GPa & & meV/\% & meV/GPa & & GHz/\% & GHz/GPa & & GHz/\% & GHz/GPa\\
 \hline
 $\alpha_1^{\left({}^3\!E\right)}$ & 91.75 & 7.63 & $\alpha_1^{\left({}^1\!E\right)}$ & 18.92 & 1.70 & $\chi_1$ & 0.61 & 0.045 & $\chi_4$ &  & \\
 $\beta_1^{\left({}^3\!E\right)}$ & -54.70 & -5.47 & $\beta_1^{\left({}^1\!E\right)}$ & -9.06 & -0.98 & $\chi_1^\prime$ & -0.029 & -0.0086 & $\chi_4^\prime$ & & \\
 $\alpha_2^{\left({}^3\!E\right)}$ & -54.4 & -5.18 & $\alpha_2^{\left({}^1\!E\right)}$ & -25.3 & -2.54 & $\chi_2$ & 0.71 & 0.059 & $\chi_5$ & -0.592 & -0.0577 \\
 $\beta_2^{\left({}^3\!E\right)}$ & -2.48 & -0.442 & $\beta_2^{\left({}^1\!E\right)}$ & -35.7 & -3.59 & $\chi_2^\prime$ & 0.11 & 0.0010 & $\chi_5^\prime$ & -0.289 & -0.030\\
 $\alpha_1^{\left({}^1\!A_1\right)}$ & 37.33 & 2.695 & & & & $\chi_3$ & -0.66 & -0.0494 & $\chi_6$ & -1.55 & -0.149\\
 $\beta_1^{\left({}^1\!A_1\right)}$ & -22.62 & -2.259 & & & & $\chi_3^\prime$ & -0.367 & -0.049 & $\chi_6^\prime$ & -0.437 & -0.0485\\
 \hline\hline
\end{tabularx}
\end{table*}

\subsection{TDDFT for electron-phonon coupling}
In addition to calculating the SOCs, we require the phonon vibrational overlap $F$ in order to compute the upper ISC rates. We also need the adiabatic couplings (PJT interactions) of the singlet states to estimate the lower ISC rates. The electron-phonon calculations are performed following Ref.~\cite{jin2022vibrationally, jin2023excited}, and we only sketch the procedures here.

The phonon calculations are only performed once on a 215-atom supercell ($3\times 3\times 3$ unit cells) at zero strain. The ground state for phonons is obtained with a slightly different computational setup compared to the DFT calculations presented in the previous subsection. Here we employed a semi-local functional by Perdew, Burke, and Ernzerhof (PBE)~\cite{perdew1996generalized}. Excited states were computed using the TDDFT method~\cite{jin2023excited} within the Tamm-Dancoff approximation.

The phonon modes of the NV center were computed using the frozen phonon approach, with configurations generated from the ${}^3\!A_2$ and ${}^1\!A_1$ states using the Phonopy package~\cite{togo2015first}. Note that phonon modes were also extrapolated to the dilute limit, approximated by a 13,823-atom supercell cell ($12 \times 12 \times 12$ unit cells). The vibrational overlap $F$ was then computed using the Huang-Rhys theory~\cite{jin2021photoluminescence, alkauskas2014first} at 300~K with only $a_1$ phonon modes, as we have discussed in Sec.~S4.1.1 [Fig.~\ref{fig:vibration_overlap}(a)].

The parameters contained in the PJT interaction in Eq.~\ref{eq:PJT_Hamiltonian} were fitted from two (orthogonal) artificial adiabatic potential energy curves crossing the high-symmetry point of ${}^1\!E$ using TDDFT with the PBE functional, also following Ref.~\cite{jin2022vibrationally}. We assumed that both the phonons and these adiabatic coupling parameters remained unchanged under strain/stress.

\subsection{Rate model for spin contrast}
Thus far, we have presented how we computed the ZPLs, VEEs, SOCs, SSCs, phonons, and electron-phonon couplings using different levels of theory. Now we integrate these pieces together to compute the ISC rates and the spin contrast. We start by discussing the ambient condition case to lay a foundation for the stressed cases.

\subsubsection{Ambient conditions}
The upper and lower ISC rates are computed based on Fermi's Golden rule, outlined in Sec.~S4.1.1 and S4.2.1. It is worth mentioning here that we used the experimentally measured energy gaps in determining the vibrational overlaps. The ISC rates near ambient conditions have also been carefully investigated in the past, both from an experimental approach via measuring the excited state lifetimes, and from first principles, as recorded in Table~\ref{tab:rates_ambient}. Let us compare our results with the literature before we address contrast.

The spontaneous emission rate $\Gamma_{\text{rad}}$ has been approximated from the lifetime of the $E_{x,y}$ sublevels of ${}^3\!E$ to be $82.9\pm3.1$ MHz~\cite{goldman2015phonon}, and it is consistent with previous experimental findings~\cite{robledo2011spin, tetienne2012magnetic}. Our calculation predicts this rate to be $81.7$~MHz, which reaches a perfect agreement with Ref.~\cite{goldman2015phonon}. The ISC rates are more involved. Experiments showed that the net upper ISC rate $\Gamma_{\text{ave}} = 50\sim80$~MHz from $|m_s = \pm\rangle$ with a $\frac{\Gamma_{E_{1,2}}}{\Gamma_{A_1}} \sim 0.52$ split amongst the $A_1, E_{1,2}$ sublevels~\cite{goldman2015state}, while $\Gamma_z \lesssim 10$~MHz for the $|m_s=0\rangle$ sublevel. Our prediction, however, is $\Gamma_{\text{ave}} = 6.26$~MHz and $\Gamma_z = 0$. The latter is because we have neglected the PJT effect on the upper ISCs and the former is mainly due to a significant underestimation of $\lambda_\perp$ from our CASSCF computation. This underestimation is also reported in Ref.~\cite{li2024excited}. Finally, the lower ISCs were experimentally found to be orders of magnitude smaller than their upper counterparts, with values $\Gamma_\perp, \Gamma_z \lesssim 3$~MHz, exhibiting little spin selectivity. Our calculations obtain a reasonable agreement with these experiments, with $\Gamma_\perp=0.11$ MHz, $\Gamma_z=0.53$~MHz.

\renewcommand{\arraystretch}{1.8}
\begin{table*}
\caption{Comparison of various rates of the NV center computed from first principles and measured from experiments.}
\label{tab:rates_ambient}
\centering
\setlength{\tabcolsep}{4pt}
\begin{tabularx}{\textwidth}{c|ccccccc}
 \hline\hline
 MHz & $\Gamma_{\text{rad}}$ & $\Gamma_{A_1}$ & $\Gamma_{E_{1,2}}$ & $\Gamma_{\text{ave}} = \frac{1}{4}\left(\Gamma_{A_1} + 2\Gamma_{E_{1,2}}\right)$ & $\Gamma_{E_x}$ & $\Gamma_\perp = \Gamma_\pm + \Gamma_\mp$ & $\Gamma_z$ \\
 \hline
 Exp.~\cite{goldman2015phonon} (6K) & $82.9\pm3.1$ & $100.5\pm3.8$ & $52.3\pm6.5$ & $51.2\pm3.4$ & $3.9\pm1.3$ & & \\
 Exp.~\cite{tetienne2012magnetic} (300K)~\textsuperscript{a} & $63.2\pm4.6$ & & & $60.7\pm6.6$ & $10.8\pm4.1$ & $0.4\pm0.2$ & $0.8\pm0.6$\\
 Exp.~\cite{robledo2011spin} (300K)~\textsuperscript{b} & $65.2\pm1.7$ & & & $79.8\pm1.5$ & $10.5\pm1.5$ & $2.6\pm0.1$ & $3.0\pm0.2$\\
 Comp.~\cite{thiering2017ab, thiering2018theory}~\textsuperscript{c} & & & & $\gg$ Expt. & $\gg$ Expt. & $0.9$ & $4.95$ \\
 Comp.~\cite{li2024excited} (0K)~\textsuperscript{d} & & 2.49 & & & 0.15 & 0.83 & 1.96\\
 Comp.~\cite{jin2025first} (0K) & & $66 \pm 12$ & $31.0 \pm 5.8$ & $32\pm4.2$ & & & \\
 This work (300K) & $81.7$ & $9.25$ & $7.89$ & $6.26$ & $0.0$ & $0.11$ & $0.53$\\
 \hline\hline
\end{tabularx}

\vspace{1mm}
\begin{minipage}{\textwidth}
\raggedright
\hspace{0mm}\textsuperscript{a} These rates are from experiments with an external magnetic field applied at $\theta=74^\circ$ relative to the NV axis.\\
\hspace{0mm}\textsuperscript{b} These rates are extracted from measured excited-state life times and branching probabilities.\\
\hspace{0mm}\textsuperscript{c} No calculated rates are explicitly reported but Thiering et al.~\cite{thiering2017ab} claimed their results were an order of magnitude larger than experimental values.\\
\hspace{0mm}\textsuperscript{d} Ref.~\cite{li2024excited} associated a geometry degeneracy prefactor $g=3$ to the ISC rates. We remove it here to make the comparison consistent with other literature.\\
\end{minipage}
\end{table*}

The rate model requires a few other rates to be complete, including the laser excitation rate and microwave driving rate [Fig.~\ref{fig:optical_cycle}(a)]. These rates affect the ODMR contrast and the linewidth, leading to varying results from different experimental setups. The laser pumping rate is typically on the order of $0.1\sim10$ MHz in ensemble experiments~\cite{robledo2011spin, tetienne2012magnetic}. Since our experiments carefully calibrated the microwave (see Sec.~S1.2.), we expect that a similar fraction of NVs are excited during contrast measurements throughout the pressure range studied, and so the measured contrasts faithfully resemble the behavior of a single NV. Therefore, we neglect microwave power broadening and will use $\Gamma_{\text{exc}} = 0.1$~MHz, $\Gamma_{\text{MW}} = 1$~MHz in all simulations independent of the stress. This produces a good agreement with our measured absolute contrast (with $\sim20\%$ difference). Since we focused on the trend of contrast, we only compared relative contrast change between theory and experiments. The contrast is obtained by solving for the steady state from the rate model, as outlined in Sec.~S4.4. Specifically, we chose $\tau = 10$ ns as one timestep, and simulated a total of $10\sim100$ ms. The steady state condition in the ambient case is obtained before 1 ms, which leads to a $\sim 5\%$ contrast.

\subsubsection{Effects of symmetry-preserving stress}
We now proceed to analyze the contrast under symmetry-preserving stress. Stress alters the ISC rates via both SOC and the vibrational overlap $F$. It is worth emphasizing here that the detuning $\Delta$ between ${}^3\!E$ and ${}^1\!A_1$ in this work was evaluated as $\Delta(\sigma) = \Delta_0 + \delta^{\text{DFT}}_{{}^3\!E}(\sigma) - \delta^{\text{CAS}}_{{}^1\!A_1}(\sigma)$, where $\delta_\Phi^\text{X}(\sigma)$ represents the change of the energy level $\Phi$ relative to the ground state due to stress from X level of theory, and $\Delta_0$ is the gap at ambient conditions which is obtained from experiments. The use of $\delta^{\text{DFT}}_{{}^3\!E}(\sigma)$ here instead of $\delta^{\text{CAS}}_{{}^3\!E}(\sigma)$ is to account for the non-negligible Condon shift of ${}^3\!E$. By comparison, the lower ISCs do not have this concern~\cite{jin2023excited, li2024excited}, and we simply use $\delta^{\text{CAS}}_{{}^1\!E}(\sigma)$.

We observe a common competing effect between the contributions of SOC and of vibrational overlap to the upper ISC rates versus stress. Under hydrostatic stress, $\lambda_\perp$ gets enhanced while $\Delta$ gets enlarged, leading to diminishing $F$. The opposite happens for the [111] case. By observing the trend of $\Gamma_{\text{ave}}$ [Fig.~\ref{fig:cas_symm_preserve}(a-d)], we see that the effect of vibrational overlap dominates over the effect of SOC in the large stress limit. The lower ISC is once again slightly more complicated than its upper counterpart. Nevertheless, we still see that the the vibrational overlap dominates, as hydrostatic stress enlarges $\Sigma$, leading to decreasing $\Gamma_\pm, \Gamma_\mp, \Gamma_z$ [Fig.~\ref{fig:cas_symm_preserve}(g)]. The opposite is true for the [111] stress case.

\begin{figure}
    \centering
    \includegraphics[width=\textwidth]{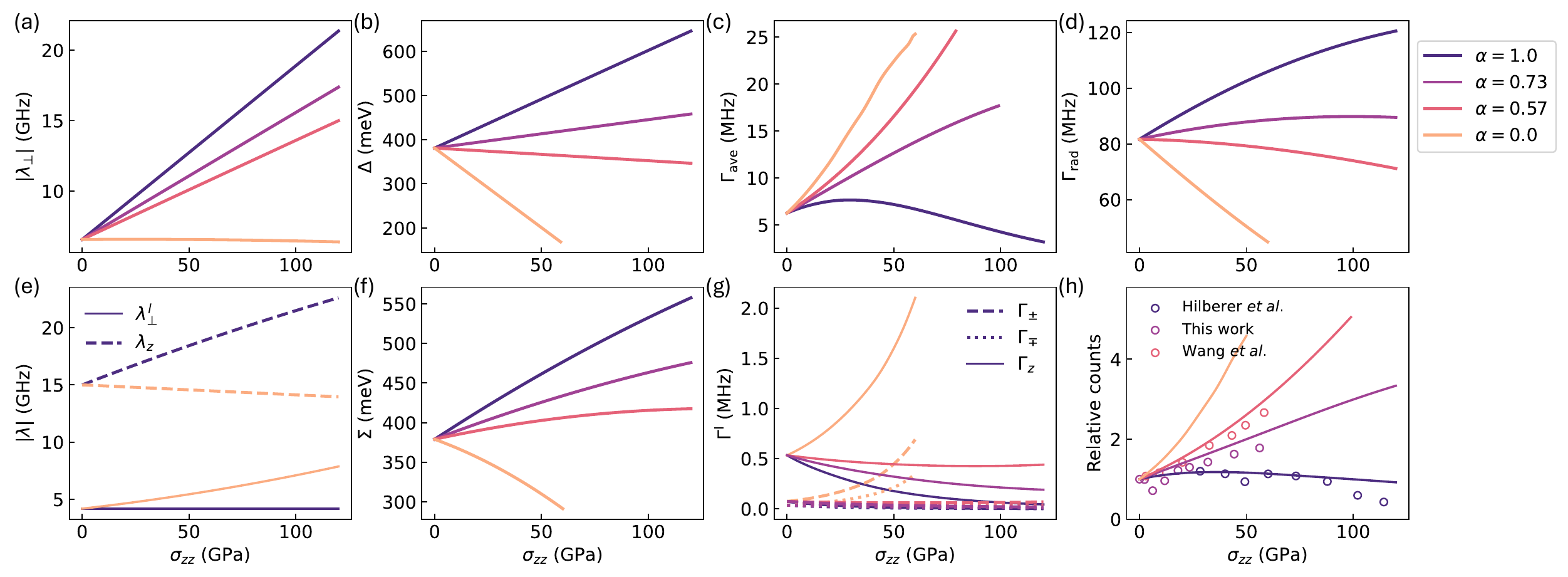}
    \caption{Simulated ISC rates and contrast under symmetry-preserving stress---(a-c) plot the transverse SOC, detuning between ${}^3\!E$ and ${}^1\!A_1$, and the upper ISC rates consisting of these two components. The definition of these rates can be found in Fig.~\ref{fig:optical_cycle}(a). (d) shows the spontaneous emission rates. The lower ISC counterparts are plotted in (e-g), with the relative contrast compared against experimental measurements~\cite{hilberer2023enabling, wang2024imaging} shown in (h).}
    \label{fig:contrast_symm_preserve}
\end{figure}

\begin{figure}
    \centering
    \includegraphics[width=\textwidth]{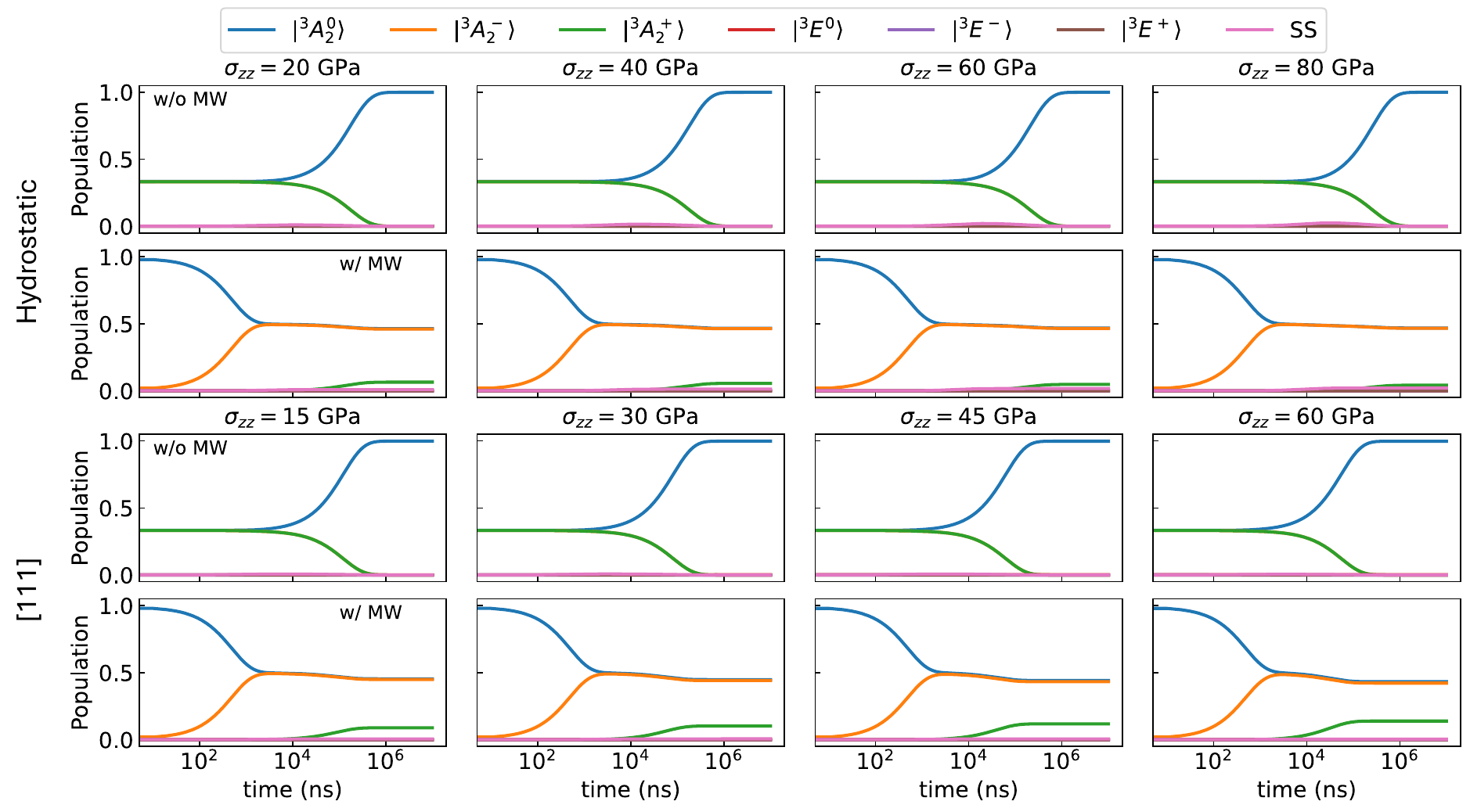}
    \caption{Dynamics of the NV center upon laser excitation under symmetry-preserving stress. The first two rows correspond to hydrostatic and the last two rows correspond to uniaxial [111] stress. The first and third row describe the initialization process w/o microwave (microwave) drive, while the second and last row describe simultaneous driving by the laser and microwave, after the initialization is completed. Each column represents a specific stress condition, denoted by the $\sigma_{zz}$ component.}
    \label{fig:dynamics_symm_preserve}
\end{figure}

We calculated the trends of contrast with stress by solving for the steady state of the rate model, with rates determined under stress as input parameters. The dynamics of the NV's population among the seven states (the singlet states are simplified here as a single shelving state) under hydrostatic and uniaxial [111] stress are plotted in Fig.~\ref{fig:dynamics_symm_preserve}. Four stress conditions ranging from small to large are picked. For each stress, the first row depicts the initialization procedure, with only the laser pumping the system. The NV centers are initialized into the $|m_s=0\rangle$ sublevel regardless of the magnitude of stress, i.e., $n_0 \approx 100\%$, which is within our expectation. The second row depicts the dynamics under co-driving between laser and microwave drive. We see that these results show a qualitatively similar behavior, converging to the steady states before 1 ms.

The relative contrast is shown in Fig.~\ref{fig:contrast_symm_preserve}(h). We observe a strong correlation between the trend of contrast versus stress and that of $\Gamma_{\text{ave}}$ versus stress. For hydrostatic stress, the contrast is stable and slowly decaying after 40 GPa while for uniaxial [111] stress, it increases monotonically. Experimentally, more accurate measurements have been carried out in recent years to calibrate contrast versus stress. Ref.~\cite{hilberer2023enabling} engineered a micro-structured anvil to achieve hydrostaticity $\alpha > 95\%$. Our predictions almost perfectly reproduce the trend of their measured contrasts. Note that reported contrasts beyond 100 GPa decay rapidly due to experimental artifacts~\cite{hilberer2023enabling}. Ref.~\cite{bhattacharyya2024imaging, wang2024imaging} measured the contrast of the (111)-cut anvil (under a mixture of hydrostatic and uniaxial [111] stress, with $\alpha\sim 57\%$), and we also performed measurements in this work with $\alpha\sim 73\%$. To compare between different experiments\footnote{These experiments have applied external magnetic fields along the NV axis that change the spin basis into $|m_s = 0, \pm1\rangle$. It is easy to verify that $B_z$ does not change the rate model or ISC rates, so we did not consider the $B$-field in our simulations and the comparisons between our calculations and experiments are still valid.}, we normalize all measured contrasts to their respective (near) ambient counterpart. Our simulations correctly capture qualitative trends, with the predicted contrast enhancements slightly overestimated. These small deviations could be due to the other three non-[111] groups of NV centers, which we will comment on in more detail in Sec.~\ref{sec:errors}.

\subsubsection{Effects of symmetry-breaking stress}
\begin{figure}
    \centering
    \includegraphics[width=\textwidth]{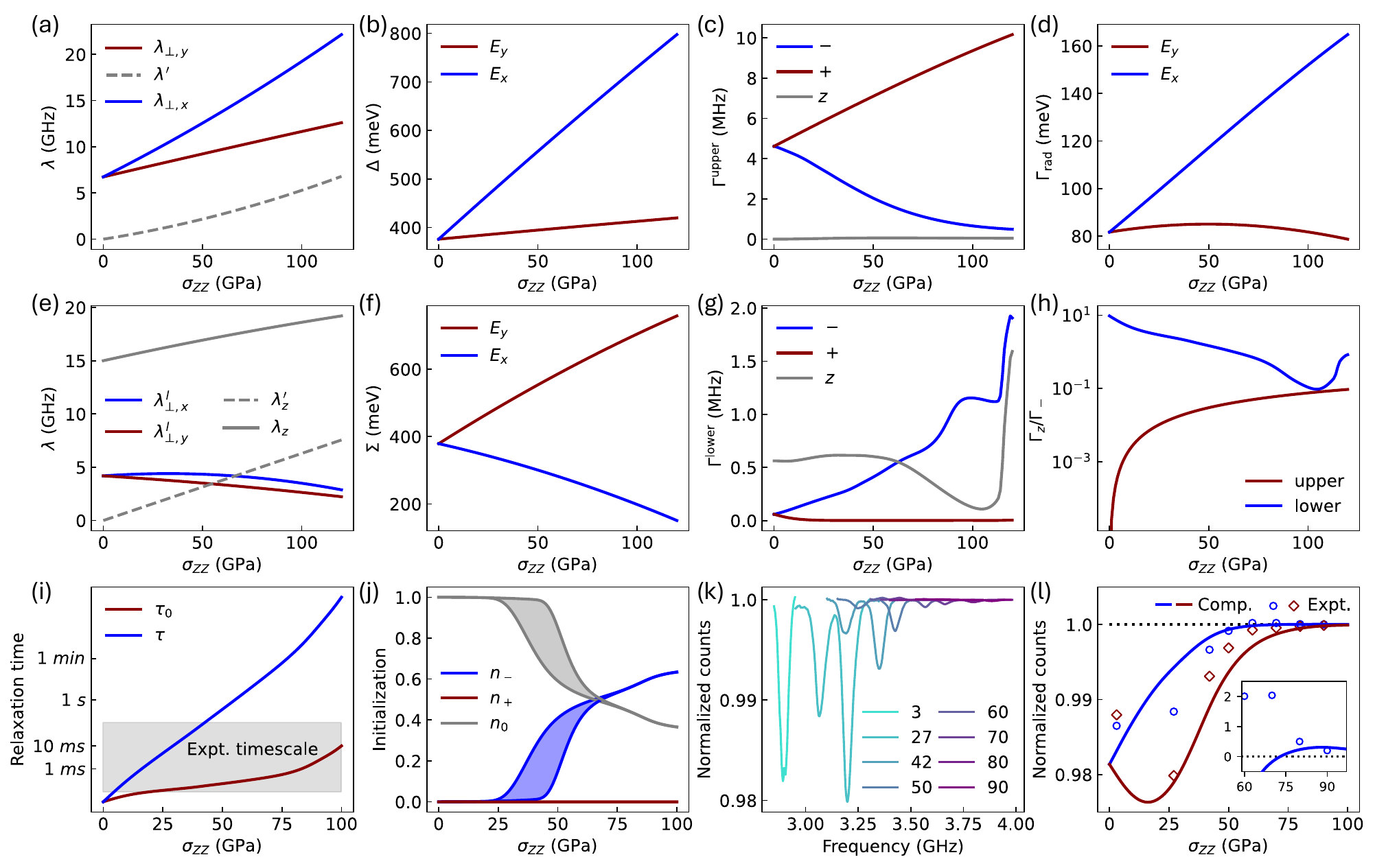}
    \caption{Simulated ISC rates and contrast under a mixture of hydrostatic and uniaxial [100] stresses, with hydrostaticity $\alpha\sim57\%$. (a-c) plots the transverse SOC (with a new matrix element $\lambda'$ for $\Gamma^\text{upper}_z$ denoted by the grey dashed curve), detuning between ${}^3\!E$ and ${}^1\!A_1$ (with the blue (red) curve denoting the $E_x$ ($E_y$) branch), and the upper ISC rates consisting of these two components. The definition of these rates can be found in Fig.~\ref{fig:optical_cycle}(b). (d) plots the spontaneous emission rates. The lower ISC counterparts are plotted in (e-g) (with another new matrix element $\lambda_z'$ \textit{destructively interfering} with $\lambda_z$ denoted by grey dashed line). (h) depicts $\Gamma_z/\Gamma_-$ for the upper and lower ISCs, which we used as an indicator of positive contrast at the steady state. 
    (i-l) presents other relevant quantities for the simulation of contrast, including a comparison of two estimated timescales (red and blue curves) for obtaining the steady-state solutions against typical experimental values (grey shaded region); the polarization outcome within these experimental timescales; and finally a comparison of contrast between experimental measurements~\cite{bhattacharyya2024imaging} (with numbers in the legend denoting the loading stress in units of GPa) and our simulations.}
    \label{fig:contrast_symm_break}
\end{figure}

The ISC rates under symmetry-breaking stress can be computed according to Sec.~S4.1.2 and S4.2.2, keeping only the leading order terms. Recent experiments on (100)-cut diamond anvil observed a mixture of roughly $\sim57\%$ hydrostatic stress and $\sim43\%$ uniaxial [100] stress~\cite{bhattacharyya2024imaging}. Therefore, we use this stress condition in our calculations. Note also that SSCs have been ignored in our simulations due to their magnitude being small\footnote{The spin mixing parameter $|\beta|^2$ is found to be no larger than 2\% after extrapolation to $\sigma_{ZZ}\sim120$ GPa experimental stress.}. The ISC rates are shown in Fig.~\ref{fig:contrast_symm_break}(c,g).

The upper ISC [Fig.~\ref{fig:optical_cycle}(a)] shows qualitatively different behavior for the three spin sublevels [Fig.~\ref{fig:contrast_symm_break}(c)]. $\Gamma_+^\text{upper}$ monotonically increases with stress, not only because its SOC is enhanced by stress, but also because it corresponds to a transition from the lower branch $E_y$, resulting in a slower decaying of the phonon vibrational overlap. $\Gamma_-^\text{upper}$ behaves oppositely: despite the enhancement of its corresponding SOC, $F(\Delta_x)$ monotonically decreases due to growing $\Delta_x$. Meanwhile, $\Gamma_z^\text{upper}$ slowly increases from zero [Fig.~\ref{fig:optical_cycle}(b)]. We also see that $|m_s=0\rangle$ remains as the brightest spin.

The lower ISC, however, exhibits counterintuitive behaviors. As we discussed in Sec.~S4.2, PJT interactions play a crucial role in the lower ISC processes. Due to increasing [100] stresses, the ${}^1\!E$ sublevels split and the effects of the PJT interaction are vanishingly small. The lower branch $E_x$ becomes more relevant to the optical cycle due to its progressively shrinking gap with respect to ${}^3\!A_2$.
This explains the reduction in $\Gamma_+^\text{lower}$, and the enhancement of $\Gamma_-^\text{lower}$, since otherwise $\lambda_{\perp, x/y}^l$ remain rather stable under stress. 
The most counter-intuitive behavior appears in $\Gamma_z^\text{lower}$, which slowly increases up to $\sim50$ GPa, and then declines until $\sim110$ GPa, after which it increases again. 
We re-emphasize here that there exist two SOCs for $\Gamma_z^\text{lower}$ from vibronic wavefunctions that transform as $E$, namely $\lambda_z$ and $\lambda_z'$, as we discussed in Sec.~S4.2.2. 
We find that $2d_i\lambda_z$ and $\frac{f_i}{\sqrt{2}}\lambda_z'$ in Eqs.~\ref{eq:low_isc_gammaz_symmetrybreaking_e} have opposite signs and they therefore \emph{destructively interfere} with each other. At small stresses, $\lambda_z$ dominates and $\Gamma_z^\text{lower}$ increases. As stress increases, $\frac{f_i}{\sqrt{2}}\lambda_z'$ becomes comparable to $2d_i\lambda_z$, so the destructive interference significantly reduces the transition rate. In the large stress limit, $\lambda_z'$ finally dominates and $\Gamma_z^\text{lower}$ becomes comparable to $\Gamma_-^\text{lower}$.

One of the puzzles from experiments is that contrast inversion was observed at loading pressure $\sigma_{ZZ} \geq 60$ GPa, as shown in Fig.~S1(b) of Ref.~\cite{bhattacharyya2024imaging}.
We have outlined the underlying mechanism for ODMR contrast in Sec.~S4.4, and one possible explanation for positive contrast is that the brightest spin is no longer the polarized-to state in the ground-state manifold. 
[100] stress mainly drives population transfer between $|m_s=0\rangle$ and $|m_s=-\rangle$. 
To see which spin gets polarized into at the steady state, we can compare the ratio $\frac{\Gamma_z}{\Gamma_-}$ for the upper and lower ISCs. $\left(\frac{\Gamma_z}{\Gamma_-}\right)_{\text{upper}} > \left(\frac{\Gamma_z}{\Gamma_-}\right)_{\text{lower}}$ indicates polarization into $|m_s=-\rangle$. However, we do not see such a crossing from our calculations [Fig.~\ref{fig:contrast_symm_break}(h)], as the two ratios only get very close around $\sigma_{ZZ}\sim100$~GPa.

However, one must ask whether experiments truly reach the steady state. We may estimate how long relaxation should take. In our optical cycle model, only the upper branch of ${}^3\!E$ allows the $|m_s=0, -\rangle$ spin states to undergo ISCs [Fig.~\ref{fig:optical_cycle}(b)]. Therefore, the timescale for relaxation can be roughly estimated by,
\begin{equation} \label{eq:relaxation_timescale_0}
    \tau_0 \sim \frac{1}{\Gamma_{\text{rds}} \left[\left(\frac{\Gamma_z}{\Gamma_-}\right)_{\text{lower}} - \left(\frac{\Gamma_z}{\Gamma_-}\right)_{\text{upper}}\right] \left(\frac{\Gamma^\text{u}_- + \Gamma^\text{u}_z}{2\Gamma_{\text{rad}, x} + \Gamma^\text{u}_- + \Gamma^\text{u}_z}\right)},
\end{equation}
assuming that the two branches of ${}^3\!E$ are equally populated. $\Gamma_{\text{rds}}$ stands for rate-determining step rate (either the laser excitation rate or lower ISC rates), and $\frac{\Gamma^\text{u}_- + \Gamma^\text{u}_z}{2\Gamma_{\text{rad}} + \Gamma^\text{u}_- + \Gamma^\text{u}_z}$ is the ISC branching ratio. This estimation still lies within typical experimental measurement timescales [Fig.~\ref{fig:contrast_symm_break}(i)]. However, phonon-assisted downward transitions~\cite{ulbricht2016jahn, ernst2023modeling} in ${}^3\!E$ could lead to the lower branch $E_y$ being favored by the excited state population. If we adopt an ad-hoc Boltzmann distribution of the excited state population among the two branches, it would significantly increase the estimated relaxation timescale by an exponential prefactor
\begin{equation} \label{eq:relaxation_timescale}
    \tau = \tau_0 \exp\left(\frac{\Pi_x}{k_B T}\right),
\end{equation} 
making it beyond typical experimental timescales past a loading stress of $\sigma_{ZZ} \sim 50$ GPa.

\begin{figure}
    \centering
    \includegraphics[width=\textwidth]{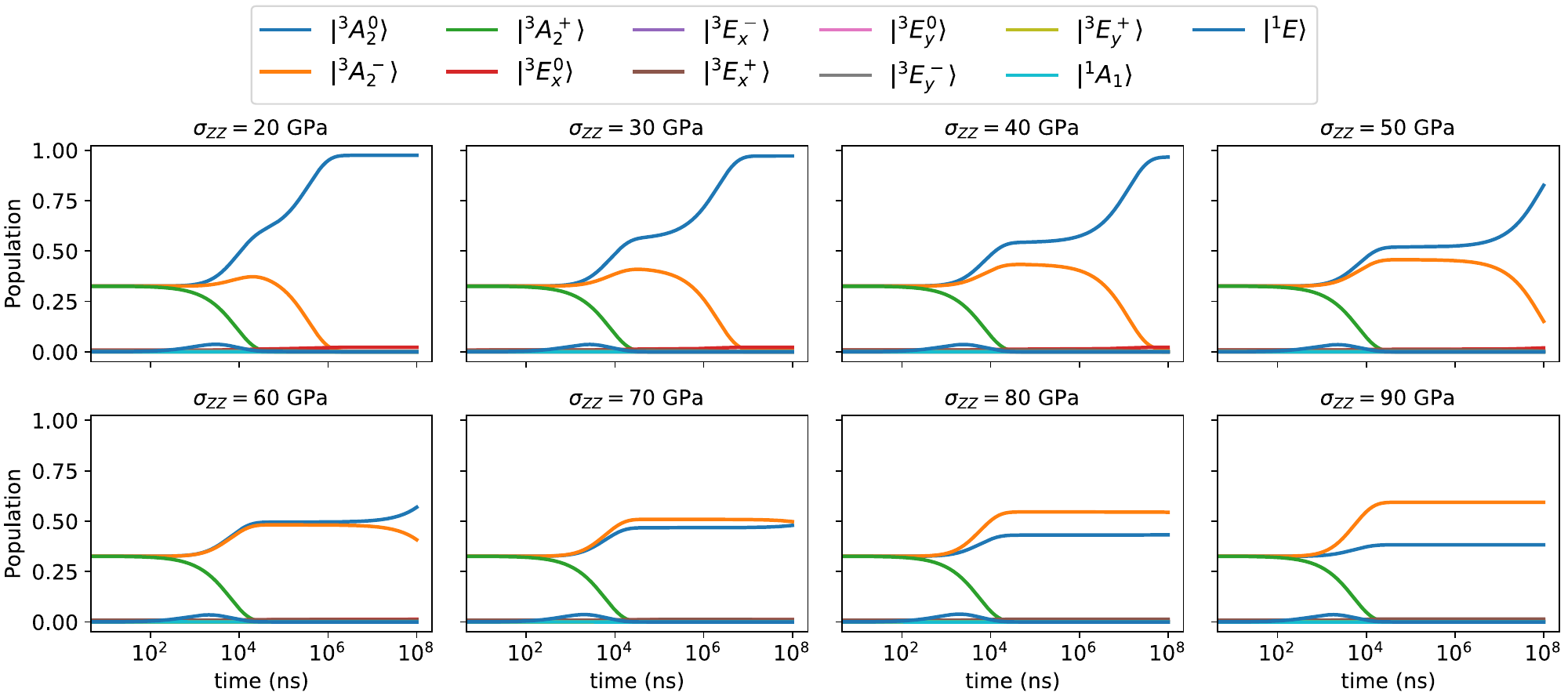}
    \caption{Dynamics of the NV center upon laser excitation under experimental [100] stress. Only the initialization process is shown, running $10^7$ time steps with 10 ns per timestep. Within this timescale, only NV with loading pressure $\sigma_{ZZ} \leq 40$ GPa can reach steady states, because the relaxation time scales exponentially with loading pressure. NVs with a loading pressure $\sigma_{ZZ} \geq 70$ GPa only arrive at metastable states.}
    \label{fig:dynamics_symm_break}
\end{figure}

We simulate the population dynamics under experimental [100] stress, with polarization shown in Fig.~\ref{fig:dynamics_symm_break}. We run a total of $10^7$ timesteps with 10 ns per timestep. We see that NVs with loading pressure $\sigma_{ZZ} \lesssim 40$~GPa can reach steady states within simulated timescales, while those with pressure $\sigma_{ZZ} \gtrsim 70$~GPa arrive only at metastable states, with $n_- > n_0$. This is within our expectation due to the exponential scaling of relaxation times. This polarization serves as the basis for the positive contrast observed from subsequent measurements of photoluminescence intensity with on-resonance microwaves. 
Therefore, we see that the positive contrast phenomenon under [100] stress, based on our model, is a reflection of meta-stable states rather than steady states. If the experimental timescales can be extended to arbitrary length, we would not expect to see positive contrast. 
We also acknowledge that our conclusion is sensitive to the inaccuracies of these calculations. We comment on the approximations made in our simulations and how they would affect our conclusions in the next section.

\section{Comments on errors} \label{sec:errors}
As a concluding remark, we summarize all the approximations made in this work and discuss their potential impact on the accuracy of our results and conclusions.

\subsection{Model-level approximations}
We begin by addressing the model-level approximations. Our optical cycle model (Fig.~\ref{fig:optical_cycle}) describes ISC processes approximately, by only retaining the leading order transitions. Under symmetry-preserving stress, we have neglected the DJT effect on the ${}^3\!E$ manifolds. This effect, as we have discussed in Sec.~S3.4.1, serves to mix the original $A_1, A_2, E_{1,2}$ states, enabling $A_2$ to inter-system cross as well. However, we note that these couplings only happen within the $|m_s = \pm\rangle$ sector, and would not induce $\Gamma_z^\text{upper}$. We therefore claim that neglecting the DJT effect would not qualitatively change our results or conclusions.

Another JT effect we have ignored is the impact of the PJT interactions of the singlet states on the upper ISCs. We have extensively discussed in previous sections how the PJT effect affects the ${}^1\!E$ states and enables the lower ISCs. However, we have not considered how it might affect ${}^1\!A_1$. Mixing ${}^1\!E$ into ${}^1\!A_1$ allows a direct coupling between $E_{1,2}$ sublevels of ${}^3\!E$ and the singlet states, which is a qualitatively second-order ISC route. In addition, it also enables a third-order ISC from $E_{x, y}$, which is the reason why experimentally measured $\Gamma_z^\text{upper}$ is non-zero\footnote{$E_{x,y}$ couples to the higher ${}^1\!E'$ states [Fig.~\ref{fig:spin_orbit_coupling}(a)], which couples to the lower lying ${}^1\!E$ states via a multi-configurational effect, and the latter finally couple to ${}^1\!A_1$ through the PJT interaction.}. This affects the optical initialization process, since any non-zero $\Gamma_z^\text{upper}$ would break the otherwise perfect spin polarization. 
Strictly speaking, the PJT effect only slightly perturbs ${}^1\!A_1$~\cite{thiering2018theory, jin2022vibrationally}, as shown in Fig.~\ref{fig:Jahn_Teller}(c). So, it is reasonable to ignore the effect of the PJT interaction on ${}^1\!A_1$, at least under symmetry-preserving stress.

The case of symmetry-breaking stress is, as always, more complicated. Considering the case of [100] stress as an example, the $y$ branches of both ${}^3\!E$ and ${}^1\!E$ manifolds move closer to ${}^1\!A_1$. Therefore, the upper ISC from $E_y$ would benefit from a larger vibrational overlap. The PJT interaction would enable $\Gamma_z^\text{upper}, \Gamma_-^\text{upper}$ from the $E_y$ branch [not shown in Fig.~\ref{fig:optical_cycle}(b)] that is complementary to those from $E_x$. This would introduce another ratio $\frac{\Gamma_z}{\Gamma_-}$ from the upper ISC, which is important for determining the polarization in the steady state solution. 
We have ignored these transitions since the effective SOC from the PJT effect will be small. We leave the computation of these rates for future investigations.

Finally, we have not considered the effect of SSCs under symmetry-breaking stress, due to their magnitude being small. As we discussed in Sec.~S3.3, $\Pi_x$ stress mixes $|m_s = 0\rangle$ and $|m_s=-\rangle$. Interestingly, it is these two spins that compte for dominance in the initialization process. From our calculations shown in Fig.~\ref{fig:cas_symm_break}, we see that the SSC is most pronounced in the $y$ branch of ${}^3\!E$, followed by the ground state, and that it is minuscule in the $x$ branch of ${}^3\!E$ in the large stress limit. The effect of SSC tends to average $\Gamma_-$ and $\Gamma_z$. In the current optical cycle model, including SSC would only increase $\left(\frac{\Gamma_z}{\Gamma_-}\right)_\text{lower}$, slightly accelerating the initialization process, but it would not change the outcome. We re-emphasize that the magnitude of SSCs under [100] stress is small, therefore the impact on our results and conclusions is negligible.

\subsection{Computation-level approximation}
Next, we examine the computational approximations made in this work. In determining the SOCs, we adopted the CASSCF method applied to the ground state geometries of the NV, for which we have assumed i). the Condon approximation that atoms do not move even in the excited states and ii). the ground and excited states' wavefunctions consist of the same set of molecular orbitals. The Herzberg-Teller effect has therefore been neglected in SOCs, and Ref.~\cite{jin2025first} shows how much of an error such a treatment could lead to for the ISC rates.

Besides, the CASSCF approach significantly underestimates the value of $\lambda_\perp$ (Table~\ref{tab:VEE_ambient}). We suspect that, by comparing the weights of different configurations in the many-body wavefunctions from Ref.~\cite{li2024excited} and Ref.~\cite{jin2025first}, this underestimation stems from the large weight of the $e_x^2 e_y^2$ configuration in the ${}^1\!A_1$ wavefunction from CASSCF\footnote{For CASSCF, the weight is $\sim20\%$~\cite{li2024excited}, while for quantum defect embedding theory it is only $\sim2.5\%$~\cite{jin2025first}.}. Nevertheless, we are more concerned with the susceptibilities of VEEs and SOCs. The susceptibilities of VEEs can be slightly overestimated from CASSCF, by comparing the very few available experimental and computational references~\cite{davies1976optical, lopez2024quantum}. This could lead to the variation of ISC rates (i.e. $\delta \Gamma_\text{ISC}$) being overestimated under stress. 
Comparatively, the potential errors associated with the susceptibilities of SOCs are difficult to discuss, as there are no available references to compare them to. These susceptibility errors would not change our explanations about how contrast varies under symmetry-preserving stress, but could have an impact on our conclusions about the positive contrast phenomenon under [100] stress, since comparing $\left(\frac{\Gamma_z}{\Gamma_-}\right)_\text{lower}$ and $\left(\frac{\Gamma_z}{\Gamma_-}\right)_\text{upper}$ requires quantitative rather than just qualitative accuracy. We leave refining the numerical estimation of these susceptibilities for future work.

Another approximation we made is for the phonon vibrational overlap of the lower ISCs. We adopted a two-effective-phonon approximation for the description of the PJT effect, and further approximated the summation of phonons $\sum_j \left|\langle \chi_j|\chi_i(\Gamma)\rangle\right|^2 \delta(\Sigma - n_j\hbar\omega_e)$ to be the phonon-occupation-number-dependent spectral density $S_E^{(n_i)}(\Sigma)$ by broadening the $\delta$ functions, regardless of the irreducible representation $\Gamma$ of the initial vibronic phonon wavefunctions $\chi_i(\Gamma)$. This leads to an overestimation of the vibrational overlap and therefore overestimation of the lower ISC rates.

Apart from the ISC rates, we also approximated the phonon-induced transition rates within the ${}^3\!E/{}^1\!E$ branches by a Boltzmann distribution of population under [100] stress. This approximation is also crucial for our explanations for positive contrast, since it is this approximation that caused the relaxation time for steady states to scale exponentially with stress and surpass typical experiment timescales.

Finally, we have been using rate equations to solve for the NV ODMR dynamics rather than the master equation. Ref.~\cite{ernst2023modeling} has demonstrated that at room temperature, rate equation results gave a perfect agreement with results derived from the master equation. Therefore, we solely employed rate equations in our work. However, we also acknowledge that the ODMR linewidth cannot be considered within the rate model. As we have discussed in Sec.~S1.2, the broadening comes from a variety of sources, including microwave broadening, dephasing, and local charge noise. We leave the exploration of linewidth using the master equation approach for future investigations.

\subsection{Simulations' deviation from experimental setup}
A majority of the attention in this work has been devoted to the contrast change under [111] stress. We attribute the enhancement of contrast with [111] stress to the reduction of detuning between the ${}^3\!E$ and ${}^1\!A_1$ states. However, one key difference between our simulations and the experiments is worth discussing---the non-[111] NVs have been ignored in our simulations. In a perfect (111)-cut diamond anvil, there are effectively two possible NV orientations\footnote{Strictly speaking, there are 8 orientations of NV. But, one can easily verify there are only two meaningfully distinct NVs in the (111)-cut.}, namely the [111] and the non-[111] orientations, with a population ratio of $\sim1:3$. By applying a magnetic field in the $Z$ direction, (which is also the $z$ direction of [111] NVs), we are able to distinguish the [111] NV's ODMR peaks from the peaks of non-[111] groups. 
However, the contrast of the [111] NV would still be affected by those non-[111] NVs, via their spontaneous emission shifting the background signal of ODMR spectrum. So, how is ignoring the non-[111] NVs justified? First, the ratio of fluorescence contributions between these two groups of NVs is in fact $3:5$, and not the more intuitive but smaller $1:3$. This is because the laser is only applied in the $Z$ direction, and its $E$-field projection (which drives optical excitation) in the normal plane of the non-[111] NVs' axes is $2/3$ smaller. So the impact from the non-[111] NVs is not as large as it may seem. Second, the experimental stress environment is always a mixture of hydrostatic and uniaxial stress, and hydrostatic portion is typically dominating\footnote{Upon pressurization, the non-[111] NVs are under effectively $\left[\overline{11}1\right]$ stress, breaking the $C_{3v}$ symmetry. So the two orbital branches have different spontaneous emission rates.}, which would partially reconcile the impact of non-[111] NVs. 
Because of the above two reasons, we believe the enhancement of [111] NV contrast should mainly be credited to its intrinsic properties under [111] stress. Taking the non-[111] NVs into consideration would slightly slow the enhancement of [111] NV's contrast, leading to a better agreement between our simulations and experiments~\cite{wang2024imaging}.

\bibliography{bibliography}